\documentclass[12pt]{elsart}
\usepackage{graphicx}
\usepackage{amssymb}
\usepackage{amsmath}
\usepackage{hyperref}

\journal{Annals of Physics}

\newcommand{\tr}{\mathop{\rm tr}\nolimits}

\begin{document}
%%%%%%%%%%%%%%%%%%%%%%%%%%%%%%%%%%%%%%%%%%%%%%%%%%%%%%%%%%%%%%%%%%%%%%%%%%%%%%%
%  TITLE AND ABSTRACT
%
%%%%%%%%%%%%%%%%%%%%%%%%%%%%%%%%%%%%%%%%%%%%%%%%%%%%%%%%%%%%%%%%%%%%%%%%%%%%%%%
\begin{frontmatter}
\title{The instanton vacuum of generalized $CP^{N-1}$ models}
\author[A]{A.M.M.~Pruisken\corauthref{lc1}}
\corauth[lc1]{Corresponding author. Fax: +31 20 525 57 88}
\ead{pruisken@science.uva.nl}
\author{and\,\,\,}
\author[A,I]{I.S.~Burmistrov}
\address[A]{Institute for Theoretical Physics, University of
Amsterdam, Valckenierstraat 65, 1018 XE Amsterdam, The
Netherlands}
\address[I]{L.D. Landau Institute for Theoretical
Physics, Kosygina str. 2, 117940 Moscow, Russia}
\ead{burmi@itp.ac.ru }

\begin{abstract}
It has recently been pointed out that the existence of massless
chiral edge excitations has important strong coupling consequences
for the topological concept of an instanton vacuum. In the first
part of this paper we elaborate on the effective action for ``edge
excitations'' in the Grassmannian $U(m+n)/U(m) \times U(n)$
non-linear sigma model in the presence of the $\theta$ term. This
effective action contains complete information on the low energy
dynamics of the system and defines the renormalization of the
theory in an unambiguous manner. In the second part of this paper
we revisit the instanton methodology and embark on the
non-perturbative aspects of the renormalization group including
the anomalous dimension of mass terms. The non-perturbative
corrections to both the $\beta$ and $\gamma$ functions are
obtained while avoiding the technical difficulties associated with
the idea of {\em constrained} instantons. In the final part of
this paper we present the detailed consequences of our
computations for the quantum critical behavior at $\theta = \pi$.
In the range $0 \leq m,n \lesssim 1$ we find quantum critical
behavior with exponents that vary continuously with varying values
of $m$ and $n$. Our results display a smooth interpolation between
the physically very different theories with $m=n=0$ (disordered
electron gas, quantum Hall effect) and $m=n=1$ ($O(3)$ non-linear
sigma model, quantum spin chains) respectively, in which cases the
critical indices are known from other sources. We conclude that
instantons provide not only a {\em qualitative} assessment of the
singularity structure of the theory as a whole, but also
remarkably accurate {\em numerical} estimates of the quantum
critical details (critical indices) at $\theta = \pi$ for varying
values of $m$ and $n$.
\end{abstract}


\end{frontmatter}
\clearpage\tableofcontents \clearpage
%%%%%%%%%%%%%%%%%%%%%%%%%%%%%%%%%%%%%%%%%%%%%%%%%%%%%%%%%%%%%%%%%%%%%%%%%%%%%%%%%%%%%%
%   INTRODUCTION
%
%%%%%%%%%%%%%%%%%%%%%%%%%%%%%%%%%%%%%%%%%%%%%%%%%%%%%%%%%%%%%%%%%%%%%%%%%%%%%%%%%%%%%%
\section{\label{Intro}Introduction}

%====================================================================================
\subsection{\label{Intro.SU} Super universality}

The quantum Hall effect has remained one of the most beautiful and
outstanding experimental realizations of the {\em instanton
vacuum} concept in non-linear sigma models~\cite{QHE,Pruisken4}.
Although originally introduced in the context of Anderson
(de-)localization in strong magnetic
fields~\cite{LevineLibbyPruisken,Pruisken1}, the topological ideas
in quantum field theory have mainly been extended in recent years
to include a range of physical phenomena and applications that are
much richer and broader than what was previously anticipated. What
remarkably emerges is that the aforementioned topological concepts
retain their significance also when the Coulomb interaction
between the disordered electrons is taken into
account~\cite{PruiskenBaranov}. A detailed understanding of
interaction effects is vitally important not only for conducting
experiments on {\em quantum criticality} of the plateau
transitions~\cite{WeiTsuiPalaanenPruisken,SchaijkdeVisserOlsthoornWeiPruisken},
but also for the long standing quest for a {\em unified action}
that incorporates the low energy dynamics of both the integral and
fractional quantum Hall
states~\cite{Unify1,Unify2,Unify3,Unify4,Unify5}.

Perhaps the most profound advancement in the field has been the
idea which says that the instanton vacuum generally displays {\em
massless chiral edge
excitations}~\cite{Unify3,PruiskenBaranovVoropaev,PruiskenBaranovBurmistrov}.
These provide the resolution of the many {\em strong coupling}
problems that historically have been associated with the instanton
vacuum concept in scale invariant
theories~\cite{LargeN2,LargeN3,LargeN4}. The physical significance
of the {\em edge} is most clearly demonstrated by the fact that
the instanton vacuum theory, unlike the phenomenological
approaches to the fractional quantum Hall effect based on Chern
Simons gauge theory~\cite{Wilczek}, can be used to derive from
first principles the complete {\em Luttinger liquid} theory of
edge excitations in disordered abelian quantum Hall
systems~\cite{Unify3}. Along with the physics of the {\em edge}
came the important general statement which says that the {\em
fundamental features} of the quantum Hall effect should all be
regarded as {\em super universal} features of the topological
concept of an instanton vacuum, i.e. independent of the number of
field components in the theory~\cite{PruiskenBaranovVoropaev}.

{\em Super universality} includes not only the appearance of {\em
massless chiral edge excitations} but also the existence of {\em
gapless bulk excitations} at $\theta=\pi$ in general as well as
the dynamic generation of {\em robust topological quantum numbers}
that explain the {\em precision} and {\em observability} of the
quantum Hall
effect~\cite{PruiskenBaranovVoropaev,PruiskenBaranovBurmistrov}.
Moreover, the previously unrecognized concept of {\em super
universality} provides the basic answer to the historical
controversies on such fundamental issues as the {\em quantization
of topological charge}, the exact significance of having {\em
discrete topological sectors} in the theory, the precise meaning
of {\em instantons} and {\em instanton
gases}~\cite{LargeN2,LargeN3}, the validity of the {\em replica
method}~\cite{RM} etc. etc. One can now state that many of these
historical problems arose because of a complete lack of any
physical assessment of the theory, both in general and in more
specific cases such as the exactly solvable large $N$ expansion of
the $CP^{N-1}$ model.

%======================================================================
\subsection{\label{Intro.BFM} The background field methodology}

In 1987, one of the authors introduced a renormalization group
scheme in replica field theory (non-linear sigma model) that was
specifically designed for the purpose of extracting the
non-perturbative features of the quantum Hall regime from the
instanton angle $\theta$~\cite{Pruisken2,Pruisken3}. This
procedure was motivated, to a large extend, by the Kubo formalism
for the conductances which, in turn, has a natural translation in
quantum field theory, namely the {\em background field
methodology}.

Generally speaking, the background field procedure expresses the
renormalization of the theory in terms of the {\em response} of
the system to a change in the {\em boundary conditions}. It has
turned out that this procedure has a quite general significance in
asymptotically free field theory that is not limited to {\em
replica limits} and condensed matter applications alone. It
actually provides a general, conceptual framework for the
understanding of the strong coupling aspects of the theory that
otherwise remain inaccessible. For example, such non-perturbative
features like {\em dynamic mass generation} are in one-to-one
correspondence with the renormalized parameters of the theory
since they are, by construction, a probe for the {\em sensitivity}
of the system to a change in the {\em boundary conditions}.

The background field procedure has been particularly illuminating
as far as the perturbative aspects of the renormalization group is
concerned. First of all, it is the appropriate generalization of
Thouless' ideas on localization~\cite{Thouless}, indicating that
the physical objectives in condensed matter theory and those in
asymptotically free field theory are in many ways the same.
Secondly, it provides certain technical advantages in actual
computations and yields more relevant results. For example, it has
led to an exact solution (in the context of an $\epsilon$
expansion) of the the AC conductivity in the {\em mobility edge}
problem or {\em metal-insulator} problem in $2+\epsilon$
dimensions~\cite{Unify5}. The physical significance of these
results is not limited, once more, to the theory in the {\em
replica limit} alone. They teach us something quite general about
the statistical mechanics of the Goldstone phase in low
dimensions.

%=================================================================
\subsection{\label{Intro.SCP} The strong coupling problem}

With hindsight one can say that the background field
procedure~\cite{Pruisken3}, as it stood for a very long time, did
not provide the complete conceptual framework that is necessary
for general understanding of the quantum Hall effect or, for that
matter, the instanton vacuum concept in quantum field theory.
Unlike the conventional theory where the precise details of the
``edge'' do not play any significant role, in the presence of the
instanton parameter $\theta$ the choice in the boundary conditions
suddenly becomes an all important conceptual issue that is
directly related to the definition of a fundamental quantity in
the theory, the {\em Hall conductance}.

The physical significance of {\em boundary conditions} in this
problem has been an annoying and long standing puzzle that has
fundamentally complicated the development of a microscopic theory
of the quantum Hall effect~\cite{QHE}. In most places in the
literature this problem has been ignored
altogether~\cite{LargeN2,LargeN3,LargeN4}. In several other cases,
however, it has led to a mishandling of the
theory~\cite{Mishandling}.

The discovery of {\em super universality} in non-linear sigma
models~\cite{PruiskenBaranovVoropaevN} has provided the physical
clarity that previously was lacking. The existence of {\em
massless chiral edge excitations}, well known in studies of
quantum Hall systems, implies that the instanton vacuum concept
generally supports distinctly different modes of excitation, those
describing the {\em bulk} of the system and those associated with
the {\em edge}. It has turned out that each of these modes has a
fundamentally different topological significance, and a completely
different behavior under the action of the renormalization group.

The existence of massless chiral edge excitations forces one to
develop a general understanding of the instanton vacuum concept
that is in many ways very different from the conventional ideas
and expectations in the field. It turns out that most of the
physics of the problem emerges by asking how the two dimensional
{\em bulk} modes and one dimensional {\em edge} modes can be
separated and studied individually. At the same time, a
distinction ought to be made between the physical {\em
observables} that are defined by the {\em bulk} of the system and
those that are associated with the {\em edge}.

%..............................................................
\subsubsection{\label{Intro.SCP.EM} Effective action for the edge modes}

The remarkable thing about the problem with {\em edge} modes is
that it {\em automatically} provides all the fundamental
quantities and topological concepts that are necessary to describe
and understand the low energy dynamics of the system. Much of the
resolution to the problem resides in the fact that the theory can
generally be written in terms of {\em bulk} field variables that
are embedded in a {\em background} of the topologically different
{\em edge} field configurations. This permits one to formulate an
{\em effective action} for the {\em edge} modes, obtained by
formally eliminating all the {\em bulk} degrees of freedom from
the
theory~\cite{PruiskenBaranovVoropaev,PruiskenBaranovBurmistrov,PruiskenBaranovVoropaevN}.
It now turns out that the {\em effective action} procedure for the
{\em edge} field variables proceeds along exactly the same lines
as the {\em background field} methodology~\cite{Pruisken3} that
was previously introduced for entirely different physical reasons!
This remarkable coincidence has a deep physical significance and
far reaching physical consequences. In fact, the many different
aspects of the problem (Kubo formulae, renormalization, edge
currents etc.) as well as the various disconnected pieces of the
puzzle (boundary conditions, quantization of topological charge,
quantum Hall effect etc.) now become simultaneously important.
They all come together as fundamental and distinctly different
aspects of a single new concept in the problem that has emerged
from the instanton vacuum itself, the effective action for the
massless chiral edge
excitations~\cite{PruiskenBaranovVoropaev,PruiskenBaranovBurmistrov,PruiskenBaranovVoropaevN}.

%.............................................................
\subsubsection{\label{Intro.SCP.QHE} The quantum Hall effect}

The effective action for massless edge excitations has direct
consequences for the strong coupling behavior of the theory that
previously remained concealed. It essentially tells us how the
instanton vacuum {\em dynamically generates} the aforementioned
{\em super universal} features of the quantum Hall effect, in the
limit of large distances.

\begin{figure}
\centerline{\includegraphics[width=110mm]{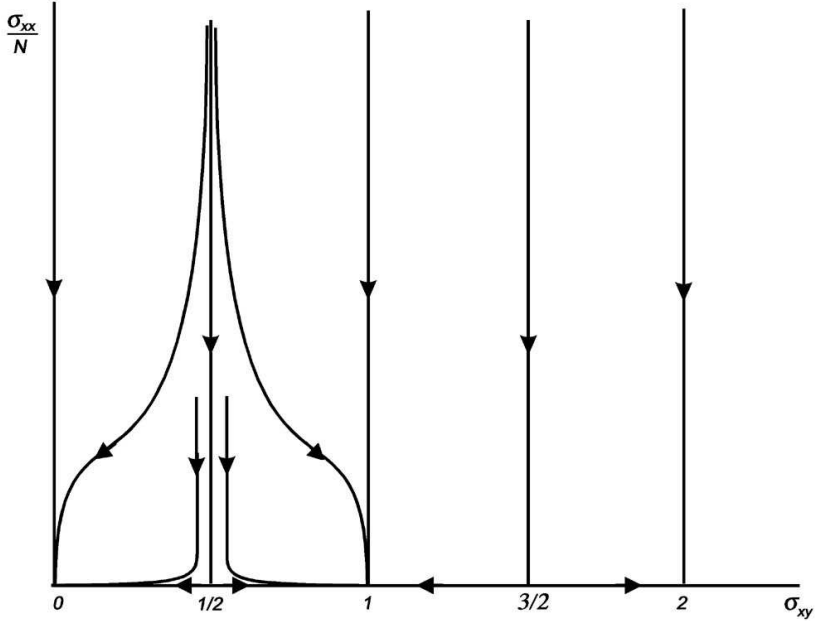}} \caption{Large N
renormalization group flow diagram for $\sigma_{xx}/N$ and the
Hall conductance $\sigma_{xy} = \theta/2\pi$. The arrows indicate
the direction toward the infrared.}\label{FIG1}
\end{figure}

The large $N$ expansion of the $CP^{N-1}$ model can be used as an
illuminating and exactly solvable example that sets the stage for
the {\em super universality} concept in asymptotically free field
theory~\cite{PruiskenBaranovVoropaevN}. The most significant
quantities of the theory are the renormalization group $\beta$
functions for the {\em response} parameters $\sigma_{xx}$ and
$\sigma_{xy}$ that appear in the effective action for massless
chiral edge excitations, (see also Fig.~\ref{FIG1})
\begin{gather}
\frac{d\sigma_{xx}}{d\ln \mu} = \beta_{\sigma} (\sigma_{xx} ,
\sigma_{xy})
\label{bxxxy} \\
\frac{d\sigma_{xy}}{d\ln \mu} = \beta_{\theta} (\sigma_{xx} ,
\sigma_{xy}).
\end{gather}
Here, the parameters $\sigma_{xx}$ and $\sigma_{xy}$ are precisely
analogous to the Kubo formulae for {\em longitudinal conductance}
and {\em Hall conductance} in quantum Hall systems. They stand for
the (inverse) {\em coupling constant} and $\theta/2\pi$
respectively, both of which appear as {\em running} parameters in
quantum field theory.

The infrared {\em stable fixed points} in Fig.~\ref{FIG1}, located
at integer values of $\sigma_{xy}$, indicate that the Hall
conductance is {\em robustly quantized} with corrections that are
exponentially small in the size of the
system~\cite{PruiskenBaranovVoropaevN}. The {\em unstable fixed
points} at half-integer values of $\sigma_{xy}$ or $\theta = \pi$
indicate that the large $N$ system develops a {\em gapless phase}
or a {\em continuously divergent correlation length} $\xi$ with a
critical exponent $\nu$ equal to
$1/2$~\cite{PruiskenBaranovVoropaevN},
\begin{equation}
\xi \propto |\theta -\pi|^{-1/2} .
\end{equation}

%/////////////////////////////////////////////////////////////////
%/////////////////////////////////////////////////////////////////

%=================================================================
\subsection{\label{Intro.LNE} The large $N$ expansion}

One of the most impressive features of the large $N$ expansion is
that it is exactly solvable for {\em all} values of $\theta$. This
is unlike the $O(3)$ non-linear sigma model, for example, which is
known to be integrable for $\theta = 0$ and $\pi$ only and the
exact information that can be extracted is rather
limited~\cite{O3}. Nevertheless, both cases appear as outstanding
limiting examples in the more general context of {\em replica
field theory} or, equivalently, the Grassmannian $U(m+n)/U(m)
\times U(n)$ non-linear sigma model. This Grassmannian manifold is
a generalization of the $CP^{N-1}$ manifold that describes, as is
well known, the Anderson localization problem in strong magnetic
fields~\cite{Pruisken1}.

It is extremely important to know, however, that \emph{none} of the
{\em super universal} features of the instanton vacuum where
previously known to exist, neither in the historical papers on the
large $N$ expansion~\cite{LargeN2,LargeN3,LargeN4,LargeN1} nor in
the intensively studied $O(3)$ case~\cite{O3}. In fact, the large
$N$ expansion, as it now stands, is in many ways an onslaught on
the many incorrect ideas and expectations in the field that are
based on the historical papers on the
subject~\cite{LargeN2,LargeN3,LargeN1}. These historical papers
are not only in conflict with the basic features of the quantum
Hall effect, but also present a fundamentally incorrect albeit
misleading picture of the instanton vacuum concept as a whole.

%..............................................................
\subsubsection{\label{Intro.LNE.GE} Gapless excitations regained}

One of the most important results of the large $N$ expansion, the
aforementioned {\em diverging correlation length} at $\theta =
\pi$, has historically been overlooked. This is one of the main
reasons why it is often assumed incorrectly that the excitations
of the Grassmannian $U(m+n)/U(m) \times U(n)$ non-linear sigma
model with $m,n \gtrsim 1$ always display a {\em gap}, also at
$\theta = \pi$.

Notice that general arguments, based on 't Hooft's duality
idea~\cite{t'Hooft1}, have already indicated that the theory at
$\theta = \pi$ is likely to be different. The matter has important
physical consequences because the lack of any {\em gapless}
excitations in the problem (or, for that matter, the lack of {\em
super universality} in non-linear sigma models) would seriously
complicate the possibility of establishing a microscopic theory of
the quantum Hall effect that is based on general topological
principles.

It now has turned out that the large $N$ expansion is one of the
very rare examples where 't Hooft's idea of using {\em twisted
boundary conditions}~\cite{t'Hooft1} can be worked out in great
detail, thus providing an {\em explicit demonstration} of the
existence of gapless excitations at $\theta =
\pi$~\cite{PruiskenBaranovVoropaevN}. Besides all this, the large
$N$ expansion can also be used to demonstrate the general {\em
nature} of the transition at $\theta = \pi$ which is otherwise is
much harder to establish. For example, {\em complete scaling
functions} have been obtained that set the stage for the {\em
transitions} between adjacent {\em quantum Hall
plateaus}~\cite{PruiskenBaranovVoropaevN}. In addition to this,
{\em exact} expressions have been derived for the {\em
distribution functions} of the mesoscopic conductance fluctuations
in the problem~\cite{PruiskenBaranovVoropaev}. These fluctuations
render anomalously large (broadly distributed) as $\theta$
approaches $\pi$, a well known phenomenon in the theory of
disordered metals. These results clearly indicate that the
instanton vacuum generally displays richly complex physics that
cannot be tapped if one is merely interested in the numerical
value of the critical exponents alone.

The large $N$ expansion is itself a good example of this latter
statement. For example, the historical results on the large $N$
expansion already indicated that the vacuum free energy with
varying $\theta$ displays a {\em cusp} at $\theta = \pi$, i.e. a
first order phase transition. This by itself is sufficient to
establish the existence of a scaling exponent $\nu =1/d$ with
$d=2$ denoting dimension of the system. However, super
universality as a whole remains invisible as long as one is
satisfied with the merely heuristic arguments that historically
have spanned the subject~\cite{LargeN2,LargeN3,LargeN4}. The
discovery of a new aspect of the theory, the {\em massless chiral
edge excitations}, was clearly necessary before the appropriate
questions could be asked and {\em super universality} be finally
established.

Given the new results on the large $N$ expansion of the $CP^{N-1}$
model, it may no longer be a complete surprise to know that the
instanton vacuum at $\theta = \pi$ is {\em generically gapless},
independent of the number of field components in the theory. Since
all members of the Grassmannian $U(m+n) / U(m) \times U(n)$
manifold are {\em topologically equivalent}, have important
features in common such as {\em asymptotic freedom}, {\em
instantons}, {\em massless chiral edge excitations} etc., it is
imperative that the same basic phenomena are being displayed,
independent of $m$ and $n$. This includes of course the theory of
actual interest, obtained by putting $m=n=0$ ({\em replica
limit}).

Unlike {\em super universality}, the details of the critical
singularities at $\theta = \pi$ ({\em critical indices}) may in
principle be different for different values of $m$ and $n$. The
situation is in this respect analogous to what happens to the
classical Heisenberg ferromagnet in $2+\epsilon$ spatial
dimensions. Like in two dimensions, the basic physics is
essentially the same for any value of $m$ and $n$. The quantum
critical behavior, however, strongly varies with a varying number
of field components in the theory, each value of $m$ and $n$
representing a different {\em universality} class.

%...............................................................
\subsubsection{\label{Intro.LNE.I} Instantons regained}

Besides the {\em strong coupling} aspects of the instanton vacuum,
the {\em effective action for massless chiral edge excitations}
also provides a fundamentally new outlook on the {\em weak
coupling} features of the theory that cannot be obtained in any
different way. Topological excitations ({\em instantons}) have
made a spectacularly novel entree, in the renormalization behavior
of theory, especially after they have been totally mishandled and
abused in the historical papers on the large $N$
expansion~\cite{LargeN2,LargeN3,LargeN4}.

Within the recently established renormalization
theory~\cite{PruiskenBaranovVoropaevN} of the $CP^{N-1}$ model
with large $N$ (see Fig.~\ref{FIG1}), {\em instantons} emerge as
non-perturbative topological objects that facilitate the {\em
cross-over} between the {\em Goldstone phase} at weak coupling or
short distances ($\sigma_{xx} \gg 1$), and the super universal
{\em strong coupling phase} of the instanton vacuum ($\sigma_{xx}
\ll 1$) that generally appears in the limit of much larger
distances only. A detailed knowledge of {\em instanton effects} is
generally important since it provides a fundamentally new concept
that the theory of ordinary perturbative expansions could never
give, namely {\em $\theta$ renormalization} or, equivalently, the
renormalization of the Hall conductance $\sigma_{xy}$.

%//////////////////////////////////////////////////////////////////
\begin{figure}
\centerline{\includegraphics[width=110mm]{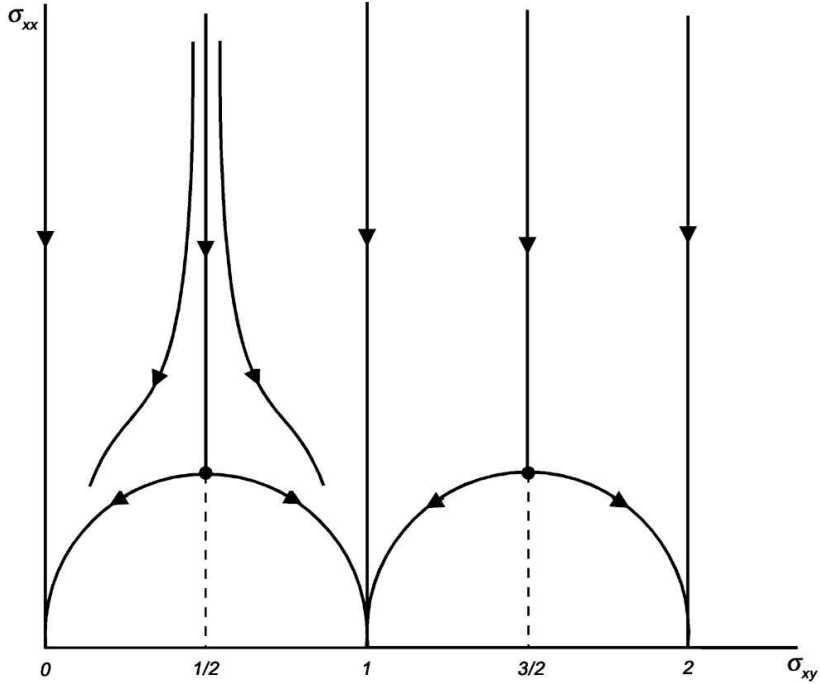}} \caption{
Renormalization group flow diagram for the conductances. The
arrows indicate the scaling toward the infrared.}\label{FIG2}
\end{figure}
%//////////////////////////////////////////////////////////////

The concept of $\theta$ renormalization originally arose in a
series of detailed papers on instantons, based on the {\em
background} field methodology, that were primarily aimed at a
microscopic understanding of the quantum Hall
effect~\cite{Pruisken2,Pruisken3}. Until to date these pioneering
papers have provided most of our insights into the singularity
structure of the Grassmannian $U(m+n) / U(m) \times U(n)$ theory
at $\theta =\pi$, in particular the case where the number of field
components is `small', $0\leqslant m,n \lesssim 1$. Under these
circumstances the instanton vacuum at half-integer values of
$\sigma_{xy}$ (or $\theta = \pi$) develops a critical fixed point
with a {\em finite} value of $\sigma_{xx}$ of order unity (see
Fig.~\ref{FIG2}). This indicates that transition at $\theta =\pi$
becomes a true {\em second order} quantum phase transition with a
non-trivial critical index $\nu$ that changes continuously with
varying values of $m$ and $n$ in the range $0\leq m,n \lesssim 1$.
This situation is distinctly different from the overwhelming
majority of Grassmannian non-linear sigma models with $m,n \gtrsim
1$ for which the scaling diagram is likely to be the same as the
one found in the large $N$ expansion (see Fig.~\ref{FIG1}). In
that case one expects a {\em first order} phase transition but
with a {\em diverging} correlation length and a {\em fixed}
exponent $\nu =1/2$.

\begin{figure}
\centerline{\includegraphics[width=110mm]{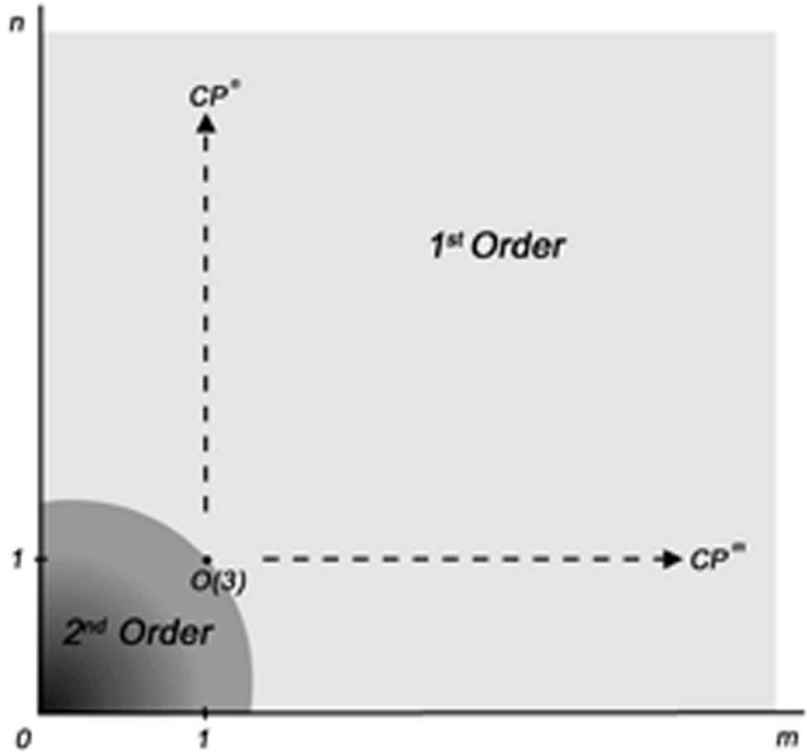}} \caption{Nature
of the transition at $\theta=\pi$ for different values of $m$ and
$n$. }\label{FIG4}
\end{figure}

The instanton vacuum with $m=n=1$ (the $SU(2)/U(1)$ or O(3) model)
is in many ways special. This case is likely to be on the
interface between a {\em large $N$-like} scaling diagram (see
Fig.~\ref{FIG1}) with a {\em first order} transition at $\theta =
\pi$, and an {\em instanton driven} scaling diagram (see
Fig.~\ref{FIG2}) where the transition is of {\em second order}.
The expected $m$ and $n$ dependence of quantum criticality is
illustrated in Fig.~\ref{FIG4} which is the main topic of the
present paper (see also Ref.~\cite{PruiskenBaranovVoropaevN}).

%/////////////////////////////////////////////////////////////////
%/////////////////////////////////////////////////////////////////

%================================================================
\subsection{\label{Intro.QPT} Quantum phase transitions}

As is well known, the instanton angle $\theta$ in replica field
theory was originally discovered in an attempt to resolve the
fundamental difficulties of the scaling theory of Anderson
localization in dealing with the quantum Hall
effect~\cite{LevineLibbyPruisken}. However, it was not until the
first experiments~\cite{WeiTsuiPalaanenPruisken} on the {\em
plateau transitions} had been conducted that the prediction of
{\em quantum criticality} in the quantum Hall
systems~\cite{Pruisken4} became a well recognized and extensively
pursued research objective in physics.

Quantum phase transitions in disordered systems are in many
respects quite unusual, from ordinary critical phenomena point of
view. For example, such unconventional phenomena like {\em
multifractality} of the density fluctuations are known to appear
as a peculiar aspect of the theory in the {\em replica limit}
$m=n=0$~\cite{Multifractality}. These subtle aspects of disordered
systems primarily arose from the non-linear sigma model approach
to the Anderson localization problem ({\em mobility edge problem})
in $2 +\epsilon$ spatial dimensions~\cite{Wegner1,Pruisken5}.

The instanton vacuum theory of the quantum Hall effect essentially
predicts that the {\em plateau transitions} in the two dimensional
electron gas behave in all respects like the {\em metal-insulator
transition} in $2+\epsilon$ dimensions. The reduced dimensionality
of the quantum Hall system offers a rare opportunity to perform
numerical work on the {\em mobility edge problem} and extract
accurate results on quantum critical behavior. By now there exists
an impressive stock of numerical data on the {\em critical
indices} of the plateau transitions, including the correlation or
{\em localization} length exponent
($\nu$)~\cite{NumNu1,NumNu2,NumNu3,NumNu4,NumNu5,NumNu6}, the {\em
multifractal} $f(\alpha)$
spectrum~\cite{NumGA1,NumGA2,NumPi2,NumLW,NumAlpha01,NumAlpha02}
and even the leading {\em irrelevant} exponent ($y_\sigma$) in the
problem~\cite{NumY1,NumY2}.

%................................................................
\subsubsection{\label{Intro.QPT.QA} Quantitative assessments}

It is important to know that the {\em laboratory} experiments and
later the {\em numerical} simulations on the {\em plateau
transitions} in the quantum Hall regime have primarily been guided
and motivated by the renormalization group ideas that were
originally obtained on the basis of the $\theta$ parameter replica
field theory as well as instanton
calculus~\cite{Pruisken4,Pruisken2,Pruisken3}. In addition to
this, the more recent discovery of {\em super universality} in
non-linear sigma models, along with the completely revised
insights in the large $N$ expansion, has elucidated the much
sought after {\em strong coupling} features of the instanton
vacuum, notably the {\em quantum Hall effect} itself, that
previously remained concealed~\cite{PruiskenBaranovVoropaevN}.
Both these strong coupling features and the renormalization group
results based on instantons have put the theory of an instanton
vacuum in a novel physical perspective. Together they provide the
complete conceptual framework that is necessary for a detailed
understanding of the quantum Hall effect as well as the $\theta$
dependence in the Grassmannian $U(m+n)/U(m) \times U(n)$
non-linear sigma model, for all non-negative values of $m$ and
$n$.

In our previous papers we have already presented rough outlines on
how the theory manages to interpolate between a {\em large
$N$-like} scaling diagram for large values of $m,n$
(Fig.~\ref{FIG1}) and an {\em instanton - driven} renormalization
behavior for small $m,n$ as indicated in Fig.~\ref{FIG2}. At
present we take the theory several steps further and extend the
instanton methodology in several ways. Our main objective is to
make detailed predictions on the {\em quantum critical} behavior
of the theory at $\theta=\pi$ with varying values of $m$ and $n$.
We benefit from the fact that this quantum critical behavior is
bounded by the theory in the replica limit ($m=n=0$) for which the
aforementioned numerical data are available, and the distinctly
different $O(3)$ non-linear sigma model ($m=n=1$) for which the
critical indices are known exactly. A detailed comparison between
our general results and those known for specific examples should
therefore provide a stringent and interesting test of the
fundamental significance of instantons in the problem.

Our most important results are listed in Table~\ref{Tab.results}
where we compare the critical exponents of the theory with $m=n=0$
with those obtained from numerical simulations on the electron
gas. These results clearly demonstrate the validity of a general
statement made in
Ref.~\cite{PruiskenBaranovVoropaev,PruiskenBaranovBurmistrov,PruiskenBaranovVoropaevN}
which says that the fundamental significance of the instanton gas
is primarily found in the renormalization behavior of the theory
or, equivalently, the effective action for chiral edge
excitations. This leads to a conceptual understanding of the
non-perturbative aspects of the theory that cannot be obtained in
any different manner.

%..........................................................................
\subsubsection{\label{Intro.QPT.Out} Outline of this paper}

We start out in Section~\ref{Form} with an introduction to the
formalism, a brief summary of the effective action procedure for
massless chiral edge excitations as well as a few comments
explaining the super universal features of the instanton vacuum.

The bulk of this work mainly follows the formalism that was
introduced in the original papers on
instantons~\cite{Pruisken2,Pruisken3}. However, the main focus at
present is on several important aspects of the theory that
previously remained unresolved. The first aspect concerns the
ambiguity in the numerical factors that arises in the computation
of the instanton determinant. These numerical factors are vitally
important since they eventually enter into the renormalization
group $\beta$ functions and, hence, determine the critical fixed
point properties of the theory. In Section~\ref{QFAI} we show how
the theory of observable parameters can be used to actually
resolve this problem. The final expressions for the $\beta$
functions that we obtain are {\em universal} in the sense that
they are independent of the specific regularization scheme that is
being used.

Secondly, the procedure sofar did not include the effect of {\em
mass terms} in the theory and, hence, the multifractal aspects of
the quantum phase transition have not yet been investigated. The
technical difficulties associated with mass terms are quite
notorious, however. These have historically resulted in the
construction of highly non-trivial extensions of the methodology
such as working with {\em constrained instantons}~\cite{Affleck1}.
We briefly introduce the subject matter in Section~\ref{Inst} and
illustrate the main ideas by means of explicit examples.

It has sofar not been obvious, however, whether the concept of a
{\em constrained instanton} is any useful in the development a
quantum theory. In an earlier paper on the subject
~\cite{PruiskenBaranov} it was pointed out that mass terms in the
theory generally involve a different {\em metric tensor} than the
one that naturally appears in the harmonic oscillator problem,
i.e. their geometrical properties are incompatible. These and
other complications are avoided in the methodology of {\em
spatially varying masses} which is based on the various tricks
introduced by 't Hooft in evaluating instanton determinants. In
Section~\ref{QFAI} we elaborate further on the methodology of
Ref.~\cite{PruiskenBaranov} and show that the idea of {\em
spatially varying masses} generally facilitates explicit
computations and yields more relevant results.

In Section~\ref{RGE.RGE} we present the detailed predictions of
quantum criticality as a function of $m$ and $n$ and make a
comparison with the results known from other sources. We end this
paper with a conclusion, Section~\ref{Conc}.

%%%%%%%%%%%%%%%%%%%%%%%%%%%%%%%%%%%%%%%%%%%%%%%%%%%%%%%%%%%%%%%%%%%%%%
%  SECTION 2
%
%%%%%%%%%%%%%%%%%%%%%%%%%%%%%%%%%%%%%%%%%%%%%%%%%%%%%%%%%%%%%%%%%%%%%%
\section{\label{Form}Formalism}
%
%

%%%%%%%%%%%%%%%%%%%%%%%%%%%%%%%%%%%%%%%%%%%%%%%%%%%%%%%%%%%%%%%%%%%
%
%
\subsection{\label{Form.NLS}Non-linear sigma model}
%
%%%%%%%%%%%%%%%%%%%%%%%%%%%%%%%%%%%%%%%%%%%%%%%%%%%%%%%%%%%%%%%%%%%

First we recall the non-linear sigma model defined on the
Grassmann manifold $G/H = {SU(m+n)}/{S(U(m)\times U(n))}$ and in
the presence of the $\theta$ term. We prefer to work in the
context of the quantum Hall effect since this provides a clear
{\em physical} platform for discussing the fundamental aspects of
the theory that have previously remained unrecognized. It is easy
enough, however, to make contact with the conventions and
notations of more familiar models in quantum field theory and
statistical mechanics such as the $O(3)$ formalism, obtained by
putting $m=n=1$, and the $CP^{N-1}$ model which is obtained by
taking $m=N-1$ and $n=1$.

The theory involves matrix field variables $Q(\textbf{r})$ of size
$(m+n)\times (m+n)$ that obey the non-linear constraint
\begin{equation}\label{constraint}
Q^{2}(\textbf{r})=1_{m+n}.
\end{equation}
A convenient representation in terms of ordinary unitary matrices
$T(\textbf{r})$ is obtained by writing
\begin{equation}\label{Qrep}
Q(\textbf{r}) = T^{-1}(\textbf{r}) \Lambda T(\textbf{r}),\qquad
\Lambda = \begin{pmatrix}
 1_{m} & 0 \\
  0 & - 1_{n}
\end{pmatrix}.
\end{equation}
The action describing the low energy dynamics of the two
dimensional electron system subject to a static, perpendicular
magnetic field is given by~\cite{Pruisken1}
\begin{equation}\label{Ssigma}
S = -\frac{\sigma _{xx}}{8} \int d \textbf{r} \tr(\nabla
Q)^{2}+\frac{\sigma _{xy}}{8} \int d \textbf{r} \tr \varepsilon
_{ab} Q \nabla_{a} Q \nabla_{b} Q + \omega \rho_0 \int d
\textbf{r} \tr Q \Lambda .
\end{equation}
Here the quantities $\sigma_{xx}$ and $\sigma_{xy}$ represent the
{\em meanfield} values for the {\em longitudinal} and {\em Hall}
conductances respectively. The $\rho_0$ denotes the {\em density}
of electronic levels in the bulk of the system, $\omega$ is the
{\em external frequency} and $\varepsilon_{ab}= -\varepsilon_{ba}$
is the antisymmetric tensor.

%%%%%%%%%%%%%%%%%%%%%%%%%%%%%%%%%%%%%%%%%%%%%%%%%%%%%%%%%%%%%%%
%
%
\subsection{Boundary conditions}
%
%%%%%%%%%%%%%%%%%%%%%%%%%%%%%%%%%%%%%%%%%%%%%%%%%%%%%%%%%%%%%%%%%

The second term in Eq. \eqref{Ssigma} defines the topological
invariant $\mathcal{C}[Q]$ which can also be expressed as a one
dimensional integral over the edge of the system~\cite{Pruisken2},
\begin{equation}\label{topcharge}
\mathcal{C}[Q] = \frac{1}{16 \pi i} \int d \textbf{r} \tr
\varepsilon _{ab} Q\nabla_{a}Q \nabla_{b} Q = \frac{1}{4 \pi
i}\oint d x \tr T \partial_x T^{-1} \Lambda.
\end{equation}
Here, $\mathcal{C}[Q]$ is integer valued provided the field
variable $Q$ equals a constant matrix at the edge. It formally
describes the mapping of the Grassmann manifold onto the plane
following the homotopy theory result
\begin{equation}\label{homotopy}
\pi_{2} \left (G/H \right ) = \pi_1 \left ( H \right ) = \Zset.
\end{equation}
To study the $\sigma_{xy}$ dependence of the theory it is in many
ways natural to put $\omega =0$ in Eq. \eqref{Ssigma} and let the
infrared of the theory be regulated by the finite system size $L$.
The conventional way of defining the $\theta$ vacuum is as follows
\begin{equation}\label{theta2}
Z = \int\limits_{\partial V} \mathcal{D} Q \exp \left [
-\frac{1}{g} \int d \textbf{r} \tr(\nabla Q)^{2} +i \theta\,
\mathcal{C}[Q]\right ] ,
\end{equation}
where the subscript $\partial V$ indicates that the functional
integral has to be performed with $Q(\textbf{r})$ kept fixed and
constant at the boundary, say $Q=\Lambda$. Under these
circumstances the topological charge $\mathcal{C}[Q]$ is strictly
integer quantized. The parameter $\theta$ in Eq. \eqref{theta2} is
equal to $2\pi\sigma_{xy}$ modulo $2\pi$ and the coupling constant
$g$ is identified as $8/\sigma_{xx}$,
\begin{equation}
\theta = 2\pi \sigma_{xy}\, {\rm mod}(2\pi), \qquad g =
\frac{8}{\sigma_{xx}}.
\end{equation}
The theory of Eq. \eqref{theta2}, as it stands, is one of the rare
examples of an asymptotically free field theory with a vanishing
mass gap. What has remarkably emerged over the years is that the
theory at $\theta=\pi$ develops a {\em gapless} phase that
generally can be associated with the {\em transitions} between
adjacent quantum Hall plateaus. This is unlike the theory with
$\theta = 0, 2\pi$ where the low energy excitations are expected
to display a {\em mass gap}. The significance of the theory in
terms of quantum Hall physics becomes all the more obvious if one
recognizes that the mean field parameter $\sigma_{xy}$ for strong
magnetic fields is precisely equal to the filling fraction ($\nu$)
of the disordered Landau bands
\begin{equation}
\sigma_{xy} = \nu.
\end{equation}
This means that the plateau transitions occur at {\em
half-integer} filling fractions $\nu = k+1/2$ with integer $k$. On
the other hand, $\theta = 0, 2\pi$ corresponds to {\em integer}
filling fractions $\nu = k$ which generally describe the center of
the quantum Hall plateaus.

As was already mentioned in the introduction, the physical
objectives of the quantum Hall effect have been - from early
onward - in dramatic conflict with the ideas and expectations with
which the $\theta$ parameter in quantum field theory was
originally perceived. Such fundamental aspects like the existence
of {\em robust topological quantum numbers}, for example, have
previously been unrecognized. This is just one of the reasons why
the quantum Hall effect primarily serves as an outstanding
laboratory where the controversies in quantum field theory can be
explored and investigated in detail.

%...................................................................
%
%
\subsubsection{Massless chiral edge excitations}
%
%...................................................................

It has turned out that the theory of Eq. \eqref{theta2} is not yet
the complete story. By {\em fixing} the boundary conditions in
this problem (or by discarding the {\em edge} all together) one
essentially leaves out fundamental pieces of physics, the {\em
massless chiral edge excitations}, that eventually will put the
strong coupling problem of an instanton vacuum in a novel
perspective.

To see how the physics of the edge enters into the problem we
consider the case where the Fermi energy of the electron gas is
located in an energy gap ({\em Landau gap}) between adjacent
Landau bands. This is represented by Eq. \eqref{Ssigma} by putting
$\sigma_{xx} = \rho_0 =0$ and $\sigma_{xy} = k$, i.e. the {\em
meanfield} value of the Hall conductance is an integer $k$ (in
units of $e^2 /h$) and precisely equal to the number of completely
filled Landau bands in the system. Eq. \eqref{Ssigma} can now be
written as follows
\begin{eqnarray}
\hspace{2cm}S_{\rm edge} [Q] &=& 2\pi i k\, \mathcal{C}[Q] +
\omega \rho_{\rm edge} \oint d x \tr Q \Lambda \notag\\ &=& \oint
d x \tr \left[ \frac{k}{2}  T
\partial_x T^{-1} \Lambda + \omega \rho_{\rm edge} Q \Lambda
\right].\label{Sedge}
\end{eqnarray}
We have added a symmetry breaking term proportional to $\rho_{\rm
edge}$ indicating that although the Fermi energy is located in the
Landau gap, there still exists a finite density of edge levels
($\rho_{\rm edge}$) that can carry the Hall current. Eq.
\eqref{Sedge} is exactly solvable and describes long ranged ({\em
critical}) correlations along the edge of the system. Some
important examples of edge correlations are given by~\cite{Unify3}
\begin{eqnarray}
\langle Q\rangle_{\rm edge} &=& \Lambda,  \notag\\
\hspace{2cm} \langle Q_{+-}^{\alpha\beta} (x)
Q_{-+}^{\beta\alpha}(x^\prime) \rangle_{\rm edge} &=&
\vartheta(x^\prime-x) \exp\left [-(x^\prime-x) \omega \rho_{\rm
edge}\right ].\label{edcor}
\end{eqnarray}
Here the expectation is with respect to the one dimensional theory
of Eq. \eqref{Sedge}. The symbol $\vartheta(x)$ stands for the
Heaviside step function. These results are the same for all $m$
and $n$, indicating that the massless chiral edge excitations are
a generic feature of the instanton vacuum concept.

Eqs \eqref{theta2} and \eqref{Sedge} indicate that field
configurations with an {\em integral} and {\em fractional}
topological charge ${\mathcal C}[Q]$ describe fundamentally
different physics and have fundamentally different properties. In
the following Sections we show in a step by step fashion how Eq.
\eqref{Sedge} generally appears as the {\em fixed point action} of
the strong coupling phase.

%%%%%%%%%%%%%%%%%%%%%%%%%%%%%%%%%%%%%%%%%%%%%%%%%%%%%%%%%%%%%%%%%%%
%
%
\subsection{\label{Form.Inst} Effective action for the edge}
%
%%%%%%%%%%%%%%%%%%%%%%%%%%%%%%%%%%%%%%%%%%%%%%%%%%%%%%%%%%%%%%%%%%%

We specialize from now onward to systems with an edge. The main
problem next is to see how in general we can separate the {\em
bulk} pieces of the action from those associated with the {\em
edge}. The resolution of this problem provides fundamental
information on the low energy dynamics of the system (Section
\ref{lrf}) that is intimately related to the Kubo formalism of the
conductances (Section \ref{lbfm}). In the end (Sections \ref{ltc}
and \ref{lcflc}) this leads to the much sought after Thouless'
criterion for the existence of robust topological quantum numbers
that explain the precision and observability of the quantum Hall
effect.

%.............................................................
%
%
\subsubsection{Bulk and edge field variables}
%
%.............................................................

It is convenient to introduce a change of variables
\begin{equation}
Q = t^{-1} Q_0 t.
\end{equation}
Here, $Q_0$ is generally defined as a matrix field with $Q_0 =
\Lambda$ at the edge of the system ({\em spherical boundary
conditions}). The matrix field $t$ describes the {\em
fluctuations} about these special spherical boundary conditions.
It is easy to see that the {\em topological charge}
$\mathcal{C}[Q]$ can be written as the sum of two separate pieces
\begin{equation}
\mathcal{C}[Q] = \mathcal{C}[t^{-1} Q_0 t] = \mathcal{C}[Q_0] +
\mathcal{C}[q], \qquad q = t^{-1} \Lambda t.
\end{equation}
Here, the first part $\mathcal{C}[Q_0]$ is {\em integer} valued
whereas the second part $\mathcal{C}[q]$ describes a {\em
fractional} topological charge,
\begin{equation}
\mathcal{C}[Q_0] = k, \qquad  -\frac{1}{2} < \mathcal{C}[q]  \leq
\frac{1}{2}.
\end{equation}
It is easy to see that the action for the edge, Eq.~\eqref{Sedge},
only contains the field variable $t$ or $q$ which we shall refer
to as the {\em edge} fields in the problem. Similarly, the $Q_0$
are the only matrix field variables that enter into the usual
definition of the $\theta$ vacuum, Eq. \eqref{theta2}. We will
refer to them as the {\em bulk} field variables.

%...................................................................
%
%
\subsubsection{Bulk and edge observables}
%
%...................................................................

\begin{figure}
\centerline{\includegraphics[width=110mm]{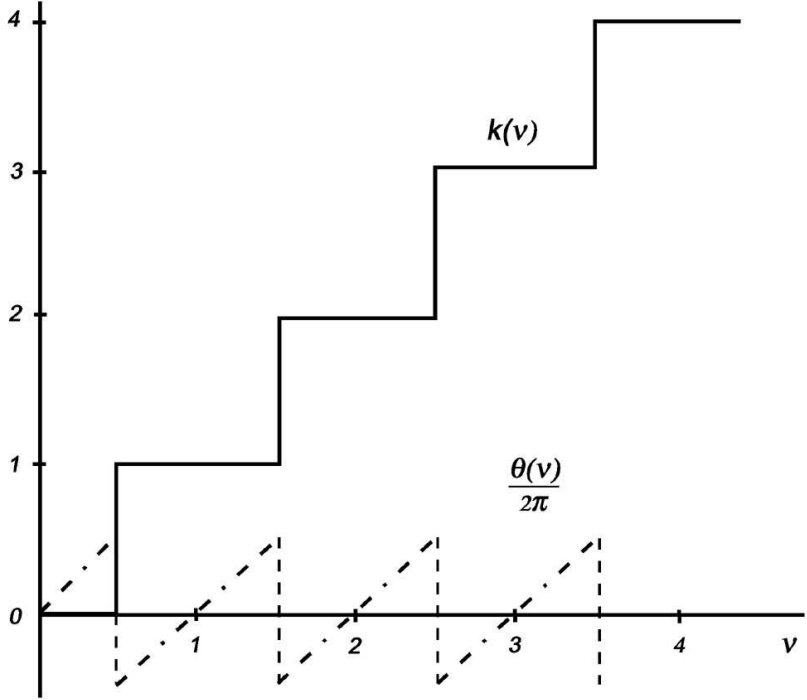}} \caption{The
quantity $\sigma_{xy}^0 = \nu$ is the sum of a {\em quantized}
edge part $k(\nu)$ and an {\em unquantized} bulk part
$\theta(\nu)$ .}\label{FIG3}
\end{figure}

Next, from Eqs \eqref{theta2} and \eqref{Sedge} we infer that the
bare or meanfield Hall conductance $\sigma_{xy}$ should in general
be split into a {\em bulk} piece $\theta(\nu)$ and a distinctly
different {\em edge} piece $k(\nu)$ as follows (see also
Fig.~\ref{FIG3})
\begin{equation}\label{split}
\sigma_{xy} = \nu = \frac{\theta(\nu)}{2\pi} + k(\nu) .
\end{equation}
Here, $k(\nu)$ is an integer whereas $\theta(\nu)$ is restricted
to be in the interval
\begin{equation}
-\pi < \theta(\nu) \leq \pi.
\end{equation}
The new symbol $\nu$ indicates that the meanfield parameter
$\sigma_{xy}$ is the same as the {\em filling fraction} of the
Landau bands. Eq. (\ref{split}) can also be obtained in a more
natural fashion and it actually appears as a basic ingredient in
the microscopic derivation of the theory.

We use the results to rewrite the topological piece of the action
as follows
\begin{equation}
2\pi i \sigma_{xy} \mathcal{C}[Q] = i (\theta(\nu) + 2\pi k(\nu))
(\mathcal{C}[q] + \mathcal{C}[Q_0]) = 2\pi i k(\nu) \mathcal{C}[q]
+ i \theta(\nu) \mathcal{C}[Q] .
\end{equation}
In the second equation we have left out the term that gives rise
to unimportant phase factor. Finally we arrive at the following
form of the action
\begin{equation}
S = S_{\rm bulk} [t^{-1} Q_0 t] + S_{\rm edge}[q], \label{wrong}
\end{equation}
where
\begin{gather}\label{bulkpart}
S_{\rm bulk}[Q]= -\frac{\sigma _{xx}}{8} \int d \textbf{r}
\tr(\nabla Q)^{2}
+i{\theta(\nu)} \mathcal{C}[Q] \\
\label{edgepart} S_{\rm edge} [q] = 2\pi i k(\nu)\,
\mathcal{C}[q].
\end{gather}

%//////////////////////////////////////////////////////////////////////
%//////////////////////////////////////////////////////////////////////
%//////////////////////////////////////////////////////////////////////
%//////////////////////////////////////////////////////////////////////

%..............................................................................
%
%
\subsubsection{\label{lrf}Response formulae}
%
%...............................................................................

After these preliminaries we come to the most important part of
this Section, namely the definition of the {\em effective action
for chiral edge modes} $S_{\rm eff}$. This is obtained by formally
eliminating the {\em bulk} field variable $Q_0$ from the theory
\begin{equation}\label{response}
{S_{\rm eff}[q]} = S_{\rm edge} [q] + S^{\,\prime}[q], \qquad
 \exp S^{\,\prime}[q] = \int\limits_{\partial V} \mathcal{D}[Q_0]
\exp S_{\rm bulk} [t^{-1} Q_0 t].
\end{equation}
The subscript ${\partial V}$ reminds us of the fact that the
functional integral has to be performed with a fixed value $Q_0 =
\Lambda$ at the edge of the system. Next, from symmetry
considerations alone it is easily established that $S'$ must be of
the general form~\cite{Unify3}
\begin{equation}\label{seff}
S^{\,\prime}[q] = -\frac{\sigma^{\,\prime}_{xx}}{8} \int \limits
d\textbf{r}\tr(\nabla q)^{2} + i\theta^{\,\prime} \mathcal{C}[q] .
\end{equation}
We have omitted the higher dimensional terms which generally
describe the properties of the electron gas at {\em mesoscopic}
scales. The most important feature of this result is that the
quantities $\sigma^\prime_{xx} = \sigma^\prime_{xx} (L)$ and
$\theta^{\,\prime}=\theta^{\,\prime}(L)$ can be identified with the Kubo
formulae for the {\em longitudinal} and {\em Hall} conductances
respectively ($L$ denoting the linear dimension of the system).
Notice that these quantities are by definition a measure for the
{\em response} of the bulk of the system to an infinitesimal
change in the boundary conditions (on the matrix field variable
$Q_0$).

 At the same time we can regulate the infrared of the
system in a different manner by introducing $U(m) \times U(n)$
invariant mass terms in Eqs \eqref{bulkpart}-\eqref{edgepart}
\begin{gather}\label{freq}
S_{\rm edge} [q] \rightarrow S_{\rm edge} [q] +
\omega_{\rm edge} \rho_{\rm edge} \oint d x \tr \Lambda q,  \\
S_{\rm bulk} [t^{-1} Q_0 t] \rightarrow S_{\rm bulk} [t^{-1} Q_0
t] + \omega_0 \rho_0 \int d \textbf{r} \tr \Lambda Q_0 .
\end{gather}
The different symbols $\omega_{\rm edge}$ and $\omega_0$ indicate
that the frequency $\omega$ plays a different role for the {\em
bulk} fields and {\em edge} fields respectively. Notice that the
response parameters $\sigma^\prime_{xx}$ and $\sigma^\prime_{xy}$ in Eq.
\eqref{seff}, for $L$ large enough, now depend on frequency
$\omega_0$ rather than $L$.
\begin{equation}\label{chiralfreq}
\sigma^\prime_{xx} \rightarrow \sigma^\prime_{xx} (\omega_0),\qquad
\sigma^\prime_{xy} \rightarrow \sigma^\prime_{xy} (\omega_0) = k(\nu) +
\frac{\theta^{\,\prime}(\omega_0)}{2\pi}.
\end{equation}
%..................................................................
%
%
\subsubsection{\label{lbfm}Background field methodology}
%
%..................................................................
Let us next go back to the background field methodology and notice
the subtle differences with the effective action procedure as
considered here. In this methodology we consider a slowly varying
but fixed background matrix field $b$ that is applied directly to
the original theory of Eq. \eqref{Ssigma},
\begin{eqnarray}\label{back}
\exp S_{\rm eff}[b^{-1}\Lambda b] &=& \int \mathcal{D} Q \exp
\left ( S_\sigma[b^{-1} Q b]
+ \omega_0 \rho_0 \int d \textbf{r} \tr Q \Lambda\right ), \\
S_\sigma [Q] &=& -\frac{\sigma _{xx}}{8} \int d \textbf{r}
\tr(\nabla Q)^{2} + 2\pi i \sigma_{xy}\, \mathcal{C}[Q].\nonumber
\end{eqnarray}
Here, $S_{\rm eff}$ can again be written in the following general
form
\begin{equation}\label{backeff}
S_{\rm eff} [q_b] = -\frac{\sigma^\prime_{xx}(\omega_0 )}{8} \int d
\textbf{r} \tr(\nabla q_b)^{2} + 2\pi i \sigma^\prime_{xy} (\omega_0
) \mathcal{C}[q_b],
\end{equation}
where now $q_b = b^{-1} \Lambda b$. Notice that an obvious
difference with the situation before is that the functional
integral in Eq. \eqref{back} is now performed for an arbitrary
matrix field $Q$ whereas in Eq. \eqref{response} the $Q_0$ is
always restricted to have $Q_0 = \Lambda$ at the edge. However, as
long as one works with a finite frequency $\omega_0$ the boundary
conditions on the $Q_0$ field are immaterial and the response
parameters in Eq. \eqref{chiralfreq} and those in Eq.
\eqref{backeff} should be identically the same. As an important
check on these statements we consider $\sigma_{xx} = 0$ and
$\sigma_{xy} = k$, i.e. $S_\sigma$ equals the action for the edge
$S_\textrm{edge}$. In this case Eq. \eqref{backeff} is obtained as
follows
\begin{eqnarray}
\exp S_{\rm eff} [q_b] &=& \int {\mathcal D}Q \exp\left [\oint dx
\left (\frac{k}{2}  \tr T b \partial_x b^{-1} T^{-1} \Lambda +\omega
\rho_{edge} \tr Q\Lambda \right ) \right] \nonumber
\\ \label{edgeeff1}
&=& \left( \int {\mathcal D}Q \exp{S_\textrm{edge} [Q]}
\right) \, \exp\left [\frac{k}{2} \oint dx \tr b
\partial_x b^{-1} \langle Q \rangle_\textrm{edge}\right ].
\end{eqnarray}
Using Eq. \eqref{edcor} we can write, discarding constants,
\begin{equation}
\label{edgeeff}
{S_{\rm eff} [q_b]} = 2\pi i k~ {\mathcal C}[q_b]
\end{equation}
Comparing Eqs \eqref{backeff} and \eqref{edgeeff} we see that
$\sigma_{xx}^{\prime} = 0$ and $\sigma_{xy}^{\prime} = k$ as it
should be. Notice that we obtain essentially the same result if
instead of integrating we fix the $Q=\Lambda$ at the edge.

Eq. \eqref{edgeeff1} clearly demonstrates why critical edge
correlations should be regarded as a fundamental aspect of the
instanton vacuum concept. If, for example, $S_\textrm{edge}$ were to
display gapped excitations at the edge then we certainly would
have $\langle Q\rangle_\textrm{edge} =0$ in Eq. \eqref{edgeeff1} and instead of Eq.
\eqref{edgeeff} we would have had a vanishing Hall conductance!

This, then, would be serious conflict with the quantum Hall effect
which says that the quantization of the Hall conductance is in
fact a {\em robust} phenomenon.
%..................................................................
%
%
\subsubsection{\label{ltc}Thouless criterion}
%
%..................................................................
We next show that a {\em Thouless criterion}~\cite{Thouless} for
the quantum Hall effect can be obtained directly, as a corollary
of the aforementioned effective action procedure. For this purpose
we notice that if the system develops a mass gap or a finite
correlation length $\xi$ in the bulk, then the theory of
$S_\textrm{bulk}$, Eq. \eqref{bulkpart}, should be insensitive to any
changes in the boundary conditions, provided the system size $L$
is large enough. Under these circumstances the response quantities
$\sigma^\prime_{xx} (L)$ and $\theta^{\,\prime}(L)$ in Eq. \eqref{seff}
should vanish. In terms of the conductances we can write
\begin{equation}
\sigma^\prime_{xx} (L) = \mathcal{O}\left (e^{-L/\xi}\right ),\qquad
\sigma^\prime_{xy} (L) = k(\nu) + \frac{\theta^\prime(L)}{2\pi} = k(\nu)
+ \mathcal{O}\left( e^{-L/\xi}\right ).
\end{equation}
This important result indicates that the {\em quantum Hall effect}
is in fact a {\em universal, strong coupling} feature of the
$\theta$ vacuum, independent of the number of field components $m$
and $n$. The {\em fixed point action} of the quantum Hall state is
generally given by the one dimensional action~\cite{Unify3}
\begin{equation}\label{response1}
S_{\rm eff} [q] = \oint d x \tr \left[ \frac{k(\nu)}{2}  t
\partial_x t^{-1} \Lambda + \omega \rho_{\rm edge} q \Lambda \right]
,
\end{equation}
which is none other than the aforementioned action for {\em
massless chiral edge excitations}.

In summary we can say that the {\em background field} methodology
that was previously introduced for renormalization group purposes
alone, now gets a new appearance in the theory and a fundamentally
different meaning in term of the {\em effective action} for
massless edge excitations. This effective action procedure emerges
from the theory itself and, unlike the background field
methodology, it provides the much sought after {\em Thouless
criterion} which associates the exact quantization of the Hall
conductance with the insensitivity of the bulk of the system to
changes in the boundary conditions.

%.................................................................
\subsubsection{\label{lcflc}Conductance fluctuations, level crossing}

The definition of the effective action, Eq. (\ref{response}),
implies that the exact expression for $S_{\rm eff}$ is invariant
under a change in the matrix field $t \rightarrow Ut$, i.e. the
replacement
\begin{equation}\label{gauge}
S_{\rm eff} [t^{-1} \Lambda t] \rightarrow S_{\rm eff} [t^{-1}
U^{-1} \Lambda U t]
\end{equation}
should leave the theory invariant. Here, the $U=U(\textbf{r})$
represents a unitary matrix field with an {\em integer} valued
topological charge, i.e. $U(\textbf{r})$ reduces to an arbitrary
$U(m) \times U(n)$ gauge at the edge of the system.

Notice that the expressions for $S_{\rm eff}$, Eqs
\eqref{response} and \eqref{seff}, in general violate the
invariance under Eq. \eqref{gauge}. This is so because we have
expressed $S^{\,\prime}[q]$ to lowest order in a series expansion in
powers of the derivatives acting on the $q$ field. If the response
parameters $\sigma^\prime_{xx}(L)$ and $\theta^\prime(L)$ are finite
then the invariance under Eq. \eqref{gauge} is generally {\em
broken} at each and every order in the derivative expansion. The
invariance is truly {\em recovered} only after the complete series
is taken into account, to infinite order. This situation typically
describes a {\em gapless} phase in the theory. The infinite series
(or at least an infinite subset of it) can in general be
rearranged and expressed in terms of {\em probability
distributions} for the response parameters $\sigma^\prime_{xx}(L)$
and $\theta^\prime(L)$. Quantum critical points, for example, seem to
generically display {\em broadly} distributed response parameters.
The large $N$ expansion has provided, once more, a lucid and exact
example of these statements~\cite{PruiskenBaranovVoropaevN}.

On the other hand, if the theory displays a {\em mass gap} then
each term in the series for $S^\prime[q]$ should vanish, i.e. the
response parameters are all exponentially small in the system
size. It is therefore possible to employ Eq. \eqref{gauge} as an
alternative criterion for the quantum Hall effect, namely by
demanding that the theory be invariant under Eq. \eqref{gauge}
{\em order by order} in the derivative expansion. This criterion
immediately demands that $\sigma^\prime_{xx}$ and $\theta^\prime$ in Eq.
\eqref{seff} be zero and the effective action be given by Eq.
\eqref{response1}. Equations \eqref{response1} and \eqref{gauge} imply
\begin{equation}\label{adiab}
S_{\rm eff} [t^{-1} U^{-1} \Lambda U t] = S_{\rm eff} [t^{-1}
\Lambda t] + 2\pi i k(\nu) {\mathcal C} [U^{-1} \Lambda U].
\end{equation}
Although the matrix $U$ merely gives rise to a phase factor that
can be dropped, physically it corresponds to an integer number
$n_e$ of (edge) electrons, equal to $k(\nu) {\mathcal C} [U^{-1}
\Lambda U]$, that have crossed the Fermi level. Eq. \eqref{adiab}
is therefore synonymous for the statement which says that the
quantization of the Hall conductance $\sigma_{xy} = k(\nu) \left[
e^2/h \right]$ is related to the quantization of flux
$\Phi = {\mathcal C} [U^{-1}\Lambda U] \left[ h/e \right]$
and quantization of charge $q_e = n_e \left[ e \right]$ according
to
\begin{equation}
q_e = \sigma_{xy} \Phi .
\end{equation}
%%%%%%%%%%%%%%%%%%%%%%%%%%%%%%%%%%%%%%%%%%%%%%%%%%%%%%%%%%%%%%%%%%
%
%
\subsection{\label{PO}Physical observables}
%
%%%%%%%%%%%%%%%%%%%%%%%%%%%%%%%%%%%%%%%%%%%%%%%%%%%%%%%%%%%%%%%%%%%

The response parameters $\sigma^\prime_{xx}$ and $\theta^\prime$ as well
as other observables, associated with the mass terms in the
non-linear sigma model, are in many ways the most significant
physical quantities in the theory that contain complete
information on the low energy dynamics of the instanton vacuum.
These quantities can all be expressed in terms of correlation
functions of the {\em bulk} field variables $Q_0$ alone. In this
Section we give a summary of the {\em physical observables} in the
theory which then completes the theory of massless chiral edge
excitations.

To start we introduce the following theory for the {\em bulk}
matrix field variables (dropping the subscript ``$0$'' on the
$Q_0$ from now onward)
\begin{equation}\label{bulkgen}
Z = \int\limits_{\partial V} \mathcal{D} Q \exp \left (
S_\sigma [Q] + S_{h}[Q]+S_{a}[Q]+S_{s}[Q]\right ).
\end{equation}
The subscript $\partial V$ indicates, as before, that the $Q$ is
kept fixed at $Q=\Lambda$ at the edge of the system. $S_\sigma$
stands for
\begin{equation}
S_\sigma [Q] =  -\frac{\sigma _{xx}}{8} \int d
\textbf{r}\tr(\nabla Q)^{2} +\frac{\theta}{16\pi} \int d
\textbf{r} \tr \varepsilon_{ab} Q \nabla_a Q \nabla_b Q
.\label{Ssig}
\end{equation}
The quantities $S_{h}$, $S_{a}$ and $S_{s}$ are the $U(m) \times
U(n)$ invariant mass terms that will be specified below.

%.............................................................
\subsubsection{Kubo formula}

Explicit expressions for response quantities $\sigma^\prime_{xx}$ and
$\theta^\prime$ can be derived following the analysis of
Ref.~\cite{Pruisken3}. The following $U(m) \times U(n)$ invariant
results have been obtained
\begin{eqnarray}\label{sdef}
\sigma_{xx}^\prime = \sigma_{xx} &+& \frac{\sigma_{xx}^2}{16 m n L^2}
\int d \textbf{r} d \textbf{r}^\prime \tr \langle Q
(\textbf{r})\nabla Q (\textbf{r}) Q (\textbf{r}^\prime)\nabla Q
(\textbf{r}^\prime)\rangle, \\
\theta^\prime = \theta &-& \frac{(m+n)\pi}{4 m n L^2} \sigma_{xx}
\int d\textbf{r} \tr \langle \Lambda Q \varepsilon_{ab} r_a
\partial_b Q \rangle \notag\\
&+ & \frac{\pi\sigma_{xx}^2}{8 m n L^2} \int d \textbf{r}
d\textbf{r}^\prime \tr \varepsilon_{ab} \langle Q (\textbf{r})
\nabla_a Q (\textbf{r}) Q (\textbf{r}^\prime)\nabla_b Q
(\textbf{r}^\prime) \Lambda \rangle.\label{tdef}
\end{eqnarray}
Here and from now onward the expectations are defined with respect
to the theory of Eq. \eqref{bulkgen}.

%.........................................................................
\subsubsection{Mass terms}

We shall be interested in traceless $U(m) \times U(n)$ invariant
operators that are linear in $Q$ $(O_h)$ and bilinear in $Q$
$(O_{s,a})$,
\begin{equation}\label{Shas}
S_{h}[Q] = z_h \int d\textbf{r}\, O_{h}[Q],\qquad S_{a,s}[Q] =
z_{a,s}\int d\textbf{r}\, O_{a,s}[Q].
\end{equation}
Here,
\begin{equation}\label{Oh}
O_{h}[Q] =  \tr \left [\Lambda -\frac{m-n}{m+n} 1_{m+n} \right] Q
.
\end{equation}
The bilinear operators generally involve a {\em symmetric} and an
{\em antisymmetric} combination~\cite{Pruisken5}
\begin{equation}\label{Osa}
O_{s,a}[Q] =  \sum \limits_{p,q}^{\alpha,\beta} K_{s,a}^{pq} \left
[ Q_{p p}^{\alpha \alpha} Q_{qq}^{\beta \beta} \pm Q_{p q}^{\alpha
\beta} Q_{qp}^{\beta \alpha} \right ],
\end{equation}
where $K_{s,a}$ is given as a $2 \times 2$ matrix
\begin{equation}\label{Ksa}
K_{s,a} = \begin{pmatrix}
 -\displaystyle \frac{m}{n\pm 1}  & 1 \\
 1 & - \displaystyle\frac{n}{m\pm 1}
\end{pmatrix}.
\end{equation}
These quantities permit us to define physical observables $z_h^{\prime
}$ and $z^\prime_{s,a}$ that are associated with the $z_h$ and
$z_{s,a}$ fields respectively. Specifically,
\begin{equation}\label{rendef}
z_h^{\prime} = z_h \frac{\langle O_h [Q]\rangle}{O_h [\Lambda]},
\quad z^{\prime}_s = z_s \frac{\langle O_s [Q]\rangle }{O_s
[\Lambda]},\quad z^{\prime}_a = z_a \frac{\langle O_a [Q]\rangle
}{O_a [\Lambda]}.
\end{equation}
The ratio on the right hand side merely indicates that the
expectation value of the operators is normalized with respect to
the classical value.

\subsubsection{$\beta$ and $\gamma$ functions}

The {\em observable parameters} of the previous Sections
facilitate a renormalization group study that can be extended to
include the non-perturbative effects of instantons. Since much of
the analysis is based on the theory in $2+\epsilon$ spatial
dimensions~\cite{Pruisken5} we shall first recapitulate some of
the results obtained from ordinary perturbative expansions. Let
$\mu^\prime$ denote the momentum scale associated with the
observable theory then the quantities $\sigma^\prime_{xx} =
\sigma_{xx}(\mu^\prime)$, $z_i^\prime = z_i (\mu^\prime)$ can be
expressed in terms of the renormalization group $\beta$ and
$\gamma$ functions according to (see Appendix ~\ref{App0})
\begin{eqnarray}\label{rgsigma}
 \hspace{2.5cm}\sigma^\prime_{xx} &=& \sigma_{xx}(\mu^\prime) = \sigma_{xx}(\mu_0) +
 \int_{\mu_0}^{\mu^\prime} \frac{d\mu}{\mu}~ \beta_{\sigma} (\sigma_{xx}) \\\label{rgz}
 z_i^\prime &=& z_i (\mu^\prime) =z_i (\mu_0)
 -\int_{\mu_0}^{\mu^\prime} \frac{d\mu}{\mu}~\gamma_i (\sigma_{xx})
 z_i (\mu),
\end{eqnarray}
where~\cite{BZG}
\begin{eqnarray}
\hspace{3cm}\beta_{\sigma} (\sigma_{xx}) &=&  \frac{m+n}{2\pi} + \frac{m n+1}{2\pi^2
\sigma_{xx}} +
{\mathcal O}(\sigma_{xx}^{-2})\label{pb1}\\
\gamma_h (\sigma_{xx}) &=& -\frac{m+n}{2\pi \sigma_{xx}} +{\mathcal O}(\sigma_{xx}^{-2})\label{pb2}\\
\gamma_s (\sigma_{xx}) &=& -\frac{m+n+2}{2\pi \sigma_{xx}}+{\mathcal O}(\sigma_{xx}^{-2})\label{pb3}\\
\gamma_a (\sigma_{xx}) &=& -\frac{m+n-2}{2\pi
\sigma_{xx}}+{\mathcal O}(\sigma_{xx}^{-2})\label{pb4}.
\end{eqnarray}
The effects of instantons have been studied in great detail in
Refs~\cite{LevineLibbyPruisken,Pruisken1} where the idea of the
$\theta$ renormalization was introduced. The main objective of the
present paper is to extend the results of Eqs
\eqref{rgsigma}-\eqref{pb4} to include the effect of instantons.
Recall that the $\gamma_i$ functions are of very special physical
interest. For example, the quantity $\gamma_h$ should vanish in
the limit $m,n \rightarrow 0$ indicating that the density of
levels of the electron gas is in general unrenormalized. At the
same time one expects the anomalous dimension $\gamma_a$ to become
positive as $m,n \rightarrow 0$ since it physically describes the
singular behavior of the (inverse) participation ratio of the
electronic levels. These statements serve as an important physical
constraint that one in general should impose upon the
theory~\cite{Pruisken5}.

%%%%%%%%%%%%%%%%%%%%%%%%%%%%%%%%%%%%%%%%%%%%%%%%%%%%%%%%%%%%%%%%%%%
%%%%%%%%%%%%%%%%%%%%%%%%%%%%%%%%%%%%%%%%%%%%%%%%%%%%%%%%%%%%%%%%%%%
%
\section{\label{Inst} Instantons}
%
%%%%%%%%%%%%%%%%%%%%%%%%%%%%%%%%%%%%%%%%%%%%%%%%%%%%%%%%%%%%%%%%%%%
%%%%%%%%%%%%%%%%%%%%%%%%%%%%%%%%%%%%%%%%%%%%%%%%%%%%%%%%%%%%%%%%%%%

%%%%%%%%%%%%%%%%%%%%%%%%%%%%%%%%%%%%%%%%%%%%%%%%%%%%%%%%%%%%%%%%%%%
%
%
\subsection{\label{Inst.FT}Introduction}
%
%%%%%%%%%%%%%%%%%%%%%%%%%%%%%%%%%%%%%%%%%%%%%%%%%%%%%%%%%%%%%%%%%%%

In the absence of symmetry breaking terms, the existence of finite
action solutions (instantons) follows from the Schwartz inequality
\begin{equation}\label{Schwartz}
\tr \left [ \nabla_{x} Q \pm i Q \nabla_{y} Q \right ]^{2} \geq 0,
\end{equation}
which implies
\begin{equation}\label{ineqS}
\frac{1}{8} \int d \textbf{r} \tr(\nabla Q)^{2} \geq 2 \pi
|\mathcal{C}[Q]| .
\end{equation}
Matrix field configurations that fulfill inequality (\ref{ineqS})
as an equality are called {\em instantons}. The classical action
becomes
\begin{equation}\label{SsigmaInst}
S_{\sigma}^{\rm inst} = - 2 \pi \sigma_{xx} |\mathcal{C}[Q] | +\,
i \theta\mathcal{C}[Q] .
\end{equation}
In this paper we consider single instantons only with a
topological charge $\mathcal{C}[Q] =\pm 1$. A convenient
representation of the single instanton solution is given by
~\cite{Pruisken2,Pruisken3}
\begin{equation}\label{InstSol}
Q_{\rm inst}(\textbf{r}) = T^{-1} {\Lambda}_{\rm inst}
(\textbf{r}) T, \qquad \Lambda_{\rm inst} (\textbf{r}) = \Lambda +
\rho(\textbf{r}) .
\end{equation}
Here, the matrix $\rho^{\alpha \beta}_{p q}(\textbf{r})$ has four
non-zero matrix elements only
\begin{equation}\label{rho1}
\rho^{11}_{11} = - \rho^{11}_{-1-1} = - \frac{2
\lambda^{2}}{|z-z_{0}|^{2}+\lambda^{2}}, \qquad \rho^{11}_{1-1} =
\bar{\rho}^{11}_{-11} =  \frac{2 \lambda
(z-z_{0})}{|z-z_{0}|^{2}+\lambda^{2}},
\end{equation}
with $z=x+iy$. The quantity $z_{0}$ describes the {\em position}
of the instanton and $\lambda$ is the scale size. These
parameters, along with the global unitary rotation $T \in U(m+n)$,
describe the manifold of the single instanton. The {\em
anti-instanton} solution with topological charge $\mathcal{C}[Q]=
-1$ is simply obtained by complex conjugation.

Next, to discuss the mass terms in the theory, we substitute Eq.
\eqref{InstSol} into Eqs \eqref{Oh} and \eqref{Osa}. Putting
$T=1_{m+n}$ for the moment then one can split the result for the
operators $O_i$ into a topologically trivial part and an instanton
part as follows
\begin{equation}\label{Oipm}
O_i [Q_{\rm inst}]=O_i [\Lambda_{\rm inst}] = O_i [\Lambda] +
O_i^{\rm inst}(\textbf{r}),
\end{equation}
where
\begin{equation}\label{Ohas1}
O_h [\Lambda] = \frac{4mn}{m+n},\quad O_a [\Lambda] = -4mn,\quad
O_s [\Lambda] = -4mn
\end{equation}
and
\begin{gather}
O_h^{\rm inst} (\textbf{r}) = 2 \rho_{11}^{11} (\textbf{r}),
\qquad O_a^{\rm inst}
(\textbf{r}) = -{4(m+n-1)} \rho_{11}^{11} (\textbf{r}) \notag \\
O_s^{\rm inst} (\textbf{r}) = -{4(m+n+1)} \rho_{11}^{11}
(\textbf{r}) \left( 1+ \frac{2 \rho_{11}^{11} (\textbf{r})}{m+n+2}
\right) .\label{Oinst00}
\end{gather}
Similarly we can write the free energy as the sum of two parts
\begin{equation}
\mathcal{F }^{class} = \mathcal{F}_0^{class} + \mathcal{F}_{\rm
inst}^{class} ,
\end{equation}
where $\mathcal{F}_0^{class}$ denotes the contribution of the
trivial vacuum with topological charge equal to zero ($Q=\Lambda$)
and $\mathcal{F}_{\rm inst}^{class}$ is the instanton part
\begin{equation}\label{F00}
\mathcal{F}_0^{class} = z_h \int d \textbf{r} O_h [\Lambda] + z_s
\int d \textbf{r} O_s [\Lambda] +z_a \int d \textbf{r} O_a
[\Lambda]
\end{equation}
\begin{eqnarray}\notag
\hspace{1.5cm}\mathcal{F}_{\rm inst}^{class} = \int\limits_{\rm
inst} \exp \Bigl [ &-& 2\pi \sigma_{xx} \pm \,i \theta
 + z_h \int d \textbf{r} O_h^{\rm inst} (\textbf{r}) \\ &+& z_s \int d \textbf{r}
 O_s^{\rm inst} (\textbf{r}) + z_a
\int d \textbf{r} O_a^{\rm inst}(\textbf{r})\Bigr ].\label{exp1}
\end{eqnarray}
The subscript ``$\rm inst$'' on the integral sign indicates that
the integral is in general to be performed over the manifold of
instanton parameters. Notice, however, that the global matrix $T$
is no longer a part of the instanton manifold except for the
subgroup $U(m) \times U(n)$ only that leaves the action invariant.
In the presence of the mass terms we therefore have, instead of
Eq. (\ref{InstSol}),
\begin{equation}\label{InstSol1a}
Q_{\rm inst}(\textbf{r}) = W^{-1} {\Lambda}_{\rm inst}
(\textbf{r}) W = \Lambda + W^{-1} \rho(\textbf{r}) W
\end{equation}
with $W \in U(m) \times U(n)$. On the other hand, the spatial
integrals in the exponential of Eq. (\ref{exp1}) still display a
logarithmic divergence in the size of the system. To deal with
these and other complications we shall in this paper follow the
methodology as outlined in Ref. \cite{PruiskenBaranov}. Before
embarking on the quantum theory, however, we shall first address
the idea of working with {\em constrained instantons}.

%%%%%%%%%%%%%%%%%%%%%%%%%%%%%%%%%%%%%%%%%%%%%%%%%%%%%%%%%%%%%%%%%%%
%
%
\subsection{\label{Inst.PBT} Constrained instantons}
%
%%%%%%%%%%%%%%%%%%%%%%%%%%%%%%%%%%%%%%%%%%%%%%%%%%%%%%%%%%%%%%%%%%%

It is well known that the discrete topological sectors do not in
general have stable classical minima for finite values of
$z_{i}=0$. It is nevertheless possible construct matrix field
configurations that minimize the action under certain fixed
constraints. In this Section we are interested in finite action
field configurations $Q$ that smoothly turn into the instanton
solution of Eq. (\ref{InstSol1a}) in the limit where the symmetry
breaking fields $z_i$ all go to zero. Although the methodology of
this paper avoids the idea of {\em constrained instantons} all
together, the problem nevertheless arises in the discussion of a
special aspect of the theory, the replica limit (Section
\ref{ZeroAD}). To simplify the analysis we will consider the
theory in the presence of the $S_h$ term only.

\subsubsection{Explicit solution}

To obtain finite action configurations for finite values of $z_h$
it is in many ways natural to start from the original solution of
Eq. (\ref{InstSol1a}) and minimize the action with respect to a
{\em spatially varying} scale size $\lambda(\textbf{r})$ rather
than a spatially independent parameter $\lambda$. Write
\begin{equation}\label{lx}
\lambda^2 \rightarrow \lambda^2(\textbf{r}) = \lambda^2 f\left (x,
\tilde{h}^2 \right ),
\end{equation}
where $f(x,\tilde{h}^2)$ is a dimensionless function of the
dimensionless quantities $x={r^2}/{\lambda^2}$ and $\tilde{h}^2=4
z_h \lambda^2 / \sigma_{xx}$ respectively. The strategy is to find
an optimal function $f$ with the following constraints
\begin{eqnarray}\label{bc1}
\hspace{5cm}f(x,0) &=& 1 \\
\label{bc2}
f(0,\tilde{h}^2) &=& 1 \\
\label{bc3} f(x \rightarrow \infty , \tilde{h}^2 > 0) &=& 0 .
\end{eqnarray}
The first of these equations ensures that in the absence of mass
terms ($z_h = \tilde{h}^2 = 0$) we regain the original instanton
solution. The second and third ensure that the classical action is
finite for finite values of $z_h$ or $\tilde{h}^2$.

\begin{figure}
\centerline{\includegraphics[width=110mm]{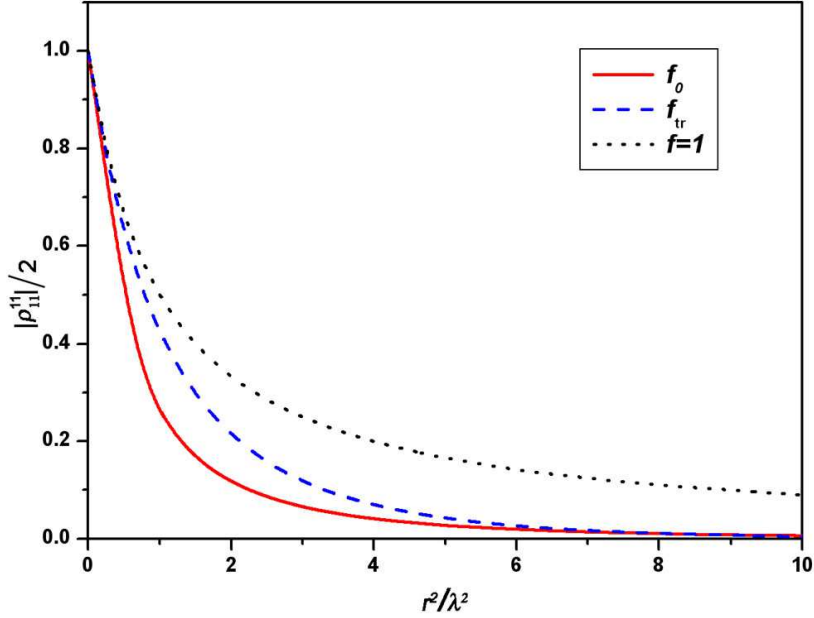}} \caption{The
matrix element $|\rho_{11}^{11}|/2$ as the function of
$r^2/\lambda^2$ for $\tilde h^2=0.3$.} \label{FIG5}
\end{figure}

It is easy to see that for all functions $f(x,\tilde{h}^2)$
satisfying Eqs \eqref{bc1} - \eqref{bc3}  the topological charge
equals unity, i.e.
\begin{equation}\label{topq2}
\mathcal{C}[Q] = \int \limits_{0}^{\infty} d x \frac{f-x
\partial_x f}{(x+f)^2}= 1 .
\end{equation}
Next, the expression for the action becomes (discarding the
constant terms in the definition of $S_h$)
\begin{equation}
S_\sigma+S_h = - \pi \sigma_{xx} \int\limits_{0}^{\infty} d x
\Bigl \{ \frac{f}{(x+f)^2} \Bigl[ 1 + \Bigl ( 1- \frac{x
\partial_x f}{f}\Bigr )^2\Bigr ]+ \tilde h^2 \frac{f}{x+f} \Bigr
\}.\label{CI1}
\end{equation}
The function $f_0(x, \tilde{h}^2)$ that optimizes the action
satisfies the following equation
\begin{equation}\label{MinF}
-x [ 2 \partial^2_x f_0 - f_0 (\partial_x f_0)^2 ] + 2 \frac{f_0
\partial_x f_0}{x + f_0} [ x \partial_x f_0 - 2 f_0 ] + \tilde h^2 f^2_0 = 0 .
\end{equation}
We are generally interested in the limit $\tilde h^2 \ll 1$ only.
Eq. \eqref{MinF} cannot be solved analytically. The asymptotic
behavior of $f_0(x, \tilde{h}^2)$ in the limit of large and small
values of $x$ is obtained as follows. For $x \gg 1$, we find
\begin{equation}\label{XLD}
f_0(x, \tilde{h}^2 ) \propto \tilde h^2 x K^2_1(\tilde h
\sqrt{x})=
  \begin{cases}
    \displaystyle\frac{\pi\tilde h}{2}\sqrt x \exp(-2\tilde h\sqrt x), & \quad x\gg \tilde h^{-2}, \\
    \displaystyle 1+\tilde h^2 x\ln\frac{\tilde h\sqrt x}{2}, & \quad \tilde h^{-2}\gg x \gg
    1.
  \end{cases}
\end{equation}
Here, $K_1(z)$ stands for the modified Bessel function. In the
regime $x \ll 1$ we obtain
\begin{equation}\label{XSD}
f_0(x, \tilde h^2) \approx 1+ 4\tilde h^2 x,\qquad x \ll 1.
\end{equation}
Notice that these asymptotic results can also be written more
simply as follows

\begin{equation}\label{XLD1}
f_0(x, \tilde{h}^2 ) \approx
  \begin{cases}
    \displaystyle\frac{\pi}{2} hr \exp(-2 hr), & \quad r \gg h^{-1} \\
    \displaystyle 1+ (hr)^2 \ln\frac{hr}{2}, & \quad h^{-1} \gg r \gg
    \lambda \\
    1+ 4 (hr)^2 , & r \ll \lambda.
  \end{cases}
\end{equation}
In the intermediate regimes $r \approx \lambda$ and $r \approx
h^{-1}$ we have obtained the solution numerically. In
Fig.~\ref{FIG5} we plot the results in terms of the matrix element
$\rho^{11}_{11} (r)$, Eq. \eqref{rho1},
\begin{equation}
\rho^{11}_{11} (r) = -\frac{2\lambda}{r^2 + \lambda^2} \rightarrow
-\frac{2f}{x+f} .
\end{equation}
Comparing the result with $\tilde h^2 =0.3$ with the original
instanton result, $f=1$, we see that the main difference is in the
asymptotic behavior with large $r$ where the $\rho^{11}_{11}$ for
the ``constrained instanton"  vanishes exponentially ($\propto
\exp(-2 hr)$), rather than algebraically ($\propto 1/r^2$).

%////////////////////////////////////////////////////////////
%///////////////////////////////////////////////////////////
%//////////////////////////////////////////////////////////////
%///////////////////////////////////////////////////////////////
\subsubsection{Finite action}

We conclude that in the presence of the mass term $S_h$ the
requirement of finite action forces the instanton scale size to
become spatially dependent in general such that the contributions
from large distances $r \gg h^{-1}$ are strongly suppressed in the
spatial integrals. On the other hand, from Eqs (\ref{XLD}) and
(\ref{XSD}) one can also see that for small scale sizes $\lambda$
such that
\begin{equation}
\lambda \ll \lambda_h \propto h^{-1},
\end{equation}
the dominant contribution comes from the original, unconstrained
instanton solution of Eq. \eqref{InstSol}.

In order to see the effect of the constrained instanton on the
analytic form of the action we first consider a simpler example of
a trial function $f_{\rm tr}(x)$ of the type
\begin{equation}\label{tf}
f_{\rm tr}(x)=\exp(-a x) .
\end{equation}
Here we use $a$ as a variational parameter. The classical action
becomes
\begin{equation}\label{S5}
S_\sigma+S_h = -2 \pi \sigma_{xx}\Bigl [ 1 + \frac{a}{2} +
\frac{\tilde h^{2}}{2}\Bigl (\ln \frac{1}{a}-\gamma\Bigr)\Bigr ]
+\mathcal{O}(\tilde h^4),
\end{equation}
where the constant $\gamma\approx 0.577$ stands for the Euler
constant. The optimal value of $a$ is given by
\begin{equation}\label{ep1}
a = \tilde h^{2}, \qquad f_{\rm tr}(x)=\exp(-\tilde h^2 x).
\end{equation}
Hence we find for Eq. \eqref{S5}
\begin{equation}\label{S6}
S_\sigma+S_h = -2 \pi \sigma_{xx}\Bigl [ 1 + \frac{\tilde
h^{2}}{2}\ln \frac{1.53}{\tilde h^2}\Bigr ].
\end{equation}
This simple result obtained from the trial function \eqref{ep1}
has the same asymptotic form as one determined by the much more
complicated optimal function $f_0(x)$ in the limit $\tilde h^2
\rightarrow 0$, except that the numerical constant $1.53$ is
replaced by $0.85$.

Eq. \eqref{S6} indicates that that in the limit $h \rightarrow 0$
the finite action of the constrained instanton smoothly goes over
into the finite action of the unconstrained instanton.
Furthermore, it is easy to see that the final result of Eq.
\eqref{S6} has precisely the features that one would normally
associate with mass terms. Consider for example the action of the
unconstrained instanton with a fixed scale size $\lambda$. By
putting the system inside a large circle of radius $R$ we obtain
the following result
\begin{equation}\label{S7}
S_\sigma+S_h = -2 \pi \sigma_{xx}\Bigl [ 1 - \frac{\lambda^2}{R^2}
+ \frac{\tilde h^2}{2} \ln \frac{R^2}{\lambda^2}\Bigr ].
\end{equation}
By comparing Eq. \eqref{S6} and Eq. \eqref{S7} we conclude that
the following effective size $R_h$ is induced by the $h$ field
\begin{equation}\label{Rh}
R_h = \frac{1.47}{h}.
\end{equation}

The effect of mass terms can therefore be summarized as follows.
First, the integral over scale sizes $\lambda$ (see Eq.
\eqref{exp1}) effectively proceeds over the interval $\lambda
\lesssim h^{-1}$. Secondly, the spatial integrals in the theory
are cut off in infrared and the effective sample size equals
$R_h$. The two different scales that are being induced by $S_h$,
$h^{-1}$ and $R_h$, are of the same order of magnitude.

%////////////////////////////////////////////////////////////
%///////////////////////////////////////////////////////////
\begin{figure}
\centerline{\includegraphics[width=110mm]{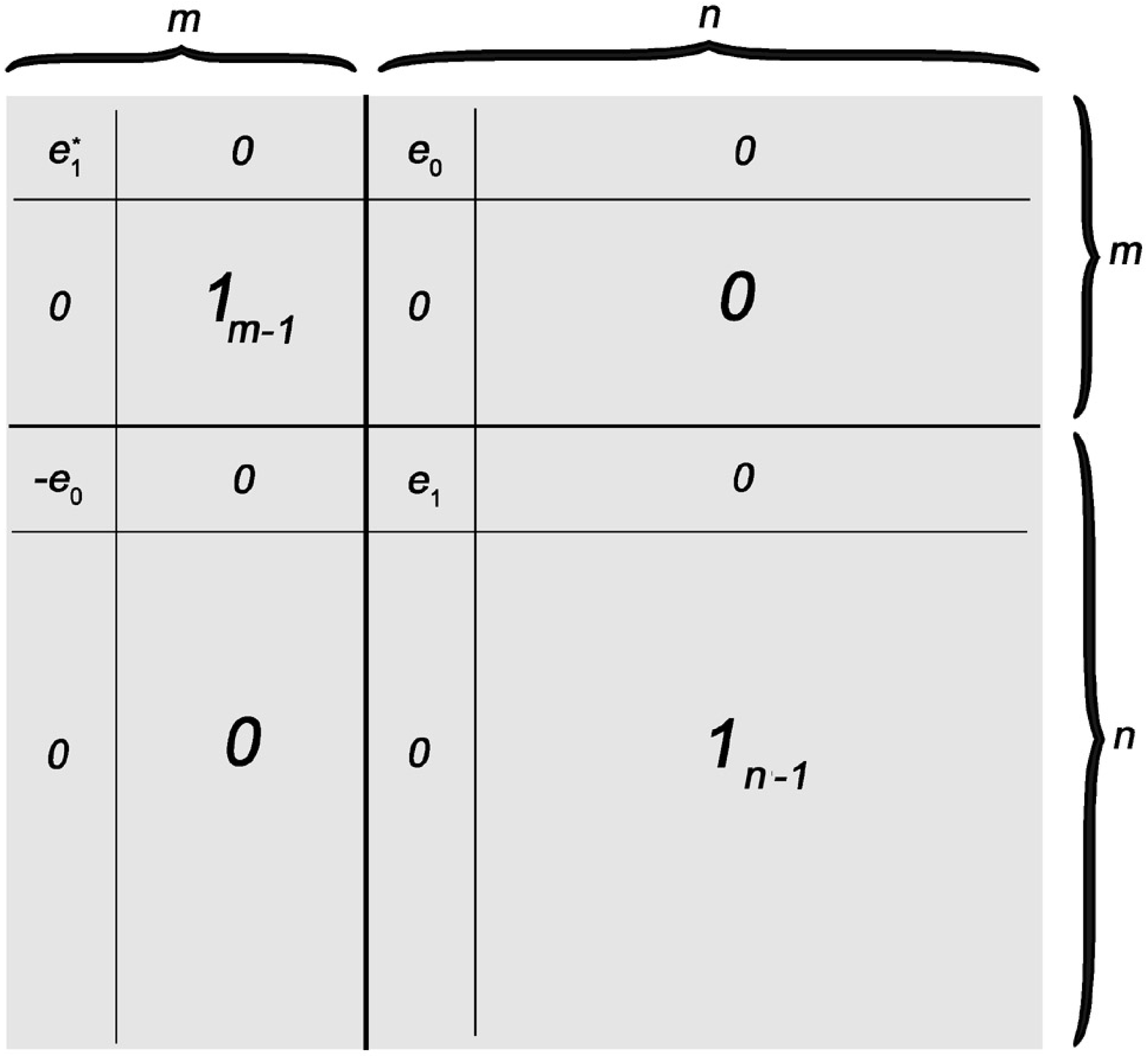}} \caption{The
matrix $R$.} \label{FIG6}
\end{figure}
%//////////////////////////////////////////////////////////////
%///////////////////////////////////////////////////////////////

%%%%%%%%%%%%%%%%%%%%%%%%%%%%%%%%%%%%%%%%%%%%%%%%%%%%%%%%%%%%%%%%%%%%%%
%%%%%%%%%%%%%%%%%%%%%%%%%%%%%%%%%%%%%%%%%%%%%%%%%%%%%%%%%%%%%%%%%%%%%%
%
%
\section{\label{QFAI}Quantum theory}
%
%%%%%%%%%%%%%%%%%%%%%%%%%%%%%%%%%%%%%%%%%%%%%%%%%%%%%%%%%%%%%%%%%%%%%%%
%%%%%%%%%%%%%%%%%%%%%%%%%%%%%%%%%%%%%%%%%%%%%%%%%%%%%%%%%%%%%%%%%%%%%%%

One of the problems with the idea of constrained instantons is
that it does not facilitate an expansion of the theory about $z_i
=0$ in any obvious fashion. This is unlike the method of {\em
spatially varying masses} which is based on the results obtained
in Ref. \cite{Pruisken2} and \cite{Pruisken3}. To start we first
recapitulate the formalism of the theory without mass terms in
Section \ref{QFAI.P}. In Section \ref{P.Metric} we introduce the
idea of {\em spatially varying masses}. Finally, in Section
\ref{QFAI.A} we present the complete action of the small
oscillator problem that will be used in the remainder of this
paper.

%%%%%%%%%%%%%%%%%%%%%%%%%%%%%%%%%%%%%%%%%%%%%%%%%%%%%%%%%%%%%%%%%%%%%%%%%
%
%
\subsection{\label{QFAI.P}Preliminaries}
%
%%%%%%%%%%%%%%%%%%%%%%%%%%%%%%%%%%%%%%%%%%%%%%%%%%%%%%%%%%%%%%%%%%%%%%%%
To obtain the most general matrix field variable $Q$ with
topological charge equal to unity we first rewrite the instanton
solution $\Lambda_{\rm inst}$ in Eqs \eqref{InstSol} and
\eqref{rho1} as a unitary rotation $R$ about the trivial vacuum
$\Lambda$
\begin{equation}
\Lambda_{\rm inst} = R^{-1} \Lambda R
\end{equation}
where
%\begin{equation}
%R^{\alpha\beta}_{pp^{\prime}} = R^{(\alpha)}_{pp^{\prime}}
%\delta^{\alpha\beta},
%\end{equation}
%where
%\begin{equation}
%R^{11} =
%\begin{pmatrix}
%\bar{e}_1 & e_0 \\
%-e_0 & e_1
%\end{pmatrix}
%\end{equation}
%and
%\begin{equation}
%R^{\alpha\beta} =\begin{pmatrix}
%1 & 0 \\
%0 & 1
%\end{pmatrix}
%\end{equation}
\begin{equation}
R =
\begin{pmatrix}
\delta^{\alpha\beta}+(\bar{e}_1-1)\delta^{\alpha 1}\delta^{\beta 1} & e_0 \delta^{\alpha 1}\delta^{\beta 1} \\
-e_0\delta^{\alpha 1}\delta^{\beta 1} &
\delta^{\alpha\beta}+(e_1-1)\delta^{\alpha 1}\delta^{\beta 1}
\end{pmatrix}.
\end{equation}
The quantities $e_0$ and $e_1$ are defined as
\begin{equation}
e_0 = \frac{\lambda}{\sqrt{|z-z_0|^2 + \lambda^2}}, \qquad e_1 =
\frac{z-z_0}{\sqrt{|z-z_0|^2 + \lambda^2}}.
\end{equation}
For illustration we have written the full matrix
$R^{\alpha\beta}_{pp'}$ in Fig.~\ref{FIG6}.
%Notice that in terms of the $e_0$, $e_1$ the
%matrix elements $\rho^{11}_{pp^{\prime}}$ can be written as
%\begin{equation}
%\rho^{11} = 2
%\begin{pmatrix}
%-e_0^2 & e_0 e_1 \\
%e_0 \bar{e}_1 & e_0^2
%\end{pmatrix}
%\end{equation}
It is a simple matter to next generalize these expressions and the
result is
\begin{equation}\label{top1}
Q=T^{-1}_0 R^{-1} q\, R\, T_0 .
\end{equation}
Here, $T_0$ denotes a global $U(m+n)$ rotation. The matrix $q$
with $q^2 = 1_{m+n}$  represents the small fluctuations about the
one instanton. Write
\begin{equation}\label{qpar}
q = w +\Lambda \sqrt{1_{m+n} - w^2}
\end{equation}
with
\begin{equation}
w=\begin{pmatrix}
0 & v \\
v^\dag & 0
\end{pmatrix}
\end{equation}
then the matrix $q$ can formally be written as a series expansion
in powers of the $m \times n$ complex matrices $v$, $v^{\dag}$
which are taken as the independent field variables in the problem.
%%%%%%%%%%%%%%%%%%%%%%%%%%%%%%%%%%%%%%%%%%%%%%%%%%%%%%%%%%%%%%%
%
%
\subsubsection{\label{P.Curved} Stereographic projection}
%
%%%%%%%%%%%%%%%%%%%%%%%%%%%%%%%%%%%%%%%%%%%%%%%%%%%%%%%%%%%%%%%

Eq. \eqref{top1} lends itself to an exact analysis of the small
oscillator problem. First we recall the results obtained for the
theory without mass terms~\cite{Pruisken2,Pruisken3},
\begin{equation}\label{Ai}
%\frac{\sigma_{xx}}{8} \int d \textbf{r}\tr \nabla_i Q \nabla_i Q
%= \frac{\sigma_{xx}}{8} \int d \textbf{r}\tr [\nabla_i +A_i , q]
%[\nabla_i +A_i , q] .
\frac{\sigma_{xx}}{8} \int d \textbf{r}\tr (\nabla Q)^2 =
\frac{\sigma_{xx}}{8} \int d \textbf{r}\tr [\nabla +\textbf{A},
q]^2,
\end{equation}
where the matrix $\textbf{A}$ contains the instanton degrees of
freedom
\begin{equation}
\textbf{A} = R T_0 \nabla T^{-1}_0 R^{-1} =R \nabla R^{-1} .
\end{equation}
By expanding the $q$ in Eq. \eqref{Ai} to quadratic order in the
quantum fluctuations $v$, $v^\dag$ we obtain the following results
\begin{gather}\notag
\frac{\sigma_{xx}}{8} \int d \textbf{r} \tr [\nabla +\textbf{A},
q]^2 = \frac{\sigma_{xx}}{4}\int d \textbf{r}\mu^2 (\textbf{r})
\Biggl[ v^{11} O^{(2)} v^{\dag 11} +\sum \limits_{\alpha=2}^{m} v^{\alpha 1} O^{(1)} v^{\dag 1 \alpha}\\
 + \sum \limits_{\beta=2}^{n} v^{1 \beta} O^{(1)} v^{\dag\beta 1} +\sum
\limits_{\alpha=2}^{m}\sum \limits_{\beta=2}^{n} v^{\alpha \beta}
O^{(0)} v^{\dag\beta \alpha} \Biggr].\label{fluct11}
\end{gather}
The three different operators $O^{(a)}$ are given as
\begin{equation}\label{Oa0}
O^{(a)} = \frac{(r^2 + \lambda^2 )^2}{4\lambda^2} \left[ \nabla_b
+ \frac{i a \varepsilon_{bc} r_c}{r^2 + \lambda^2} \right]^2
+\frac{a}{2}.
\end{equation}
%\begin{eqnarray}
%O^{(2)} &=& \frac{(r^2 + \lambda^2 )^2}{4\lambda^2} \left[
%\nabla_i + \frac{2i \epsilon_{ij} r_j}{r^2 + \lambda^2} \right]^2 +1 \\
%O^{(1)} &=& \frac{(r^2 + \lambda^2 )^2}{4\lambda^2} \left[
%\nabla_i + \frac{i \epsilon_{ij} r_j}{r^2 + \lambda^2} \right]^2 +\frac{1}{2} \\
%O^{(0)} &=& \frac{(r^2 + \lambda^2 )^2}{4\lambda^2} \left[
%\nabla_i \right]^2 .
%\end{eqnarray}
The introduction of a measure $\mu^2(\textbf{r})$ for the spatial
integration in Eq. \eqref{fluct11},
\begin{equation}\label{measure}
\mu(\textbf{r}) = \frac{2\lambda}{r^2 + \lambda^2} ,
\end{equation}
indicates that the quantum fluctuation problem is naturally
defined on a sphere with radius $\lambda$. It is convenient to
employ the stereographic projection
\begin{equation}\label{eta}
\eta = \frac{r^{2}-\lambda^{2}}{r^{2}+\lambda^{2}},\qquad -1 <
\eta < 1
\end{equation}
\begin{equation}\label{theta}
\theta = \tan^{-1} \frac{y}{x},\qquad 0 \leq \theta < 2 \pi .
\end{equation}
In terms of $\eta$, $\theta$ the integration can be written as
\begin{equation}
\int d \textbf{r}\mu^2(\textbf{r})  = \int d \eta d\theta .
\end{equation}
Moreover,
\begin{equation}\label{e0e1}
e_{0} = \sqrt{\frac{1-\eta}{2}},\qquad \qquad e_{1} =
\sqrt{\frac{1+\eta}{2}} e^{i \theta},
\end{equation}
and the operators become
\begin{equation}\label{Oa}
O^{(a)} = \frac{\partial}{\partial \eta} \left [ (1-\eta^{2})
\frac{\partial}{\partial \eta} \right ] + \frac{1}{1-\eta^{2}}
\frac{\partial^{2}}{\partial^{2} \theta} - \frac{i a}{1-\eta}
\frac{\partial}{\partial \theta}  - \frac{a^{2}}{4}
\frac{1+\eta}{1-\eta} + \frac{a}{2} ,
\end{equation}
with $a=0,1,2$. Finally, using Eq. \eqref{fluct11} we can count
the total number of fields $v^{\alpha\beta}$ on which each of the
operators $O^{(a)}$ act. The results are listed in
Table~\ref{TZM}.

%..........................................................................
%

%************************************************************************
\begin{table}
\begin{center}
\caption{\vspace{0.5cm}Counting the number of zero modes}
\begin{tabular}{||c|c|c||}\hline\hline & &\\
Operator & The number of fields $v^{\alpha\beta}$ involved & Degeneracy \\
& & \\
\hline & & \\
$O^{(0)}$ & $(m-1)(n-1)$ & $1$ \\
& &  \\
$O^{(1)}$ & $(m-1)+(n-1)$ & $2$ \\
& &  \\
 $O^{(2)}$ & $1$ & $3$ \\
 & & \\ \hline\hline
\end{tabular}
\label{TZM}
\end{center}
\vspace{0.5cm}
\end{table}
%****************************************************************

%..................................................................................
\subsubsection{\label{ES}Energy spectrum}

%...................................................................................

We are interested in the eigenvalue problem
\begin{equation}\label{EigEq} O^{(a)} \Phi^{(a)} (\eta, \theta) =
E^{(a)} \Phi^{(a)} (\eta, \theta) ,
\end{equation}
where the set of eigenfunctions $\Phi^{(a)}$ are taken to be
orthonormal with respect to the scalar product
\begin{equation}
(\bar{\Phi}^{(a)}_1 , \Phi^{(a)}_2) = \int d\eta d\theta\,
\bar{\Phi}^{(a)}_1(\eta, \theta) \Phi^{(a)}_2(\eta, \theta) .
\end{equation}
The Hilbert space of square integrable eigenfunctions is given in
terms of Jacobi polynomials,
\begin{equation}\label{Jacobi}
P^{\alpha, \beta}_{n}(\eta) =  \frac{(-1)^{n}}{2^{n} n!}
(1-\eta)^{-\alpha}(1+\eta)^{-\beta} \frac{d^{n}}{d\eta^{n}}
(1-\eta)^{n+\alpha}(1+\eta)^{n+\beta}
\end{equation}
Introducing the quantum number $J$ to denote the discrete energy
levels
\begin{equation}\label{Ev}
\begin{array}{cclrcl}
E^{(0)}_J &=& J(J+1),\qquad & J &=& 0,1, \cdots\\
E^{(1)}_J &=& (J-1)(J+1),\qquad & J &=& 1,2, \cdots\\
E^{(2)}_J &=& (J-1)(J+2), \qquad & J &=& 1,2, \cdots
\end{array}
\end{equation}
then the eigenfunctions are labelled by $(J,M)$ and can be written
as follows
\begin{equation}\label{Ef}
\begin{array}{cclrcl}
\Phi^{(0)}_{J,M}& = &C_{J,M}^{(0)}
e^{i M\theta}\sqrt{(1-\eta^2)^M}P_{J-M}^{M,M}(\eta),&\, M& = & -J,\cdots,J\\
\Phi^{(1)}_{J,M}& = &C_{J,M}^{(1)} e^{i
M\theta}\sqrt{(1-\eta^2)^M}\sqrt{1-\eta}
P_{J-M-1}^{M+1,M}(\eta),&\, M & = &-J,\cdots,J-1 \\
\Phi^{(2)}_{J,M}& = &C_{J,M}^{(2)} e^{i
M\theta}\sqrt{(1-\eta^2)^M}(1-\eta)P_{J-M-1}^{M+2,M}(\eta),& \, M
& = &-J-1,\cdots,J-1
\end{array}
%\Phi^{(a)}_{J,M} = C_{J,M}^{(a)} e^{i M \theta}
%\sqrt{(1-\eta^{2})^M(1-\eta)^a} P_{J-M-k_{a} }^{M+a,M}(\eta),
\end{equation}
where the normalization constants equal
\begin{equation}\label{Ca}
\begin{array}{ccl}
C_{J,M}^{(0)}& = &\displaystyle \frac{\sqrt{\Gamma(J-M+1)\Gamma(J+M+1)(2J+1)}}{2^{M+1}\sqrt\pi\Gamma(J+1)},\\
C_{J,M}^{(1)}& = &\displaystyle \frac{\sqrt{\Gamma(J-M)\Gamma(J+M+1)}}{2^{M+1}\sqrt\pi\Gamma(J)},\\
C_{J,M}^{(2)}& = &\displaystyle
\frac{\sqrt{\Gamma(J-M)\Gamma(J+M+2)(2J+1)}}{2^{M+2}\sqrt\pi\Gamma(J)\sqrt{J(J+1)}}.
\end{array}
%C_{J,M}^{(a)}= \frac{\sqrt{(2J+1)^{|a-1|}}}{2^{M+a-k_{a}+1}
%\sqrt{\pi}}
%\frac{\sqrt{\Gamma(J-M+1-k_{a})}}{\sqrt{\Gamma(J+1-k_{a})}} \frac{
%\sqrt{\Gamma(J+M+1+a-k_{a})}}{\sqrt{\Gamma(J+1+2 a-3 k_{a})}}.
\end{equation}

%..............................................................
\subsubsection{\label{ZM}Zero modes}

%.................................................................

From Eq. \eqref{Ev} we see that the operators $O^{(a)}$ have the
following zero modes
\begin{equation}\label{ZM.0}
\begin{array}{cclcllcllcl}
O^{(0)} & \Longrightarrow & \Phi_{0,0}^{(0)} & = & 1, \quad & & &
& &
& \\
O^{(1)} & \Longrightarrow & \Phi_{1,-1}^{(1)} & = &
\displaystyle\frac{1}{\sqrt{2 \pi}}\bar{e}_{1} ,\quad  &
\Phi_{1,0}^{(1)}&= &\displaystyle\frac{1}{\sqrt{2
\pi}}e_{0}, \quad  & & & \\
 O^{(2)}& \Longrightarrow & \Phi_{1,-2}^{(2)} & = & \displaystyle
\sqrt{\frac{3}{4\pi}} \bar{e}_{1}^{2}, \quad
&\displaystyle\Phi_{1,-1}^{(2)} & = &\displaystyle\sqrt{\frac{3}{2
\pi}} e_{0} \bar{e}_{1}, \quad & \Phi_{1,0}^{(2)} & = &
\displaystyle\sqrt{\frac{3}{4\pi}} e_{0}^{2}.
\end{array}
\end{equation}
The number of the zero modes of each $O^{(a)}$ is listed in
Table~\ref{TZM}. The total we find $2 (mn + m+n)$ zero modes in
the problem. Next, it is straight forward to express these zero
modes in terms of the instanton degrees of freedom contained in
the matrices $R$ and $T_0$ of Eq. \eqref{top1}. For this purpose
we write the instanton solution as follows
\begin{equation}
Q_{\rm inst} (\xi_i) = U^{-1} (\xi_i) \Lambda U (\xi_i).
\end{equation}
Here, $U= R\, T_0$ and the $\xi_i$ stand for the position $z_0$ of
the instanton, the scale size $\lambda$ and the generators of
$T_0$. The effect of an infinitesimal change in the instanton
parameters $\xi_i \rightarrow \xi_i + \varepsilon_i$ on the
$Q_{\rm inst}$ can be written in the form of Eq. \eqref{top1} as
follows
\begin{equation}
Q_{\rm inst} (\xi_i + \varepsilon_i) = U^{-1} (\xi_i)
q_\varepsilon U (\xi_i),
\end{equation}
where
\begin{equation}
q_\varepsilon \approx \Lambda - \varepsilon_i \left[ U
\partial_{\varepsilon_i} U^{-1} , \Lambda \right].
\end{equation}
Notice that
\begin{equation}
- \varepsilon_i \left[ U \partial_{\varepsilon_i} U^{-1} , \Lambda
\right] = 2 \varepsilon_i
\begin{pmatrix}
  0 & \left[ U\partial_{\varepsilon_i} U^{-1} \right]_{1,-1}^{\alpha\beta} \\
  -\left[ U \partial_{\varepsilon_i} U^{-1} \right]_{-1,1}^{\alpha\beta} & 0
\end{pmatrix}.
\end{equation}
By comparing this expression with Eq. \eqref{top1} we see that the
small changes $\varepsilon_i$ tangential to the instanton manifold
can be cast in the form of the quantum fluctuations $v$, $v^\dag$
according to
\begin{equation}
v^{\alpha\beta} = 2\varepsilon_i \left[ R\,T_0
\partial_{\varepsilon_i} T_0^{-1}
R^{-1} \right]_{1,-1}^{\alpha\beta}, \qquad v^{\dag\alpha\beta} =
-2\varepsilon_i \left[ R\,T_0 \partial_{\varepsilon_i} T_0^{-1}
R^{-1} \right]_{-1,1}^{\alpha\beta}.
\end{equation}
Next we wish to work out these expressions explicitly. Let $t$
denote an infinitesimal $U(m+n)$ rotation
\begin{equation}
T_0 =1_{m+n} +i\, t ,
\end{equation}
and $\delta\lambda$, $\delta z_0$ infinitesimal changes in the
scale size and position respectively,
\begin{equation}
R(\lambda+\delta\lambda, z_0 + \delta z_0) = R(\lambda, z_0 )
\left[ 1_{m+n} +\delta\lambda\, R^{-1}\partial_\lambda R + \delta
z_0\, R^{-1} \partial_{z_0} R \right] .
\end{equation}
The zero frequency modes can be expressed in terms of the
instanton parameters $t$, $\delta\lambda$ and $\delta z_0$ and the
eigenfunctions $\Phi^{(a)}_{JM}$ according to
%%%%
%OLD
%%%%
%\begin{equation}
%\begin{array}{lcccl}
%O^{(0)}& \Rightarrow & v^{\alpha \beta} & = & 2 i t_{1,-1}^{\alpha \beta} \Phi_{0,0}^{(0)} \\
%
%\begin{array}{l}
%O^{(1)} \\
%\hspace{0.1cm}
%\end{array}
%&
%\begin{array}{c}
%\Rightarrow \\
%\hspace{0.1cm}
%\end{array}
%& \left[
%\begin{array}{c}
%  v^{\alpha 1} \\
%  v^{1 \beta}
%\end{array}
%\right] &= & 2 \sqrt{2\pi} i \begin{pmatrix}
%  t_{1,-1}^{\alpha 1} & -t_{1,1}^{\alpha 1} \\
%  t_{1,-1}^{1 \beta} & ~t_{-1,-1}^{1\beta}
%\end{pmatrix} \left[
%\begin{array}{c}
%  \Phi_{1,-1}^{(1)} \\
%  \Phi_{1,0}^{(1)}
%\end{array}
%\right]
% \end{array}
%\end{equation}
%\begin{eqnarray}
%O^{(2)} & \Rightarrow &
%v^{11}  \notag\\
%&=& \displaystyle 4 \sqrt{\frac{\pi}{3}}\left[i t_{-1,1}^{1 1},
%\frac{1}{\sqrt{2}} \left( i t_{-1,-1}^{11} - i
%t_{1,1}^{11}-\frac{\delta\lambda}{\lambda} \right),\, -i
%t_{1,-1}^{1 1} +\frac{\delta \bar{z}_0}{\lambda} \right ] \left[
%\begin{array}{c}
% \Phi_{1,-2}^{(2)} \\
% \Phi_{1,-1}^{(2)} \\
% \Phi_{1,0}^{(2)}
%\end{array}
%\right]\notag
%\end{eqnarray}
%%%%%%%%%%%%
\begin{equation}
\begin{array}{lcl}
\hspace{0.3cm} v^{\alpha \beta} & = & 2 i t_{1,-1}^{\alpha \beta} \Phi_{0,0}^{(0)} \\
\left[
\begin{array}{l}
  v^{\alpha 1} \\
  v^{1 \beta}
\end{array}
\right] &= & 2 \sqrt{2\pi} i \begin{pmatrix}
  t_{1,-1}^{\alpha 1} & -t_{1,1}^{\alpha 1} \\
  t_{1,-1}^{1 \beta} & ~t_{-1,-1}^{1\beta}
\end{pmatrix} \left[
\begin{array}{c}
  \Phi_{1,-1}^{(1)} \\
  \Phi_{1,0}^{(1)}
\end{array}
\right] \\
\hspace{0.3cm}v^{11} & = & \displaystyle 4
\sqrt{\frac{\pi}{3}}\left[i t_{-1,1}^{1 1}, \left( i
t_{-1,-1}^{11} - i t_{1,1}^{11}-\frac{\delta\lambda}{\lambda}
\right),\, -i t_{1,-1}^{1 1} +\frac{\delta \bar{z}_0}{\lambda}
\right ]  \left[
\begin{array}{c}
 \Phi_{1,-2}^{(2)} \\
 \Phi_{1,-1}^{(2)} \\
 \Phi_{1,0}^{(2)}
\end{array}
\right]
\end{array}
\end{equation}

%////////////////////////////////////////////////////////////
%///////////////////////////////////////////////////////////
\begin{figure}
\centerline{\includegraphics[width=110mm]{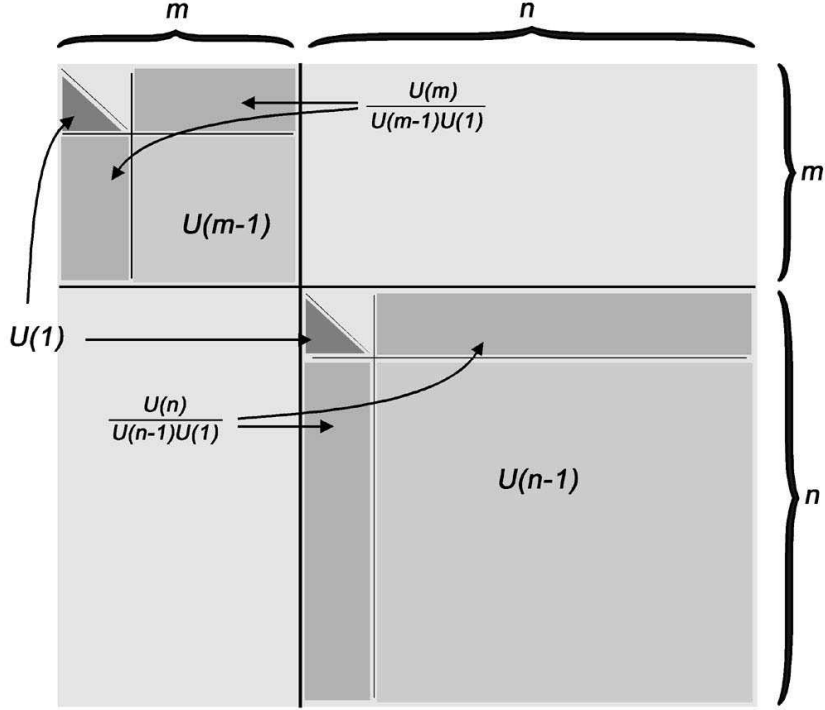}} \caption{Symmetry
breaking.} \label{FIG7}
\end{figure}
%//////////////////////////////////////////////////////////////
%///////////////////////////////////////////////////////////////

From this one can see that besides the scale size $\lambda$ and
position $z_0$ the instanton manifold is spanned by the
$t^{\alpha\beta}_{1,-1}$ and $t^{\alpha\beta}_{-1,1}$ which are
the generators of $U(m+n)/U(m) \times U(n)$. The
$t^{1\alpha}_{1,1}$ and $t^{\alpha 1}_{1,1}$ with $\alpha >1$ are
the generators of $U(n)/U(n-1) \times U(1)$ and the
$t^{1\alpha}_{-1,-1}$ and $t^{\alpha 1}_{-1,-1}$ those of
$U(m)/U(m-1) \times U(1)$. Finally, $t_{1,1}^{11} -t_{-1,-1}^{11}$
is the $U(1)$ generator describing the rotation of the $O(3)$
instanton about the $z$ axis. In total we find $2(mn + m+n)$ zero
modes as it should be. The hierarchy of symmetry breaking by the
one-instanton solution is illustrated in Fig.~\ref{FIG7}.

%................................................................

\subsection{\label{P.Metric} Spatially varying masses}

%.................................................................

We have seen that the quantum fluctuations about the instanton
acquire the metric of a {\em sphere}, Eq. \eqref{measure}. This,
however, complicates the problem of mass terms which are naturally
written in {\em flat} space. To deal with this problem we modify
the definition of the mass terms and introduce a spatially varying
momentum scale $\mu (\textbf{r})$ as follows
\begin{equation}\label{zRen}
z_i \rightarrow z_i \, \mu^2(\textbf{r}),
\end{equation}
such that the action now becomes {\em finite} and can be written
as
\begin{equation}\label{Ohmu}
S_i = z_i \int d\textbf{r}\, O_{i}[Q] \rightarrow z_i \int
d\textbf{r}\, \mu^2 (\textbf{r}) O_{i}[Q] .
\end{equation}
Several comments are in order. First of all, we expect that the
introduction of a spatially varying momentum scale
$\mu(\textbf{r})$ does not alter the singularity structure of the
theory at short distances. We shall therefore proceed and first
develop a full quantum theory for the modified mass terms in
Sections ~\ref{@@} and ~\ref{@@}. Secondly, we postpone the
problem of {\em curved} versus {\em flat} space all the way until
the end of the computation in Section {@@} where we elaborate on
the tricks developed by 't Hooft.

As we shall discuss in detail in the remainder of this paper, the
validity of the procedure with {\em spatially varying masses}
relies entirely on the statement which says that the quantum
theory of the modified instanton problem displays exactly the same
ultraviolet singularities as those obtained in ordinary
perturbative expansions. In fact, we shall greatly benefit from
our introduction of {\em observable parameters} since it can be
used to explicitly verify this statement.

Since the action is now finite one can go ahead and formally
expand the theory about $z_i =0$. To see how this works let us
first consider the operator $O_h$. Using Eq. \eqref{top1} we can
write for $S_h$
\begin{equation}
S_h = z_h \int d \textbf{r}\, \mu^2(\textbf{r}) \tr A_h q,
\end{equation}
where
\begin{equation}\label{ah1}
A_h = R\,T_0 \left(\Lambda -\frac{m-n}{m+n} 1_{m+n} \right)
T^{-1}_0 R^{-1}.
\end{equation}
We can now proceed by evaluating the expectation $\langle
q\rangle$ with respect to the theory of the previous Section, Eq.
\eqref{fluct11}. The details of the computation are presented in
Appendix~\ref{AppA}. It is important to keep in mind, however,
that the global unitary matrix $T_0$ is eventually restricted to
run over the subgroup $U(m) \times U(n)$ only. Since it is in many
ways simpler to carry out the quantum fluctuations about the
theory with $T_0 =1_{m+n}$, we shall in what follows specialize to
this simpler case. We will come back to the more general situation
in Section ~\ref{@@} where we show that the final results are in
fact independent of the specific choice made for the matrix $T_0$.

%===============================================================
\subsection{\label{QFAI.A}Action for the quantum fluctuations}
%===============================================================
Keeping the remarks of the previous Section in mind we obtain the
complete action as the sum of a classical part $S^{\rm inst}$ and
a quantum part $\delta S$ as follows
\begin{equation}\label{Sfluc1}
S = S^{\rm inst} +  \delta S,
\end{equation}
where
\begin{equation}\label{Sfluc1_1}
S^{\rm inst}=-2\pi\sigma_{xx} +i\,\theta +S^{\rm inst}_h +S^{\rm
inst}_a + S^{\rm inst}_s
\end{equation}
and
\begin{equation}\label{Sfluc2}
\delta S = \delta S_{\sigma} + \delta S_{h} + \delta S_{a}+\delta
S_{s}.
\end{equation}
Here, $S^{\rm inst}_i$ stands for the classical action of the
modified mass terms $O_i^{\rm inst}$, Eq. \eqref{Oipm}, and is
given by
\begin{eqnarray}\label{OCSM}
S_{h}^{\rm inst} &=& z_h \int  d \textbf{r}\,\mu^2(\textbf{r})
O_h^{\rm inst}
(\textbf{r}) = - 8\pi z_h, \notag \\
S_{a}^{\rm inst} &=& z_a \int d\textbf{r}\,\mu^2 (\textbf{r})
O_a^{\rm inst}
(\textbf{r}) =16 \pi (m+n-1) z_{a},  \\
S_{s}^{\rm inst} &=& z_s \int d\textbf{r}\,\mu^2 (\textbf{r})
O_s^{\rm inst} (\textbf{r}) =16 \pi (m+n+1)z_{s} \left [ 1 -
\frac{8}{3(m+n+2)} \right].\notag
\end{eqnarray}
Next, the results for $\delta S_{\sigma}$ and $\delta S_i$ can be
written up to quadratic order in $v$, $v^\dag$ as follows
%..................................................................
\begin{gather}
\delta S_{\sigma} = - \frac{\sigma_{xx}}{4}\int \limits_{\eta
\theta}\Biggl [ \sum \limits_{\alpha=2}^m\sum\limits_{\beta=2}^{n}
v^{\alpha \beta} O^{(0)} v^{\dag\beta \alpha} + \sum
\limits_{\alpha=2}^{m} v^{\alpha 1} O^{(1)} v^{\dag 1 \alpha} +
\sum\limits_{\beta=2}^{n}v^{1 \beta} O^{(1)} v^{\dag\beta 1}
\notag\\ + v^{11} O^{(2)} v^{\dag 11} \Biggr ]\label{SfSi}
\end{gather}
The operators $O^{(a)}$ with $a=0,1,2$ are given by Eq.
\eqref{Oa}. Furthermore,
%.....................................................................
\begin{gather}
\delta S_{h} =  -z_h \int \limits_{\eta \theta}\Biggl [ \sum
\limits_{\alpha=2}^{m}\sum\limits_{\beta=2}^{n} v^{\alpha \beta}
v^{\dag \beta \alpha} + (1-e_{0}^{2})\left (\sum
\limits_{\alpha=2}^{m} v^{\alpha 1} v^{\dag 1 \alpha} +
\sum\limits_{\beta=2}^{n} v^{1 \beta} v^{\dag \beta 1}\right )
\notag \\+ (1-2 e_{0}^{2}) v^{11} v^{\dag 11} +2 ( e_{0}e_{1}
v^{11} + e_{0} \bar{e}_{1} v^{\dag 11} ) \Biggr ],\label{Sfh}
\end{gather}
%...................................................................
\begin{gather}
\delta S_{a} = 2 (m+n-1) z_{a}\int \limits_{\eta \theta}\Biggl [
\frac{m+n-2-4e_{0}^{2}}{m+n-2}\sum
\limits_{\alpha=2}^{m}\sum\limits_{\beta=2}^{n} v^{\alpha \beta}
v^{\dag\beta \alpha} + (1-e_{0}^{2}) \notag
\\ \hspace{5cm}\times \Bigl ( \sum
\limits_{\alpha=2}^{m}v^{\alpha 1} v^{\dag 1 \alpha}+
\sum\limits_{\beta=2}^{n} v^{1 \beta} v^{\dag\beta 1}\Bigr ) +
(1-2 e_{0}^{2}) v^{11} v^{\dag 11} \notag \\ \hspace{0.8cm}+2(
e_{0}e_{1} v^{11} + e_{0} \bar{e}_{1} v^{\dag 11} )\Biggr
],\label{Sfa}
\end{gather}
%......................................................................
\begin{gather}
\delta S_{s} =  2 \frac{m+n+1}{m+n+2} z_{s}\int \limits_{\eta
\theta}\Biggl [(m+n+2-4e_{0}^{2}) \sum \limits_{\alpha=2}^{m}
\sum\limits_{\beta=2}^{n}v^{\alpha \beta} v^{\dag \beta \alpha}
 + (1-e_{0}^{2})\notag \\ \hspace{4.5cm}\times (m+n+2-8e_{0}^{2})  \left ( \sum
\limits_{\alpha=2}^{m}v^{\alpha 1} v^{\dag 1 \alpha} +
\sum\limits_{\beta=2}^{n} v^{1 \beta} v^{\dag \beta 1}\right
)\notag \\ \hspace{4.5cm}+\Bigl [(1-2 e_{0}^{2})(m+n+2-8e_{0}^{2})
-8 e_{0}^{2}|e_{1}|^{2}\Bigr ] v^{11} v^{\dag 11}  \notag \\
\hspace{1.1cm} - 4 (e_{0}^{2}e_{1}^{2} v^{11} v^{11}
+e_{0}^{2}\bar{e}_{1}^{2} v^{\dag 11} v^{\dag 11}) \notag \\
\hspace{4.4cm}+ 2(m+n+2-8e_{0}^{2})( e_{0}e_{1} v^{11} + e_{0}
\bar{e}_{1}v^{\dag 11} )\Biggr ].\label{Sfs}
\end{gather}

Notice that the terms linear in $v$, $v^\dag$ in Eqs
\eqref{Sfh}-\eqref{Sfs} can be written as
\begin{equation}
\int \limits_{\eta \theta} \left[ e_{0}e_{1} v^{11} + e_{0}
\bar{e}_{1} v^{\dag 11} \right] \propto \int \limits_{\eta \theta}
\left[ \Phi^{(2)}_{1,-1} v^{11} + \bar{\Phi}^{(2)}_{1,-1} v^{\dag
11} \right] \propto \frac{\delta\lambda}{\lambda} .
\end{equation}
This means that the fluctuations tangential to the instanton
parameter $\lambda$ are the only unstable fluctuations in the
problem. However, the linear fluctuations are not of any special
interest to us and we proceed by formally evaluating the quantum
fluctuations to first order in the fields $z_i$ only. The
expansion is therefore with respect to the theory with $\delta
S_\sigma$ alone and this has been analyzed in detail in
Ref.\cite{Pruisken2,Pruisken3}.

Finally, we also need the action $S^0$ for the quantum
fluctuations about the trivial vacuum. The result is given by

%%%%%%%%%%%%%%%%%%%%%%%%%%%%%%%%%%%%5
\begin{equation}\label{Sfluc10}
S^{(0)} = \delta S_{\sigma}^{(0)} + \delta S_{h}^{(0)} + \delta
S_{a}^{(0)} +\delta S_{s}^{(0)}
\end{equation}
where
%..................................................................
\begin{gather}
\delta S_{\sigma}^{(0)}= - \frac{\sigma_{xx}}{4}\int \limits_{\eta
\theta} \sum \limits_{\alpha=1}^{m}\sum \limits_{\beta=1}^{n}
v^{\alpha \beta} O^{(0)} v^{\dag\beta\alpha}, \qquad \delta
S_{h}^{(0)} =  -z_h \int \limits_{\eta \theta} \sum
\limits_{\alpha=1}^{m}\sum \limits_{\beta=1}^{n} v^{\alpha \beta}
v^{\dag\beta\alpha} ,
 \notag\\
\delta S_{a,s}^{(0)}=  2 (m+n\mp1) z_{a,s}\int \limits_{\eta
\theta} \sum \limits_{\alpha=1}^{m}\sum \limits_{\beta=1}^{n}
v^{\alpha \beta} v^{\dag \beta\alpha} .\label{Sfh0}
\end{gather}
%..........................................................................
%

%===============================================================
\section{\label{QFAI.SZMRT}Pauli-Villars regulators}
%================================================================

Recall that after integration over the quantum fluctuations one is
left with two sources of divergences. First, there are the
ultraviolet divergences which eventually lead to the
renormalization of the coupling constant or $\sigma_{xx}$. At
present we wish to extend the analysis to include the
renormalization of the the $z_i$ fields. The ultraviolet can be
dealt with in a standard manner, employing Pauli-Villars regulator
fields with masses ${\mathcal M}_{f}$ ($f=0,1,\cdots, K$) and with
an alternating metric $\hat e_{f}$~\cite{t'Hooft1}. We assume
$\hat e_{0} =1$, ${\mathcal M}_{0}=0$ and large masses
$\mathcal{M}_{f} \gg 1$ for $f>1$. The following constraints are
imposed
\begin{equation}\label{ConscM}
\sum \limits_{f=0}^{K} \hat e_{f} \mathcal{M}^{k}_{f} = 0, \qquad
0 \leq k < K,\qquad  \sum \limits_{f=1}^{K} \hat e_{f} \ln
{\mathcal M}_{f} = - \ln {\mathcal M}.\nonumber
\end{equation}
The regularized theory is then defined as
\begin{equation}\label{SReg}
\delta S_{\rm reg} = \delta S_0 + \sum \limits_{f=1}^{K} \hat
e_{f} \delta S_{f}.
\end{equation}
Here action $\delta S_{f}$ is the same as the action $\delta S$
except that the kinetic operators $O^{(a)}$ are all replaced by
$O^{(a)}+ {\mathcal M}_{f}^{2}$.

Our task is to evaluate Eq. \eqref{SReg} to first order in the
fields $z_i$. This means that $\delta S_0$ still naively diverges
due to the zero modes of the operators $O^{(a)}$. These zero modes
are handled separately in Section~\ref{Imani}, by employing the
collective coordinate formalism introduced in
Ref.\cite{Pruisken2}. The regularized theory $\delta S_{\rm reg}$
is therefore defined by omitting the contributions of all the zero
modes in $\delta S_0$.

To simplify the notation we shall work with $m=n$ in the
subsequent Sections. The final answer will be expressed in terms
of $m$ and $n$, however.
%
%=======================================================================
\subsection{\label{QFAI.ZMC}Explicit computations}
%=======================================================================
To simplify the notation we will first collect the results
obtained after a naive integration over the field variables
$v,v^\dag$. These are easily extended to include the alternating
metric and the Pauli-Villars masses which will be main topic of
the next Section. Consider the ratio
\begin{eqnarray}
\frac{Z_{\rm inst}}{Z_0} = \frac{\displaystyle \int \mathcal{D}
[v,v^\dag] \exp S}{ \displaystyle\int \mathcal{D} [v,v^\dag] \exp
S_0}
= \exp \Bigl [ &-& 2 \pi \sigma_{xx} \pm i \theta + D \nonumber \\
&+& S_h^{\rm inst} + \Delta S_{h} \nonumber \\
&+& S_a^{\rm inst} + \Delta S_{a} \nonumber \\
&+& S_s^{\rm inst} + \Delta S_{s}  \Bigr ] . \label{AD1}
\end{eqnarray}
Here, the quantum corrections $D$, $\Delta S_{h}$, $\Delta S_a$
and $\Delta S_{s}$ can be expressed in terms of the propagators
\begin{equation}\label{Ga}
\mathcal{G}_a = \frac{1}{O^{(a)}} = \sum \limits_{J M} \frac{|J
M\rangle_{(a)} {}_{(a)}\langle J M|}{E_J^{(a)}} ,\qquad a=0,1,2.
\end{equation}
The results can be written as follows
\begin{eqnarray}\label{DAv}
D &=&  \tr\left [ 2(n-1) \Bigl ( \ln \mathcal{G}_1-\ln
\mathcal{G}_0 \Bigr ) -  \Bigl (\ln \mathcal{G}_2-\ln
\mathcal{G}_0 \Bigr )\right ], \\
\Delta S_{h} &=& -\frac{4z_h}{\sigma_{xx}} \tr\Bigl[ 2 (n-1)
[(1-e_{0}^{2})\mathcal{G}_1 - \mathcal{G}_0] +
[(1-2e_{0}^{2})\mathcal{G}_2 - \mathcal{G}_0]\Bigr ],
\\ \label{OaAv}
\Delta S_{a} &=& \frac{8(2n-1)z_{a}}{\sigma_{xx}}\tr\Bigl [ (n-1)
[ (n-1-2 e_{0}^{2})\mathcal{G}_0 - (n-1)\mathcal{G}_0]  \\
&& \hspace{2.5cm}+2 (n-1) [(1-e_{0}^{2})\mathcal{G}_1 -
\mathcal{G}_0] +
[(1-2e_{0}^{2})\mathcal{G}_2 - \mathcal{G}_0]\Bigr ],\nonumber\\
\Delta S_{s} &=&  \frac{8(2n+1)z_{s}}{(n+1)\sigma_{xx}}\tr\Bigl [
(n-1)^{2} [(n+1-2 e_{0}^{2})\mathcal{G}_0 -
(n+1)\mathcal{G}_0]\label{OsAv_0} \\
&& \hspace{2.5cm}+ 2 (n-1)
[(1-e_{0}^{2})(n+1-4e_{0}^{2})\mathcal{G}_1  -(n+1)\mathcal{G}_0]\nonumber\\
&& \hspace{2.5cm}+ \Bigl([(1-2e_{0}^{2})(n+1-4e_{0}^{2})-
4e_{0}^{2}(1-e_{0}^{2})]\mathcal{G}_2 \notag \\
&& \hspace{2.5cm}- (n+1)\mathcal{G}_0\Bigr
 )\Bigr ].\nonumber
\end{eqnarray}
In these expressions the trace is taken with respect to the
complete set of eigenfunctions of the operators $O^{(a)}$. To
evaluate these expressions we need the help of the following
identities (see Appendix \ref{AppB})
\begin{gather}\label{e02}
\sum \limits_{M=-J-a+k_{a}}^{J-k_{a}}{}_{(a)}\langle J,M
|e_{0}^{2}| M,J\rangle_{(a)} = \frac{1}{2}(2J+1+a -2 k_{a}), \\
\label{e04} \sum \limits_{M=-J-a+k_{a}}^{J-k_{a}}{}_{(a)}\langle
J,M |e_{0}^{4}| M,J\rangle_{(a)} = \frac{1}{3}(2J+1+a -2 k_{a}) ,
\end{gather}
where $k_a={0,1,1}$ for $a=0,1,2$ respectively. After elementary
algebra we obtain
\begin{eqnarray}\label{DAv1}
D &=& -2(n-1)D^{(1)} - D^{(2)}, \\
\Delta S_{h} &=& \hspace{2.21cm}-z_h \frac{4}{\sigma_{xx}} \left [
(n-1) Y^{(1)} -(2n-1)Y^{(0)}\right ], \\
\Delta S_{a} &=&
\hspace{1.45cm}z_{a}\frac{8(2n-1)}{\sigma_{xx}}\left [
(n-1)Y^{(1)} -(3n-2)Y^{(0)} \right ], \\
\label{OsAv} \Delta S_{s} &=& z_{s}
\frac{8(2n+1)(3n-1)}{3(n+1)\sigma_{xx}} \left [ (n-1)Y^{(1)} - 3n
Y^{(0)}\right ].
\end{eqnarray}
We have introduced the following quantities
\begin{eqnarray}\label{D12}
D^{(r)} &=& \sum \limits_{J=1}^{\infty} (2 J+ r-1) \ln E_{J}^{(r)}
- \sum \limits_{J=0}^{\infty} (2 J +1) \ln E_{J}^{(0)}, ~~~r=1,2
\\ \label{Y}
Y^{(s)} &=& \sum \limits_{J=s}^{\infty}\frac{2 J+
1-s}{E^{(s)}_{J}},\qquad s=0,1.
\end{eqnarray}

%................................................................
\subsection{\label{QFAI.MM.PVR}Regularized expressions}
%................................................................
To obtain the regularized theory one has to include the
alternating metric $e_{f}$ and add the masses $\mathcal{M}_{f}$ to
the energies $E_J^{(a)}$ in the expressions for $D^{(r)}$ and
$Y^{(s)}$. To start, let us define the function
\begin{equation}\label{Phi}
\Phi^{(\Lambda)}(p) = \sum \limits_{J=p}^{\Lambda} 2 J \ln (J^{2}
- p^{2}) .
\end{equation}
According to Eq. (\ref{SReg}), the regularized function
$\Phi^{(\Lambda)}_{\rm reg}(p)$ is given by
\begin{equation}\label{RPhi}
\Phi^{(\Lambda)}_{\rm reg}(p) = \sum \limits_{J=p+1}^{\Lambda} 2 J
\ln (J^{2} - p^{2})+ \sum \limits_{f=1}^{K} e_{f} \sum
\limits_{J=p}^{\Lambda} 2 J \ln (J^{2} - p^{2} + {\mathcal
M}^{2}_{f}) ,
\end{equation}
where we assume that the cut-off $\Lambda$ is much larger than
$\mathcal{M}_f$. In the presence of a large mass $\mathcal{M}_{f}$
we may consider the logarithm to be a slowly varying function of
the discrete variable $J$. We may therefore approximate the
summation by using the Euler-Maclaurin formula
\begin{equation}\label{EMF}
\sum \limits_{J=p+1}^{\Lambda} g(J) = \int \limits_{p}^{\Lambda}
g(x) d x + \frac{1}{2} g(x)\Bigr |_{p}^{\Lambda}+\frac{1}{12}
g'(x)\Bigr |_{p}^{\Lambda} .
\end{equation}
After some algebra we find that Eq. (\ref{RPhi}) can be written as
follows~\cite{Pruisken2}
\begin{gather}
\Phi^{(\Lambda)}_{\rm reg}(p) = - 2 \Lambda ( \Lambda +1) \ln
\Lambda + \Lambda^{2} - \frac{\ln e \Lambda }{3} + 4 \sum
\limits_{J=1}^{\Lambda}  J \ln J  \nonumber \\
+ \frac{1-6 p}{3} \ln {\mathcal M} + 2 p^{2} - 2 \sum
\limits_{J=1}^{2 p} (J-p) \ln J.\label{RPhi1}
\end{gather}
The regularized expression for $D^{(r)}$ can be obtained as
\begin{equation}\label{Dr}
D^{(r)}_{\rm reg} = \lim \limits_{\Lambda\to \infty} \left [
\Phi^{(\Lambda)}_{\rm reg}\left (\frac{1+r}{2}\right ) -
\Phi^{(\Lambda)}_{\rm reg}\left (\frac{1}{2}\right )\right ] .
\end{equation}
From this we obtain the final results
\begin{equation}\label{D12reg1}
D^{(1)}_{\rm reg} = - \ln \mathcal{M} + \frac{3}{2} - 2 \ln 2,
\qquad D^{(2)}_{\rm reg} = - 2 \ln \mathcal{M} + 4 -3 \ln 3 - \ln
2.
\end{equation}
Next, we introduce another function
\begin{equation}\label{Ydef}
Y^{(\Lambda)}(p) = \sum \limits_{J=p}^{\Lambda} \frac{2 J}{J^{2} -
p^{2}},
\end{equation}
According to Eq. (\ref{SReg}), the regularized function
$Y^{(\Lambda)}_{\rm reg}(p)$ is given by
\begin{equation}\label{RY}
Y^{(\Lambda)}_{\rm reg}(p) = \sum \limits_{f=1}^{K} e_{f} \sum
\limits_{J=p}^{\Lambda} \frac{2 J}{J^{2} - p^{2} + {\mathcal
M}^{2}_{f}}
 + \sum \limits_{J=p+1}^{\Lambda} \frac{2 J}{J^{2} -
p^{2}} ,
\end{equation}
where as before we assume that the cut-off $\Lambda \gg
\mathcal{M}_{f}$. By using a similar procedure as discussed above
we now find
\begin{equation}\label{Yres}
Y^{(\Lambda)}_{\rm reg}(p) = 2 \ln \mathcal{M} + 2\gamma - \sum
\limits_{J=1}^{2 p} \frac{1}{J}.
\end{equation}
The regularized expressions for $Y^{(s)}$ can be written as
\begin{equation}\label{Yr}
Y^{(s)}_{\rm reg} = \lim \limits_{\Lambda\to \infty}
Y^{(\Lambda)}_{\rm reg}\left (\frac{1+s}{2}\right ),
\end{equation}
such that we finally obtain
\begin{equation}\label{Y01}
Y^{(0)}_{\rm reg}= 2 \ln \mathcal{M} + 2 \gamma - 1, \qquad
Y^{(1)}_{\rm reg}= 2 \ln \mathcal{M} + 2 \gamma - \frac{3}{2}.
\end{equation}
We therefore have the following results for the quantum
corrections
\begin{eqnarray}\label{Dregf}
D^{\rm reg} &=& 2 n \ln \mathcal{M} + n(4 \ln 2 -3) -1 + 3 \ln
\frac{3}{2}, \\
\Delta S_{h}^{\rm reg} &=& \hspace{3.26cm}z_h\frac{8 n
}{\sigma_{xx}} \left [ \ln
\mathcal{M} e^{\gamma-1/2} -\frac{2n-1}{4 n }\right ], \\
\Delta S_{a}^{\rm reg} &=& \hspace{1.48cm}-z_{a}\frac{16(2n-1)^{2}
}{\sigma_{xx}} \left [  \ln \mathcal{M} e^{\gamma-1/2}
 - \frac{3n-2}{2(2n-1)}\right ],\label{Ohregf}\\
\Delta S_{s}^{\rm reg} &=& -z_{s} \frac{16(2n+1)^{2} (3n-1)
}{3\sigma_{xx}(n+1)} \left [ \ln\mathcal{M} e^{\gamma-1/2}  -
\frac{3n}{2(2n+1)}\right ].
\end{eqnarray}
Apart from the logarithmic singularity in $\mathcal{M}$, the
numerical constants in the expression for $D^{\rm reg}$ are going
to play an important role in what follows. This is unlike the
expressions for $\Delta S_a^{\rm reg}$ where the second term in
the brackets should actually be considered as higher order terms
in an expansion in powers of $1/\sigma_{xx}$. We collect the
various terms together and obtain the following result for the
instanton contribution to the free energy
\begin{eqnarray}\label{const-reg}
\ln \left[ \frac{Z_{\rm inst}}{Z_0} \right]^{\rm reg} =  -1 + 3
\ln
\frac{3}{2} &-&(m+n)\left (\gamma +\frac{3}{2} - 2\ln 2\right ) \pm i\, \theta \\
\label{sigma-reg} - 2\pi &\sigma_{xx}& \left[1 -
\frac{m+n}{2\pi\sigma_{xx}} \ln \mathcal{M} e^{\gamma } \right]
\\ \label{zh-reg}
- 8\pi  &z_h& \left[1 - \frac{m+n}{2\pi\sigma_{xx}} \ln
\mathcal{M} e^{\gamma -1/2}  \right] \\ \label{za-reg} + 16\pi
(m+n-1) &z_a& \left[ 1 -\frac{m+n-1}{\pi\sigma_{xx}} \ln
\mathcal{M} e^{\gamma -1/2} \right] \\ \label{zs-reg}
+ \frac{16
\pi (m+n+1)(3m+3n-2)}{3(m+n+2)} &z_s& \left[ 1 -
\frac{m+n+1}{\pi\sigma_{xx}} \ln \mathcal{M} e^{\gamma -1/2}
\right]. \label{ADfins}
\end{eqnarray}

\subsection{\label{Ptheory} Observable theory in Pauli Villars regularization}

The important feature of these last expressions, as we shall see
next, is that the quantum corrections to the parameters
$\sigma_{xx}$, $z_h$, $z_a$ and $z_a$ are all identically the same
as those obtained from the perturbative expansions of the
observable parameters $\sigma'_{xx}$, $z'_h$, $z'_a$ and $z'_a$
introduced in Section~\ref{PO}. Notice that we have already
evaluated this theory in {\em dimensional regularization} in
Appendix \ref{AppA}. The problem, however, is that the different
regularization schemes ({\em dimensional} versus {\em Pauli
Villars}) are not related to one another in any obvious fashion.
Unlike dimensional regularization, for example, it is far from
trivial to see how the general form of the observable parameters,
Eq. \eqref{rgsigma}, can be obtained from the theory in Pauli
Villars regularization.

In Appendix \ref{AppC} we give the details of the computation
using Pauli-Villars regulators. Denoting the results for
$\sigma_{xx}^\prime$ and $z_i^\prime$ by
$\sigma_{xx}(\mathcal{M})$ and $z_i(\mathcal{M})$ respectively,
\begin{equation}
\sigma_{xx}^\prime = \sigma_{xx}(\mathcal{M}),~~~z_i^\prime =
z_i(\mathcal{M})
\end{equation}
then we have
\begin{eqnarray}
\hspace{3cm}\sigma_{xx}(\mathcal{M}) &=& \sigma_{xx} \left[1 -
\frac{m+n}{2\pi
\sigma_{xx} } \ln \mathcal{M} e^{\gamma} \right],\label{pr0}\\
z_i(\mathcal{M}) &=&\hspace{0.25cm} z_i \left[1
+\frac{\gamma_i^{(0)}}{\sigma_{xx}} \ln \mathcal{M} e^{\gamma
-1/2}\right ] .\label{pr}
\end{eqnarray}
The coefficients $\gamma_i^{(0)}$ are given by
\begin{equation}\label{gammai0}
\gamma_h^{(0)}=-\frac{m+n}{2\pi}, \qquad
\gamma_{a}^{(0)}=-\frac{m+n-1}{\pi}, \qquad
\gamma_s^{(0)}=-\frac{m+n+1}{\pi}.
\end{equation}

\indent{Since} so much of what follows is based on the results
obtained in this Section and the previous one, it is worthwhile to
first present a summary of the various issues that are involved.\\
First of all, it is important to emphasize that our results for
observable parameter $\sigma_{xx}^\prime$, Eq. \eqref{pr0},
resolve an ambiguity that is well known to exist, in the instanton
analysis of scale invariant theories. Given Eq. \eqref{pr0} one
uniquely fixes the quantity $\sigma_{xx}(\mathcal{M})$ (Eq.
\ref{sigma-reg}) and the constant term of order unity (Eq.
\ref{const-reg}) that is otherwise left undetermined. This result
becomes particularly significant when we address the
non-perturbative aspects of the renormalization group $\beta$
functions in Section~\ref{Curved-to-Flat}. \\
Secondly, the results for $z_i$ in the observable theory, Eq.
\ref{pr}), explicitly shows that the idea of spatially varying
masses does not alter the ultraviolet singularity structure of the
the instanton theory, i.e. Eqs \ref{zh-reg} - \ref{zs-reg}. A
deeper understanding of this problem is provided by the
computations in Appendix \ref{AppC} where we show that
Pauli-Villars regularization retains translational invariance in
the sense that the expectation of local operators like $< O_i
\left( Q(r) \right)>$ is independent of $r$. This aspect of the
problem is especially meaningful when dealing with the problem of
electron-electron interactions. As is well known, the presence of
mass terms generally alters the renormalization of the theory at
short distances in this case, i.e. the renormalization group
$\beta$ functions~\cite{Unify1}.\\
Finally, on the basis of the theory of observable parameters Eqs
~\eqref{pr0} and \eqref{pr} we may summarize the results of our
instanton computation, Eqs ~\eqref{const-reg}-\eqref{ADfins}, as
follows
\begin{eqnarray}
 \ln \left[ \frac{Z_{\rm inst}}{Z_0} \right]^{\rm reg} =  &-&1 + 3
 \ln \frac{3}{2} -(m+n)\left (\gamma +\frac{3}{2} - 2\ln 2\right ) \\
 &-& 2\pi \sigma_{xx}(\mathcal{M}) \pm \,i \theta (\nu) \\
 &+& ~~~~~z_h (\mathcal{M})\int d \textbf{r} \mu^2 (\textbf{r})
 O_h^{\rm inst} (\textbf{r}) \\
 &+& ~~~~~z_s (\mathcal{M})\int d \textbf{r} \mu^2 (\textbf{r})
 O_s^{\rm inst} (\textbf{r}) \\
 &+& ~~~~~z_a (\mathcal{M})\int d \textbf{r} \mu^2 (\textbf{r})
 O_a^{\rm inst} (\textbf{r}).
 \label{exp1a}
\end{eqnarray}
Here, the quantities $O_i^{\rm inst} (\textbf{r})$ are the
classical expressions given by Eq.\eqref{Oinst00}. On the other
hand, the parameters $\sigma_{xx}(\mathcal{M})$ and
$z_i(\mathcal{M})$ are precisely those obtained from the
observable theory.

%==================================================================
\section{\label{Imani} Instanton manifold}
%==================================================================
In this Section we first recapitulate the integration over the
zero frequency modes following Refs~\cite{Pruisken2} and
\cite{Pruisken3}. In the second part of this Section we address
the zero modes describing the $U(m+n)/U(m) \times U(n)$ rotation
of the instanton that we sofar have discarded.

\subsection{Zero frequency modes}
The complete expression for ${Z_{\rm inst}}/{Z_0}$ can be written
as follows
\begin{equation} \frac{Z_{\rm inst}}{Z_0} =\frac{\int \mathcal{D}
[Q_{\rm inst} ] \left[ Z_{\rm inst} \right]^{\rm reg} }{\int
\mathcal{D} [Q_{0} ] \left[ Z_{0} \right]^{\rm reg}}.
\end{equation}
Here, $Q_{\rm inst}$ denotes the manifold of the instanton
parameters as is illustrated in Fig.~\ref{FIG7}
\begin{eqnarray}\label{ZM2}
\int \mathcal{D} [Q_{\rm inst} ] &=& A_{\rm inst} \int d
\textbf{r}_0 \int \frac{d \lambda}{\lambda^{3}} \int {\mathcal
D}[U(1)]  \notag\\
&\times & \int {\mathcal D}\left[ \frac{U(m)}{U(1)\times U(m-1)}
\right] \int {\mathcal D}\left[ \frac{U(n)}{U(1)\times U(n-1)}
\right]
\notag \\
&\times &\int {\mathcal D} \left[ \frac{U(m+n)}{U(m)\times U(n)}
\right] .
\end{eqnarray}
The $Q_{0}$ represents the zero modes associated with the trivial
vacuum
\begin{eqnarray} \int \mathcal{D} [Q_{0} ] = A_{0} \int
{\mathcal D} \left[ \frac{U(m+n)}{U(m)\times U(n)} \right].
\end{eqnarray}
The numerical factors $A_{\rm inst}$ and $A_0$ are given by
\begin{eqnarray}
A_{\rm inst} &=&  \langle e^{4}_{0} \rangle \langle |e_{1}|^{4}
\rangle \langle e^{2}_{0} |e_{1}|^{2}\rangle \left( \langle
e^{2}_{0} \rangle \langle |e_{1}|^{2}
\rangle \right)^{m+n-2} \langle 1 \rangle^{(m-1)(n-1)} \pi^{-mn-m-n}\\
A_{0} &=&  \langle 1 \rangle^{mn} \pi^{-mn}
\end{eqnarray}
where the average $\langle \cdots \rangle$ is with respect to the
surface of a sphere
\begin{equation}\label{<>}
\langle f \rangle = \sigma_{xx} \int \limits_{-1}^{1} d \eta \int
\limits_{0}^{2 \pi} d \theta f(\eta, \theta).
\end{equation}
Notice that in the absence of symmetry breaking terms the
integration over $U(m+n)/U(m) \times U(n)$ drops out in the ratio.
We shall first discard this integration in the final answer which
is then followed by a justification in Section~\ref{Gold}. With
the help of the identity~\cite{Pruisken2}
\begin{equation}\label{vol}
\int {\mathcal D} \left[ \frac{U(k)}{U(1)\times U(k-1)} \right] =
\frac{\pi^{k-1}} {\Gamma(k)} ,
\end{equation}
we can write the complete result as follows
\begin{equation}\label{OmegaInstRes}
\frac{Z_{\rm inst}}{Z^{(0)}} = \frac{m n}{2}D_{mn} \int
d\textbf{r}_0 \int \frac{d \lambda}{\lambda^{3}}
(2\pi\sigma_{xx})^{n+m} \exp S^\prime_{\rm inst}
\end{equation}
where
\begin{equation}\label{S-prime}
S^\prime_{\rm inst} = - 2\pi  \sigma_{xx}({\mathcal M}) \pm i
\theta +  z_h({\mathcal M}) \int \limits_{\eta \theta} O_h^{\rm
inst} + z_s({\mathcal M}) \int \limits_{\eta \theta} O_s^{\rm
inst} + z_a({\mathcal M}) \int \limits_{\eta \theta} O_a^{\rm
inst} .
%\label{SW1}
\end{equation}
The numerical constant $D_{mn}$ is given by
\begin{equation}\label{Dnm}
D_{mn} = \frac{4}{\pi e} e^{-(m+n)(\gamma +3/2-\ln
2)}\Gamma^{-1}(1+m)\Gamma^{-1}(1+n).
\end{equation}
%=================================================================
\subsection{\label{Gold} The $U(m+n)/U(m) \times U(n)$ zero modes}
%=================================================================
To justify the result of Eq. \eqref{OmegaInstRes} we next consider
the full expression for $Z_{\rm inst}/Z_0$ that includes the
$U(m+n)/U(m) \times U(n)$ rotational degrees of freedom. For
simplicity we limit ourselves to the theory in the presence of the
$z_h$ field only. We now have
\begin{equation}\label{T00}
\frac{Z_{\rm inst}}{Z_0} =\frac{\int \mathcal{D} [Q_{\rm inst} ]
\exp \{-2\pi\sigma_{xx} \pm i\, \theta + z_h \int
\limits_{\eta\theta} \tr \Lambda_h T^{-1}_0 \left[ R^{-1} \langle
q \rangle R \right]^{\rm reg} T_0 \} }{\int \mathcal{D} [{Q}_{0} ]
\exp \{ z_h \int \limits_{\eta\theta} \tr \Lambda_h {t}^{-1}_0
\left[ \langle q\rangle _0 \right]^{\rm reg} {t}_0 \}} .
\end{equation}
Here, we have defined
\begin{equation}
\Lambda_h = \Lambda -\frac{m-n}{m+n} 1_{m+n}.
\end{equation}
The expectation $\langle\cdots \rangle$ is with respect to the
theory of $\delta S_\sigma$, Eq. \eqref{SfSi}, whereas
$\langle\cdots \rangle_0$ refers to the quantum theory of the
trivial vacuum which is obtained from $\delta S_\sigma$ by
replacing all operators $O^{(a)}$ by $O^{(0)}$. Let us write the
rotational degrees of freedom $T_0 \in U(m+n) $ as follows
\begin{equation}
T_0 =t_0 W, \qquad Q_0 = t_0^{-1} \Lambda t_0 .
\end{equation}
The quantity $t_0$ or $Q_0$ runs over the manifold $U(m+n)/U(m)
\times U(n)$ and $W$ stands for the remaining degrees of freedom
$U(m)/U(m-1)\times U(1)$, $U(n)/U(n-1)\times U(1)$ and $U(1)$
respectively. We can write
\begin{equation}
\int \mathcal{D} [Q_{\rm inst} ] = A_{\rm inst} \int d\textbf{r}_0
\frac{d\lambda}{\lambda^3} \mathcal{D} [Q_0] \mathcal{D} [W]
\end{equation}
The quantity $\left[ R^{-1} \langle q\rangle R \right]^{\rm reg}$,
unlike $\left[ \langle q\rangle_0 \right]^{\rm reg}$ in Eq.
\eqref{T00}, is not invariant under $U(m) \times U(n)$ rotations.
We therefore perform the integration over $W$ explicitly as
follows
\begin{gather}
\left \langle \exp \left [ z_h \int \limits_{\eta\theta} \tr
\Lambda_h t^{-1}_0 W^{-1} \left[ R^{-1} \langle q\rangle R
\right]^{\rm reg} W t_0 \right ]\right \rangle_{W} \qquad \qquad {}\\
{}\qquad \qquad\qquad  = \exp \left [ z_h \int
\limits_{\eta\theta} \frac{O_h (Q_0)}{O_h(\Lambda)} \tr \Lambda_h
\left[ R^{-1} \langle q\rangle R \right]^{\rm reg} +
\mathcal{O}(z_h^2)\right ].\notag
\end{gather}
Here,
\begin{equation}
\langle X[W]\rangle_W =\frac{\int \mathcal{D} [W] X[W] }{\int
\mathcal{D} [W]}.
\end{equation}
The $Q_0$ is now the only rotational degree of freedom left in the
terms with $z_h$ and the result can therefore be written as
follows
\begin{equation}\label{OmegaInstRes2}
\frac{Z_{\rm inst}}{Z^{(0)}} = \frac{m n}{2}D_{m n} \int
d\textbf{r}_0 \int \frac{d \lambda}{\lambda^{3}}
(2\pi\sigma_{xx})^{m+n} \exp\left [-2\pi\sigma_{xx} \pm i\,\theta
+ \delta S^\prime_{\rm inst}\right ],
\end{equation}
where
\begin{equation}
\exp \delta S^\prime_{\rm inst} = \frac{\int \mathcal{D} [Q_{0} ]
\exp \{ z_h \int \limits_{\eta\theta} \frac{O_h
(Q_0)}{O_h(\Lambda)} \tr \Lambda_h \left[ R^{-1} \langle q\rangle
R \right]^{\rm reg}
 \} }{\int \mathcal{D} [{Q}_{0} ] \exp
\{ z_h \int \limits_{\eta\theta} \frac{O_h ({Q}_0)}{O_h(\Lambda)}
\tr \Lambda_h  \left[ \langle q\rangle_0 \right]^{\rm reg} \}} .
\end{equation}
We can write the result as an expectation with respect to the
matrix field variable $Q_0$,
\begin{equation}\label{deltaSh1}
\exp\delta S^\prime_{\rm inst} = \left \langle \exp \{ z_h \int
\limits_{\eta\theta} \frac{O_h (Q_0)}{O_h(\Lambda)} \tr \Lambda_h
\left[ R^{-1} \langle q\rangle R -\langle q\rangle_0 \right]^{\rm
reg}
 \} \right\rangle_{Q_0} ,
\end{equation}
where
\begin{equation}\label{XQ0}
\langle X[Q_0]\rangle_{Q_0} = \frac{\int \mathcal{D} [Q_{0} ] \exp
\{ z_h \int \limits_{\eta\theta} \frac{O_h (Q_0)}{O_h(\Lambda)}
\tr \Lambda_h \left[ \langle q\rangle_0 \right]^{\rm reg}
 \} X[Q_0] }{ \int \mathcal{D} [{Q}_{0} ] \exp
\{ z_h \int \limits_{\eta\theta} \frac{O_h ({Q}_0)}{O_h(\Lambda)}
\tr \Lambda_h  \left[ \langle q\rangle_0 \right]^{\rm reg} \}} .
\end{equation}
Notice that by putting the classical value $\langle
q\rangle=\langle q\rangle_0=\Lambda$ in the expression $[\cdots
]^{\rm reg}$ in Eq. \eqref{deltaSh1} we precisely obtain the
quantity $O_h^{\rm inst}$. We therefore identify (see also Eq.
\eqref{S-prime})
\begin{equation}\label{zM1}
\int \limits_{\eta\theta} z_h  \tr \Lambda_h \left[ R^{-1} \langle
q\rangle R  -\langle q\rangle_0 \right ]^{\rm reg} = \int
\limits_{\eta\theta} z_h (\mathcal{M}) O_h^{\rm inst} .
\end{equation}
This is precisely the result that was obtained before, by fixing
$Q_0 = \Lambda$ at the outset. By the same token we write
\begin{equation}\label{zM2}
z_h \tr \Lambda_h \left [\langle q\rangle_0\right ]^{\rm reg} = z_h (\mathcal{M}) O_h
(\Lambda)
\end{equation}
We have already seen that the quantity $z_h (\mathcal{M})$ defined
by Eq. \eqref{zM1} differs from that of Eq. \eqref{zM2} by a
constant of order $1/\sigma_{xx}$ which is not of interest to us.
The final expression for $\delta S^{\prime}_{\rm inst}$ can now be
written in a more transparent fashion as follows
\begin{equation}\label{OmegaInstRes0}
\exp\delta S^{\prime}_{\rm inst} = \left\langle \exp \left
\{\frac{O_h (Q_0)}{O_h(\Lambda)} \int
\limits_{\eta\theta}z_h(\mathcal{M}) O_h^{\rm inst} \right \}\right
\rangle_{Q_O} ,
\end{equation}
where instead of Eq. \eqref{XQ0} we now write
\begin{equation}\label{XQ01}
\langle X[Q_0]\rangle_{Q_0} = \frac{\int \mathcal{D} [Q_{0}] \exp
\{ \int \limits_{\eta\theta} z_h (\mathcal{M}) {O_h (Q_0)}
 \} X[Q_0] }{ \int \mathcal{D} [{Q}_{0} ] \exp
\{  \int \limits_{\eta\theta} z_h (\mathcal{M}) O_h ({Q}_0)  \}} .
\end{equation}

In summary we can say that as long as one works with mass terms in
{\em curved} space, the rotational degrees of freedom are
non-trivial and the integration over the global matrix field $Q_0$
has to be performed in accordance with Eq. \eqref{XQ01}. However,
we are ultimately interested in the theory in {\em flat} space
which means that the integral over the unit sphere $\int d\eta
d\theta$ in Eq. \eqref{XQ01} is going to be replaced by the
integral over the entire plane in flat space, $\int d \textbf{r}$.
This then fixes the matrix variable $Q_0$ in Eqs
\eqref{OmegaInstRes1} and \eqref{XQ01} to its classical value $Q_0
= \Lambda$. The final results are therefore the same as those that
are obtained by putting $Q_0 =\Lambda$ at the outset of the
problem.
%
%
%
%
%.....................................................................
\section{\label{Curved-to-Flat}Transformation from curved space to flat space}
%.....................................................................
%
In this Section we embark on the various steps that are needed in
order express the final answer in quantities that are defined in
flat space. As a first step we have undo the transformation $z_i
\rightarrow z_i \mu^2 ({\bf r})$ that was introduced in Section
\ref{P.Metric} (see Eq. \eqref{zRen}). This means that the
integrals over $\eta$, $\theta$ in the expression for
$S^{\prime}_{\rm inst}$, Eq. \eqref{S-prime}, have to be replaced
as follows
\begin{equation}
\int d\eta d\theta O_i^\textrm{inst} = \int d \textbf{r} \mu^2
(\textbf{r}) O_i^\textrm{inst} ({\bf r}) \quad \rightarrow \quad \int d
\textbf{r} O_i^\textrm{inst} ({\bf r}).
\end{equation}
The complete expression for the instanton contribution to the free
energy is therefore the same as Eq. \eqref{OmegaInstRes} but with
$S^{\prime}_{\rm inst}$ now given by
\begin{eqnarray}\label{S-prime1}
 \hspace{3cm}S^\prime_{\rm inst}\quad \rightarrow\quad &-& 2\pi \sigma_{xx}({\mathcal M})
 \pm i \theta \\ \label{S-prime2}
 &+& \int^{\prime} d \textbf{r} z_h({\mathcal M}) O_h^{\rm inst}
 (\textbf{r}) \\ \label{S-prime3}
 &+& \int^{\prime} d \textbf{r} z_s({\mathcal M})  O_s^{\rm inst} (\textbf{r})
 \\ \label{S-prime4}
 &+& \int^{\prime} d \textbf{r} z_a({\mathcal M}) O_a^{\rm inst} (\textbf{r}).
\end{eqnarray}
The ``prime" on the integral signs reminds us of the fact that the
mass terms still formally display a logarithmic divergence in the
infrared. However, from the discussion on {\em constrained}
instantons we know that a finite value of $z_i$ generally induces
an infrared cut-off on both the spatial integrals $\int
d\textbf{r} O_i^{\rm inst}(\textbf{r})$ and the integral over
scale sizes $\lambda$ in the theory. Keeping this in mind, we can
proceed and evaluate the expressions for the {\em physical
observables} of the theory, introduced in Section~\ref{PO}.

%.....................................................................
\subsection{Physical observables}
%.....................................................................

According to definitions in Section~\ref{PO} we obtain the
following results for the parameters $\sigma_{xx}^{\prime}$ and
$\theta^{\prime}$ (see also Ref. ~\cite{Pruisken2,Pruisken3})
\begin{eqnarray}\label{ss1}
 \sigma_{xx}^{\prime} &=& \sigma_{xx}(\mathcal{M}) - {D_{mn}} \int^{\prime}
 \frac{d \lambda}{\lambda} (2\pi\sigma_{xx})^{m+n+2}
 e^{-2\pi\sigma_{xx}(\mathcal{M})}\cos \theta \\ \label{ss2}
 \frac{\theta^{\prime}}{2\pi} &=& \frac{\theta}{2\pi} \hspace{0.9cm} - D_{mn} \int^{\prime}
 \frac{d \lambda}{\lambda} (2\pi\sigma_{xx})^{m+n+2}
 e^{-2\pi\sigma_{xx}(\mathcal{M})}\sin \theta.
\end{eqnarray}
Similarly we obtain the $z^{\prime}_i$ parameters as follows
\begin{gather}
 z_i^{\prime} = z_i(\mathcal{M}) +  D_{mn} \int^{\prime}
 \frac{d \lambda}{\lambda} (2\pi\sigma_{xx})^{m+n}
 e^{-2\pi\sigma_{xx}(\mathcal{M})} \cos \theta \label{zi} \\
\hspace{2cm}\times z_i(\mathcal{M}) \left( \frac{m
n}{O_i[\Lambda]}
 \int^{\prime} \frac{d \textbf{r}}{\lambda^2} O_i^{\rm inst} (\textbf{r})
 \right),
 \notag
\end{gather}
where $i=a,h$ and $s$. By using the results of Eqs \eqref{Oipm}
and \eqref{Ohas1}, the expression simplifies somewhat and can be
written in a more general fashion as follows
\begin{gather}
 z_i^{\prime} = z_i(\mathcal{M})+\pi \gamma_i^{(0)} D_{mn} \int^{\prime}
 \frac{d \lambda}{\lambda} (2\pi\sigma_{xx})^{m+n}
 e^{-2\pi\sigma_{xx}(\mathcal{M})} \cos \theta\label{zi1}\\
\hspace{2cm}\times  z_i(\mathcal{M})
 \int^{\prime} d
 \textbf{r}\frac{\mu(\textbf{r})}{\lambda}.\notag
\end{gather}
The important feature of the results of this Section is that the
non-perturbative (instanton) contributions are all unambiguously
expressed in terms of the perturbative quantities $\sigma_{xx}
({\mathcal M})$, $\theta$ and $z_i ({\mathcal M})$.

\subsection{Transformation ~$\mu^{2} ({\bf r}) {\mathcal M} \rightarrow \mu_0$}

Next we wish to obtain the results in terms of a spatially {\em
flat} momentum scale $\mu_0$, rather than in the spatially varying
quantity $\mu^2(\textbf{r})\mathcal{M}$ which appears in the
Pauli-Villars regularization scheme. For this purpose we introduce
the following renormalization group counter terms
\begin{gather}\label{countersigma}
 \sigma_{xx}({\mathcal M}) \rightarrow \sigma_{xx}({\mathcal M}) \left[ 1
 + \frac{m+n}{2\pi \sigma_{xx}} \ln \frac{\mu(\textbf{r})\mathcal{M}}{\mu_{0}}
 \right]\hspace{4cm}{}\\
 = \sigma_{xx} \left[ 1
 - \frac{m+n}{2\pi \sigma_{xx}} \ln \frac{\mu_{0}}{\mu(\textbf{r})}
 e^{\gamma} \right] = \sigma_{xx} (\mu(\textbf{r})),\notag \\
 z_i ({\mathcal M}) \rightarrow z_i ({\mathcal M})
 \left[ 1 + \frac{\gamma_i^{(0)}}{\sigma_{xx}} \ln
 \frac{\mu(\textbf{r}){\mathcal M}}{\mu_{0}} \right]\hspace{4cm}{}\label{counterzi}\\
 =z_i \left[ 1 - \frac{\gamma^{(0)}_i}{\sigma_{xx}}
 \ln \frac{\mu_{0}}{\mu(\textbf{r})} e^{\gamma -1/2} \right] = z_i
 (\mu(\textbf{r})).\notag
\end{gather}

The expression for $S^{\prime}_{\rm inst}$ now becomes
\begin{eqnarray}\label{S-prime1a}
 \hspace{2.5cm}S^\prime_{\rm inst}\quad \rightarrow\quad &-& \int d\textbf{r}
 \sigma_{xx} (\mu(\textbf{r}))
 \tr (\nabla Q_{\rm inst} (\textbf{r}))^2
 \pm i \theta \\ \label{S-prime2a}
 &+& \int^{\prime} d \textbf{r} z_h
 (\mu(\textbf{r})) O_h^{\rm inst}
 (\textbf{r}) \\ \label{S-prime3a}
 &+& \int^{\prime} d \textbf{r} z_s
 (\mu(\textbf{r}))  O_s^{\rm inst} (\textbf{r})
 \\ \label{S-prime4a}
 &+& \int^{\prime} d \textbf{r} z_a
 (\mu(\textbf{r})) O_a^{\rm inst} (\textbf{r}).
\end{eqnarray}

%.....................................................................
\subsection{The $\beta$ functions}
%.....................................................................
Let us first evaluate Eq. \eqref{S-prime1a} which can be written
as
\begin{equation}
 \int d{\bf r} ~\sigma_{xx} (\mu({\bf r})) \tr \nabla_i Q_\textrm{inst}
 ({\bf r}) \nabla_i Q_\textrm{inst} ({\bf r}) =
 \int d {\bf r} \mu^2 ({\bf r}) \sigma_{xx} (\mu({\bf r}))
= 2\pi \sigma_{xx} (\zeta \lambda),  \label{newsigma}
\end{equation}
where
\begin{equation}\label{sxxzl}
\sigma_{xx} (\zeta \lambda) = \sigma_{xx}
 - \frac{m+n}{2\pi } \ln {\zeta \lambda \mu_{0}}
 e^{\gamma}, \qquad \zeta =e^2 /4.
\end{equation}
Notice that the expression for $\sigma_{xx} (\zeta \lambda)$ can
be simply obtained from $\sigma_{xx}({\mathcal M})$ by replacing
the Pauli-Villars mass $\mathcal M$ according to
\begin{equation}\label{M.vs.lambda}
{\mathcal M} \rightarrow \zeta \lambda \mu_0 .
\end{equation}
We next wish to express the quantity $\sigma_{xx}^{\prime}
({\mathcal M})$ in a similar fashion. Write
\begin{eqnarray}
\hspace{4cm}\sigma_{xx}^{\prime} ({\mathcal M}) & \rightarrow &
\sigma_{xx}^{\prime} (\mu^\prime ({\bf r})) \\\label{sxxprimezl}
\sigma_{xx}^{\prime} (\zeta\lambda^\prime) & = & \frac{1}{2\pi}
\int d {\bf r} (\mu^\prime ({\bf r}))^2 \sigma_{xx}^{\prime}
(\mu^\prime ({\bf r})) .
\end{eqnarray}
One can think of the $\mu^\prime ({\bf r}) =2\lambda^\prime / (r^2
+\lambda^{\prime 2})$ as being a background instanton with a large
scale size $\lambda^\prime$. The expressions for
$\sigma_{xx}^{\prime}$ and $\theta^{\prime}$ in flat space can now
be written as follows
\begin{eqnarray}\label{ss1a}
 \sigma_{xx}^{\prime} (\zeta \lambda^{\prime})&=&
 \sigma_{xx}(\zeta \lambda^{\prime}) - {D_{mn}} \int^{\prime}
 \frac{d \lambda}{\lambda} (2\pi\sigma_{xx})^{m+n+2}
 e^{-2\pi\sigma_{xx}(\zeta \lambda)}\cos \theta \\ \label{ss2a}
 \frac{\theta^{\prime} (\zeta \lambda^{\prime})}{2\pi} &=& ~\frac{\theta}{2\pi}
 \hspace{0.8cm}- {D_{mn}} \int^{\prime}
 \frac{d \lambda}{\lambda} (2\pi\sigma_{xx})^{m+n+2}
 e^{-2\pi\sigma_{xx}(\zeta \lambda)}\sin \theta .
\end{eqnarray}
In words, the scale size $\lambda^\prime$ has identically the same
meaning for the perturbative and instanton contributions. Notice
that $\sigma_{xx}(\zeta \lambda^{\prime})$ is the same as Eq.
\eqref{sxxzl} with $\lambda$ replaced by $\lambda^\prime$. Next,
introducing an arbitrary scale size $\lambda_0$ we can write the
perturbative expression $\sigma_{xx} (\zeta \lambda_0)$ as follows
\begin{equation}\label{ss1a0}
 \sigma_{xx} (\zeta \lambda^{\prime})=\sigma_{xx} (\zeta \lambda_0)
 -\frac{m+n}{2\pi} \ln \frac{\lambda^{\prime}}{\lambda_0} =
 \sigma_{xx} (\zeta \lambda_0) -\frac{m+n}{2\pi} \int_{\lambda_0}^{\prime}
 \frac{d \lambda}{\lambda}.
\end{equation}
On the basis of these results one obtains the following complete
expressions for the quantities $\sigma_{xx}^{\prime}$ and
$\theta^{\prime}$
\begin{eqnarray}%\label{ss1a1}
 \hspace{-0.5cm}\sigma_{xx}^{\prime} &=& \sigma_{xx}(\zeta \lambda_0) -
 %\nonumber\\
 %&&
 \int^{\prime}_{\zeta \lambda_0} \frac{d [\zeta \lambda]}{\zeta
 \lambda} \left[ \frac{m+n}{2\pi} +
 {D_{mn}} (2\pi\sigma_{xx})^{m+n+2}
 e^{-2\pi\sigma_{xx}(\zeta \lambda)}\cos \theta \right], \notag\\ \label{ss2a1}
 \frac{\theta^{\prime}}{2\pi} &=& ~\frac{\theta (\zeta \lambda_0 )}{2\pi} \hspace{0.18cm} -
\int^{\prime}_{\zeta \lambda_0} \frac{d [\zeta \lambda]}{\zeta \lambda}
 {D_{mn}} (2\pi\sigma_{xx})^{m+n+2}
 e^{-2\pi\sigma_{xx}(\zeta \lambda)}\sin \theta.
\end{eqnarray}
Several remarks are in order. First of all, we have made use of
the well known fact that the quantities $\sigma_{xx}$ in the
integral over scale sizes all acquire the same quantum corrections
and can be replaced by $\sigma_{xx} (\zeta \lambda)$. Secondly,
although the instanton contributions are finite in the
ultraviolet, they have nevertheless dramatic consequences for the
behavior of the system in the infrared. Equations \eqref{ss1a1} and
\eqref{ss2a1} determine the renormalization group $\beta$
functions as follows
\begin{eqnarray}\label{ss1a1}
 \sigma_{xx}^{\prime} &=& \sigma_{xx}(\zeta \lambda_0) -
 \int^{\prime}_{\zeta \lambda_0} \frac{d [\zeta \lambda]}{\zeta
 \lambda}  \beta_{\sigma}
 (\sigma_{xx}(\zeta\lambda),\theta(\zeta\lambda)) \\ \label{ss2a10}
 \frac{\theta^{\prime}}{2\pi} &=& ~\frac{\theta (\zeta \lambda_0 )}{2\pi} \hspace{0.18cm}-
 \int^{\prime}_{\zeta \lambda_0} \frac{d [\zeta \lambda]}{\zeta \lambda}
 \beta_{\theta} (\sigma_{xx}(\zeta\lambda),\theta(\zeta\lambda))
\end{eqnarray}
where
\begin{eqnarray}\label{betafunctions}
\beta_{\sigma} (\sigma_{xx},\theta ) &=& -\frac{d\sigma_{xx}}{d\ln
 \lambda} \hspace{0.46cm}= \frac{m+n}{2\pi} + {D_{mn}} (2\pi\sigma_{xx})^{m+n+2}
 e^{-2\pi\sigma_{xx}(\zeta \lambda)}\cos \theta,\notag
 \\
 \beta_\theta (\sigma_{xx} ,\theta ) &=& -\frac{d (\theta/2\pi)}{d\ln
 \lambda} =  {D_{mn}} (2\pi\sigma_{xx})^{m+n+2}
 e^{-2\pi\sigma_{xx}(\zeta \lambda)}\sin \theta .
\end{eqnarray}
These final results which generalize those obtained earlier, on
the basis of perturbative expansions (see Eq.~\eqref{pb1}), are universal in
the sense that they are independent of the particular
regularization scheme that is being used to define the
renormalized theory.

%.....................................................................
\subsection{Negative anomalous dimension}
%.....................................................................

Eqs \eqref{sxxzl}, \eqref{M.vs.lambda} and \eqref{sxxprimezl}
provide a general prescription that should be used to translate
the parameters $z_i ({\mathcal M})$ and $z_i^\prime ({\mathcal
M})$ into the corresponding quantities $z_i ({ \zeta \lambda})$
and $z_i^\prime ({\zeta \lambda^\prime})$ in {\em flat} space.
Analogous to Eqs \eqref{newsigma} and \eqref{sxxprimezl} we
introduce the parameters $z_i$ and $z_i^\prime$ associated with
scale sizes $\lambda$ and $\lambda^\prime$ respectively as follows
\begin{eqnarray}\label{zzl}
\hspace{3cm}z_i ({\zeta \lambda}) &=& \frac{1}{2\pi}\int d {\bf r} \mu^2 ({\bf r})  z_i (\mu
({\bf r})),  \label{zprimezl0}
\\\label{zprimezl}
z_i^\prime ({\zeta \lambda^\prime}) &=& \frac{1}{2\pi}\int d {\bf r} (\mu^\prime ({\bf
r}))^2  z_i (\mu ({\bf r}).
\end{eqnarray}
Equation \eqref{zprimezl0} implies that $z_i (\zeta\lambda)$ is
related to $z_i ({\mathcal M})$ according to the prescription of
Eq. \eqref{M.vs.lambda},
\begin{equation}\label{flatz}
z_i(\zeta\lambda) =z_i  \left[ 1
 - \frac{\gamma_i^{(0)}}{\sigma_{xx}} \ln {\zeta \lambda \mu_{0}}
 e^{\gamma -1/2} \right].
\end{equation}
It is important to emphasize that the final expressions for $z_i
({\zeta \lambda})$ and $\sigma_{xx} ({\zeta \lambda})$ are
consistent with those obtained in dimensional regularization
(Appendix ~\ref{App0}). Next we make use of Eqs \eqref{sxxzl},
\eqref{zprimezl0} and \eqref{zprimezl} and write the result for
$z_i^\prime$, Eq. \eqref{zi1}, as follows
\begin{equation}
 z_i^{\prime} (\zeta \lambda^\prime) = z_i(\zeta \lambda^\prime) + D_{mn} \int^{\prime}
 \frac{d \lambda}{\lambda}  (2\pi\sigma_{xx})^{m+n}
 e^{-2\pi\sigma_{xx}(\zeta \lambda)} {A}_i \cos \theta .
\end{equation}
The problem that remains is to find the appropriate expression for
the quantity $A_i$ which is defined as
\begin{equation}\label{Ai_1}
 A_i = \pi \gamma_i^{(0)} \int^{\prime} d\textbf{r}
 \frac{\mu(\textbf{r})}{\lambda} z_i({\mu(\textbf{r})}).
\end{equation}

\subsubsection{Amplitude $A_i$}

To evaluate $A_i$ further it is convenient to introduce the
quantity $M_i(\textbf{r})$ and write
\begin{equation}
A_i =-2\pi^2  \gamma_i^{(0)} z_i(\mu(0)) \int
\limits_{\mu(0)}^{\mu(L^\prime)} d [\ln \mu (\textbf{r}) ]
M_i(\textbf{r}), \qquad M_i (\textbf{r})=
\frac{z_i(\mu(\textbf{r}))}{z_i(\mu(0))}.
\end{equation}
In the language of the Heisenberg ferromagnet $M_i (\textbf{r})$
represents a spatially varying {\em spontaneous magnetization}
which is measured relative to the center $|\textbf{r}| =0$ of the
instanton. Notice that for small instanton sizes $\lambda$ which
are of interest to us, the associated momentum scale
$\mu(\textbf{r})$ strongly varies from {\em large} values
${\mathcal O}(\lambda^{-1})$ at short distances ($|\textbf{r}| \ll
\lambda$) to {\em small} values ${\mathcal O}(|\textbf{r}|^{-2})$
at very large distances ($|\textbf{r}| \gg \lambda$). Since a
continuous symmetry cannot be spontaneously broken in two
dimensions the results indicate that $M_i(\textbf{r})$ generally
vanishes for large $|\textbf{r}|$. We therefore expect the
amplitude $A_i$ to remain finite as $L' \rightarrow \infty$. This
is quite unlike the theory at a classical level where $A_i$
diverges and one is forced to work with the idea of constrained
instantons.

Notice that the theory in the replica limit $m=n=0$ is in many
ways special. In this case the anomalous dimension $\gamma_i$ of
mass terms can have an arbitrary sign which means that $M_a
(\textbf{r})$ can {\em diverge} as $|\textbf{r}|$ increases. In
what follows we shall first deal with the problem of ordinary
negative anomalous dimensions, including $\gamma_i =0$. This is
then followed by an analysis of the special cases.

\subsubsection{Details of computation}

To simplify the discussion of the amplitude $A_i$ we limit
ourselves to the theory with $\theta=0, \pi$ such that
$\beta_\theta =0$ and $\beta_\sigma$, $\gamma_i$ are functions of
$\sigma_{xx}$ only,
\begin{equation}\label{hAi4a}
\gamma_i ({\sigma_{xx}}, \theta) \rightarrow  \gamma_i
({\sigma_{xx}}),\qquad \beta_\sigma ({\sigma_{xx}}, \theta) \rightarrow
\beta_{\sigma} ({\sigma_{xx}}).
\end{equation}
Write
\begin{equation}\label{hAi10}
M_i(\textbf{r}) =   \exp\Bigl \{ -\int\limits_{\ln \mu(0)}^{\ln
\mu(\textbf{r})} d [\ln \mu ] \gamma_i \Bigr \},
\end{equation}
then the complete expression for $A_i$ becomes
\begin{equation}\label{hAi10_1}
A_i = -2\pi^2  \gamma_i^{(0)} z_i(\mu(0)) \int\limits_{\ln
\mu(0)}^{\ln \mu(L')} d [\ln \mu(\textbf{r})] \exp\Bigl \{
-\int\limits_{\ln \mu(0)}^{\ln \mu(\textbf{r})} d [\ln \mu
]\gamma_i \Bigr \}.
\end{equation}
As a next step we change the integrals over $\ln\mu$ into
integrals over $\sigma_{xx}$ and write
\begin{equation}\label{hAi9}
A_i =  z_i(\mu(0)) {\mathcal H}_i ( \sigma_{xx} (\mu(0))),
\end{equation}
where
\begin{equation}
\mathcal{H}_i  = - 2\pi^2  \gamma_i^{(0)}
\int\limits_{\sigma_{xx}(\mu(0))}^{\sigma_{xx}(\mu(L^\prime))}
\frac{d \sigma_{xx}} {\beta_{\sigma} (\sigma_{xx})} \exp \Bigl\{ -
\int\limits_{\sigma_{xx}(\mu(0))}^{\sigma_{xx}} \frac{d \sigma}{
\beta_{\sigma}(\sigma)} \gamma_i(\sigma) \Bigr\} .
\end{equation}
The meaning of this result becomes more transparent if we write it
in differential form. Taking the derivative of $\mathcal{H}_i$
with respect to $\ln\lambda$ we find
\begin{equation}\label{Dif1}
\left ( \beta_{\sigma}(\sigma_{xx}(\mu(0))) \frac{d}{d
\sigma_{xx}(\mu(0))} -\gamma_i(\sigma_{xx}(\mu(0)))\right
)\mathcal{H}_i = 2\pi^2 \gamma^{(0)}_i \left (1 +
M_i(L^\prime)\right ) .
\end{equation}
Since in general we have $M_i(L^\prime)\to 0$ for $\gamma^{(0)}_i
<0$ we can safely put $L^\prime = \infty$ from now onward. At the
same time one can solve Eq. \eqref{Dif1} in the weak coupling
limit $\lambda \to 0$ where $\mu(0)$, $\sigma_{xx}(\mu(0)) \to
\infty$. Under these circumstances it suffices to insert for
$\gamma_i$ and $\beta_{\sigma}$ the perturbative expressions
\begin{eqnarray}\label{hAi4}
\hspace{3cm}\gamma_i ({\sigma_{xx}}) &=& \frac{\gamma_i^{(0)}}{\sigma_{xx}} +
\mathcal{O} ( \sigma_{xx}^{-2} ) ,
\\ \label{hAi6}
\beta_{\sigma} ({\sigma_{xx}}) &=& \beta_0
+\frac{\beta_1}{\sigma_{xx}}+ \mathcal{O} (\sigma_{xx}^{-2}),
\end{eqnarray}
where
\begin{equation}\label{hAi8}
\beta_0 = \frac{m+n}{2\pi}, \qquad \beta_1 = \frac{mn + 1
}{2\pi^2}.
\end{equation}
The differential equation becomes
\begin{equation}\label{Dif2}
\left (\beta_0 + \frac{\beta_1}{\sigma_{xx}(\mu(0))} \right )
\frac{d \mathcal{H}_i}{d \sigma_{xx}(\mu(0))}
-\frac{\gamma_i^{(0)}}{\sigma_{xx}(\mu(0))}\mathcal{H}_i = 2\pi^2
\gamma^{(0)}_i .
\end{equation}
The special solution can be written as follows
\begin{equation}\label{Dif4}
\mathcal{H}_i^{(1)} =  \frac{2\pi^2
\gamma_i^{(0)}}{\beta_0-\gamma_i^{(0)}} \left (\sigma_{xx}(\mu(0))
+ \frac{\beta_1}{\gamma_i^{(0)}} \right )
\end{equation}
indicating that $\mathcal{H}_i$ can be written as a series
expansion in powers of $1/\sigma_{xx}(\mu(0))$. The special
solution $\mathcal{H}_i^{(1)}$ does not generally vanish when
$\gamma_i^{(0)} \to 0$, however. To obtain the solution with the
appropriate boundary conditions we need to solve the homogeneous
equation. The result is
\begin{equation}\label{Dif3}
\mathcal{H}_i^{(0)} = C \left (1+ \frac{\beta_0}{\beta_1}
\sigma_{xx}(\mu(0))\right )^{\gamma_i^{(0)}/\beta_0} .
\end{equation}
We obtain $\mathcal{H}_i=0$ for $\gamma_i^{(0)} =0$ provided we
choose
\begin{equation}\label{Dif5}
C = - \frac{2\pi^2 \gamma_i^{(0)}}{\beta_0-\gamma_i^{(0)}}
\frac{\beta_1}{\gamma_i^{(0)}} .
\end{equation}
The desired result for ${\mathcal H}_i(\sigma_{xx} (\mu(0)))$
therefore becomes
\begin{equation}
 {\mathcal H}_i =
\frac{2\pi^2 \gamma_i^{(0)}}{\beta_0-\gamma_i^{(0)}} \left\{
 \sigma_{xx} (\mu(0))+ \frac{\beta_1}{\gamma_i^{(0)}} \left [ 1 -
 \left (1+ \frac{\beta_0}{\beta_1} \sigma_{xx} (\mu(0))\right
 )^{\gamma_i^{(0)}/\beta_0}  \right]\right\}.\label{hAi9a}
\end{equation}
As a final step we next express $\sigma_{xx} (\mu(0))$ and
$z_i(\mu(0))$ in terms of the flat space quantities $\sigma_{xx}
(\zeta\lambda)$ and $z_i(\zeta\lambda)$ respectively. From the
definitions of Eqs \eqref{sxxzl}, \eqref{flatz},
\eqref{countersigma} and \eqref{counterzi} we obtain the following
relations
\begin{eqnarray}
\hspace{3cm}\sigma_{xx}(\mu(0)) &=& \sigma_{xx} (\zeta\lambda)
\left[
1+ \frac{\beta_0}{\sigma_{xx} (\zeta\lambda)} \ln 2\zeta \right],  \nonumber \\
z_i(\mu(0)) &=& \hspace{0.3cm}z_i (\zeta\lambda) \left[ 1+
\frac{\gamma_i^{(0)}}{\sigma_{xx} (\zeta\lambda)} \ln 2\zeta
\right] .
\end{eqnarray}
For our purposes the correction terms ${\mathcal{O}}
(\sigma_{xx}^{-1})$ are unimportant and it suffices to simply
replace the $\sigma_{xx} (\mu(0))$ and $z_i(\mu(0))$ by
$\sigma_{xx} (\zeta\lambda)$ and $z_i (\zeta\lambda)$ respectively
in the final expression for $z_i^{\prime}$,
\begin{equation}\label{ziprimezlprime}
 z_i^{\prime} (\zeta\lambda^\prime) = z_i(\zeta\lambda^\prime)
 + D_{mn} \int^{\prime}
 \frac{d \lambda}{\lambda}  (2\pi\sigma_{xx})^{m+n}
 z_i (\zeta\lambda ) {\mathcal H}_i (\sigma_{xx} (\zeta\lambda))
 e^{-2\pi\sigma_{xx}(\zeta\lambda)} \cos \theta.
\end{equation}
This result solves the problem stated at the outset which is to
express the amplitude $A_i$, Eq. \eqref{Ai}, in terms of the
quantities $z_i (\zeta\lambda)$ and $\sigma_{xx}(\zeta\lambda)$,
i.e.
\begin{equation}\label{Aifinal}
 A_i = z_i (\zeta\lambda) {\mathcal H}_i (\zeta\lambda)
\end{equation}
with the function ${\mathcal H}_i$ given by Eq. \eqref{hAi9a}.

\subsubsection{$\gamma_i$ function}

Introducing an arbitrary renormalization point $\lambda_0$ as
before one can write the perturbative expression for $z_i
(\zeta\lambda^\prime)$, Eq. \eqref{flatz}, as follows
\begin{equation}
 z_i(\zeta\lambda^\prime) =z_i(\zeta\lambda_0)
 +\int_{\zeta\lambda_0}^\prime \frac{d[\zeta\lambda]}{\zeta\lambda}
 \frac{\gamma_i^{(0)}}{\sigma_{xx} (\zeta\lambda)} z_i (\zeta\lambda)
\end{equation}
then Eq. \eqref{ziprimezlprime} can be written in terms of the
$\gamma_i$ function as follows
\begin{equation}
 z_i^\prime (\zeta\lambda^\prime) =z_i(\zeta\lambda_0)
 +\int_{\zeta\lambda_0}^\prime \frac{d[\zeta\lambda]}{\zeta\lambda}
 \gamma_i (\sigma_{xx} (\zeta\lambda), \theta(\zeta\lambda)) z_i (\zeta\lambda)
\end{equation}
where the complete expression for $\gamma_i$ equals
\begin{equation}\label{finalgamma-i}
\gamma_i (\sigma_{xx} ,\theta) =\frac{\gamma_i^{(0)}}{\sigma_{xx}}
+ D_{mn} (2\pi\sigma_{xx})^{m+n} {\mathcal H}_i (\sigma_{xx})
 e^{-2\pi\sigma_{xx}} \cos \theta ,
\end{equation}
with
\begin{eqnarray}\label{calH}
 {\mathcal H}_i (\sigma_{xx}) &=&
 \frac{2\pi^2 \gamma_i^{(0)}}{\beta_0-\gamma_i^{(0)}} \left\{
 \sigma_{xx} + \frac{\beta_1}{\gamma_i^{(0)}} \left [ 1 -
 \left (1+ \frac{\beta_0}{\beta_1} \sigma_{xx} \right
 )^{\gamma_i^{(0)}/\beta_0}  \right]\right\}.
\end{eqnarray}
This final expression generalizes the results obtained earlier, on
the basis of perturbative expansions (see Eq.~\eqref{pb2}). Moreover, it
demonstrates that the replica limit can in general be taken, at
least for all operators with $\gamma_i^{(0)} \leq 0$.

%=======================================
\subsection{\label{ZeroAD} Free energy}
%=======================================
For completeness we next discuss the part $S_h$ of the free energy
\eqref{S-prime3}
\begin{equation}\label{Shrepl}
S_h=\int^{\prime} d \textbf{r} z_h
 (\mu(\textbf{r})) O_h^{\rm inst} ({\bf r}).
\end{equation}
Notice that $S_h$ can be expressed in terms of the amplitude
$A_h$, Eq.~\eqref{Aifinal},
\begin{equation}\label{ShAi}
S_h = \frac{2 \lambda^2 A_h}{\pi\gamma_h^{(0)}} = 2 \lambda^2 z_h (\zeta\lambda)\frac{
{\mathcal H}_h (\zeta\lambda)}{\pi\gamma_h^{(0)}}.
\end{equation}
Although this result is correct for positive values of $m$ and
$n$, it does not give the right result in the replica limit
$m=n=0$ where both $\gamma_h^{(0)}$ and ${\mathcal H}_i$ go to zero.
To obtain the correct result for ${\mathcal H}_h
(\zeta\lambda)/(\pi\gamma_h^{(0)})$ in this case we write Eq. \eqref{Dif1}
as follows
\begin{equation}\label{Dif1hg}
\left ( \beta_{\sigma}(\sigma_{xx}(\mu(0))) \frac{d}{d
\sigma_{xx}(\mu(0))} -\gamma_h(\sigma_{xx}(\mu(0)))\right )
\left[\frac{ {\mathcal H}_h (\zeta\lambda)}{\pi\gamma_h^{(0)}}
\right]= - 2\pi \left (1 + M_h(L^\prime)\right ) .
\end{equation}
In the limit where $\gamma_h=0$ we have $M_h(L^\prime) = 1$ and
Eq.\eqref{Dif1hg} becomes simply
\begin{equation}\label{Dif2hg}
\frac{d}{d \ln \lambda} \frac{{\mathcal H}_h (\zeta\lambda)}{\pi\gamma_h^{(0)}} =
4\pi.
\end{equation}
The result is given by
\begin{equation}\label{Dif3gh}
\frac{{\mathcal H}_h (\zeta\lambda)}{\pi\gamma_h^{(0)}} = 2\pi\ln \tilde h^2 +
\textrm{const}
\end{equation}
and the expression for $S_h$ becomes
\begin{equation}\label{sh21}
S_h = 4\pi z_h \lambda^2 \left (\ln \tilde h^2 +
\textrm{const}\right ).
\end{equation}
This result can of course be obtained directly from Eq.
\eqref{hAi10_1} by substituting $h^{-1}$ for $L^\prime$. In this
way we recover the result on the basis of constrained instantons,
Section \ref{Inst.PBT}.

%===================================================
\subsection{\label{Pad} Positive anomalous dimension}
%===================================================

In case the anomalous dimension of the $z_a$ becomes positive
($\gamma_a^{(0)} >0$) we have to follow a slightly different
route. The quantity to consider in this case is $y_a = z_a^{-1}$
which has an ordinary negative anomalous dimension
\begin{eqnarray}
 y_a^{\prime} = y_a (\mathcal{M})-\pi \gamma_a^{(0)} D_{mn} \int^{\prime}
 \frac{d \lambda}{\lambda} && (2\pi\sigma_{xx})^{m+n}
 e^{-2\pi\sigma_{xx}(\mathcal{M})} \cos \theta \notag \\
 \label{ya1}
 && \times  y_a (\mathcal{M})
 \int^{\prime} d \textbf{r} \frac{\mu(\textbf{r})}{\lambda} .
\end{eqnarray}
In fact, the analysis for the $y_a$ field proceeds along exactly
the same lines as written in the previous Section and the $\gamma$
function is correctly given by Eq. \eqref{finalgamma-i}. Since the
$\gamma_a$ functions of the $y_a$ and $z_a$ fields are identical
except for a difference in the overall sign, one can trivially
obtain the final result for the $z_a$ field from the known
expression for the $y_a$ field. This leads to the following
generalization of Eq. \eqref{finalgamma-i}
\begin{equation}\label{genfinalgamma-i1}
\gamma_i =\frac{\gamma_i^{(0)}}{\pi\sigma_{xx}} + D_{mn}
(2\pi\sigma_{xx})^{m+n} \tilde{\mathcal H}_i (\sigma_{xx})
 e^{-2\pi\sigma_{xx}} \cos \theta
\end{equation}
where
\begin{equation}\label{tildeH}
\tilde{\mathcal H}_i = \frac{2\pi^2
\gamma_i^{(0)}}{\beta_0+|\gamma_i^{(0)}|} \left[ \sigma_{xx}
 - \frac{\beta_1}{|\gamma_i^{(0)}|} \left\{ 1 -
\left (1+ \frac{\beta_0}{\beta_1} \sigma_{xx} \right
)^{-|\gamma_i^{(0)}|/\beta_0} \right\} \right] .
\end{equation}
This final expression which holds for both positive and negative
values of $\gamma_i^{(0)}$, including $\gamma_i^{(0)} =0$, is one
of the main results of the present work. In the remainder of this
paper we shall embark on the physical consequences of our results.

%=================================================================================
\section{\label{RGE.RGE} Summary of results}
%=================================================================================
Since the theory with an ordinary integer number of field
components $m,n \gtrsim 1$ is distinctly different from the one
with $0 \leq m,n \lesssim 1$ we next discuss the two cases
separately.

\subsection{$m,n \gtrsim 1$}

In this case the results for the $\beta$ and $\gamma$ functions
are essentially the same for all values of $m,n$ and consistent
with the Mirmin-Wagner-Coleman theorem which says that a
continuous symmetry cannot be spontaneously broken in two
dimensions. The $\gamma_i$ functions are all negative and the
results can be written as
\begin{equation}
\begin{array}{lclcr}
\beta_\sigma & = & \displaystyle\frac{m+n}{2\pi} +
\displaystyle\frac{mn +1}{2\pi^2\sigma_{xx}} &+& D_{mn}
(2\pi\sigma_{xx})^{m+n+2}
e^{-2\pi\sigma_{xx}} \cos 2\pi \sigma_{xy}, \\
\beta_\theta & = &  & &  D_{mn} (2\pi\sigma_{xx})^{m+n+2}
e^{-2\pi\sigma_{xx}} \sin 2\pi \sigma_{xy},
\end{array}
\end{equation}
\vspace{-0.5cm}
\begin{equation}\label{genfinalgamma-i}
\begin{array}{lclcr}
\gamma_{s,a} &=& \displaystyle -\frac{m+n\pm 1}{\pi\sigma_{xx}}
&-&\displaystyle \frac{2\pi(m+n\pm 1)}{3(m+n) \pm 2 } D_{mn}
(2\pi\sigma_{xx})^{m+n+1} e^{-2\pi\sigma_{xx}} \cos 2\pi
\sigma_{xy},\\
\gamma_{h} &=& \displaystyle -\frac{m+n}{2\pi\sigma_{xx}}
&-&\displaystyle \frac{\pi}{2} D_{mn} (2\pi\sigma_{xx})^{m+n+1}
e^{-2\pi\sigma_{xx}} \cos 2\pi \sigma_{xy}.
\end{array}   %  \notag
\end{equation}
We see that along the $\theta=0$ line the instanton contributions
$\propto e^{-2\pi\sigma_{xx}}$ generally reinforce the results
obtained from ordinary perturbation theory. This means that the
instanton contribution generally tends to make the $\beta_\sigma$
function more {\em positive} and the $\gamma_i$ functions more
{\em negative}. Therefore, upon decreasing the momentum scale - or
in the limit of large distances - the instantons {\em enhance} the
flow of the system toward the {\em strong coupling} phase or {\em
symmetric} phase.

Notice that for $\theta = \pi$ the perturbative and
non-perturbative contributions carry an opposite sign indicating
that the $\theta = \pi$ line displays infrared properties that are
generally different from those along $\theta = 0$. The instanton
contributions indicate that the renormalization group flow is
generally controlled by the weak coupling fixed point located at
$\sigma_{xx} = \infty$ and the strong coupling fixed points
located at $\sigma_{xx} = 0$. The large $N$ expansion can be used
as a stage setting for what happens in the strong coupling
symmetric phase. The results of Ref.~\cite{PruiskenBaranovVoropaevN} indicate that although the
transition at $\theta = \pi$ is a first order one, there is
nevertheless a diverging length scale in the problem
\begin{equation}
\xi =\xi_0 |\theta - \pi|^{-1/2},
\end{equation}
where the exponent $1/2$ equals $1/D$ with $D=2$ the dimension of
the system. The following scaling results have been
found~\cite{PruiskenBaranovVoropaevN}
\begin{eqnarray}\label{scal1}
\hspace{2.5cm}\sigma_{xx} (X,Y) &=& \frac{e^{-X}}{e^{-2X} + 1 + 2Y e^{-X}}
Y\\\label{scal2} \sigma_{xy} (X,Y) &=& k(\nu) + \frac{1 +Y e^{-X}}{e^{-2X}
+ 1 + 2Y e^{-X}}
\end{eqnarray}
with
\begin{eqnarray}\label{scal3}
\hspace{3cm}X &=& \frac{L^2}{\xi_0^2} |\theta(\nu) - \pi| = \pm \left[ \frac{L}{\xi} \right]^2 \\
\label{scal4} Y &\propto&  \frac{\sinh(X)}{X} \exp \left (-\frac{L}{\xi_M}\right ).
\end{eqnarray}
These scaling results are in many ways the same as those expected
for second order transitions and indicate that the $\theta$ vacuum
generically displays all the fundamental features of the quantum
Hall effect.

%%%%%%%%%%%%%%%%%%%%%%%%%%%%%%%%%%%%%%%%%%%%%%%%%%%%%%%%%%%%%%%%%%%%%%%%%%%%%
\begin{figure}
\begin{center}
\includegraphics[width=110mm]{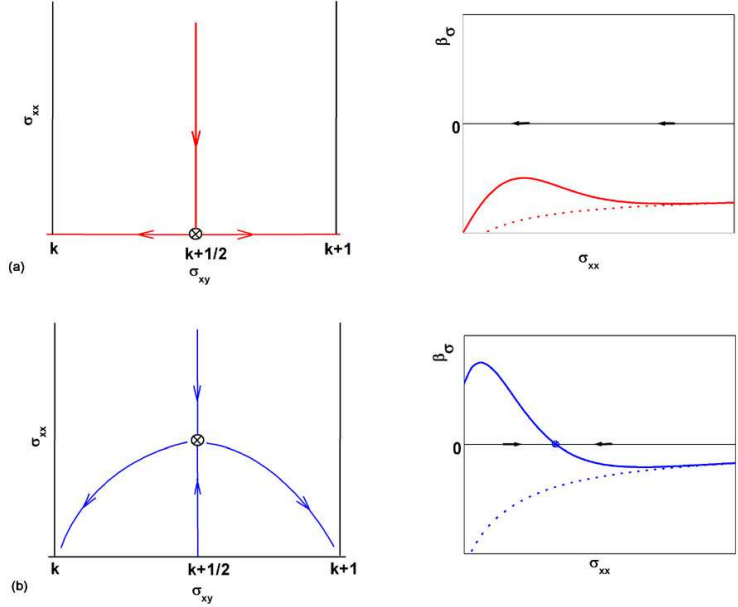}
\caption{The renormalization group flow diagram for different
values of field components $n$ and $m$. (a) The results for large
values $n,m \gtrsim 1$. (b) The results for small values $n,m
\lesssim 1$.} \label{FIG.Renorm}
\end{center}
\end{figure}
%%%%%%%%%%%%%%%%%%%%%%%%%%%%%%%%%%%%%%%%%%%%%%%%%%%%%%%%%%%%%%%%%%%%%%%%%%%%%%
%.

Equations \eqref{scal1}-\eqref{scal4} are valid in the regime
where $L \gg \xi_M$ where $\xi_M$ denotes the correlation length
that describes the cross-over between the weak coupling Goldstone
singularities at short distances (described by Eqs
\eqref{genfinalgamma-i}) and the quantum Hall singularities
(described by Eqs \eqref{scal1}-\eqref{scal4}) that generally
occur at much larger distances only. An overall sketch of the
renormalization in the $\sigma_{xx}$, $\sigma_{xy}$ conductivity
plane is given in Fig.~\ref{FIG.Renorm}a. We see that the infrared
of the system is generally controlled by the {\em stable} quantum
Hall fixed points with $\sigma_{xy} =k(\nu)$ and the {\em
unstable} fixed points located at $\sigma_{xy} =k(\nu) +1/2$ that
describe the singularities of the {\em plateau transitions}.

\subsection{$0 \leq m,n \lesssim 1$}

%%%%%%%%%%%%%%%%%%%%%%%%%%%%%%%%%%%%%%%%%%%%%%%%%%%%%%%%%%%%%%%%%%%%%%%%%%%%
\begin{figure}
\begin{center}
\includegraphics[width=110mm]{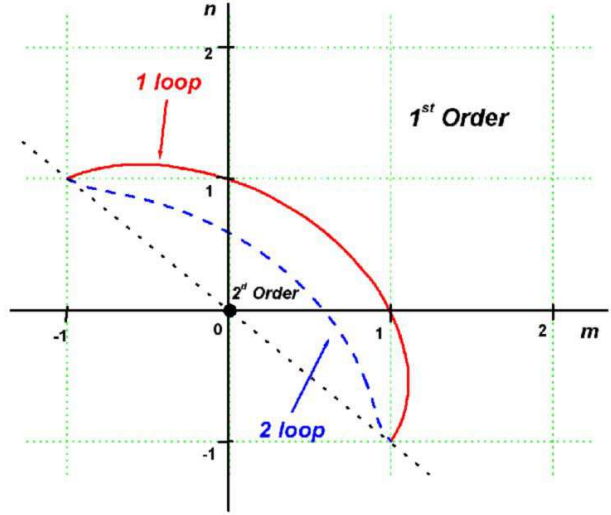}
\caption{Nature of the transition at $\sigma_{xy}=k+1/2$. The
solid red line separates the regions of the first and second order
transitions as predicted by $\beta_\sigma$, $\beta_\theta$, Eqs
\eqref{genfinalgamma-i}. The dashed blue line is obtained by
extending the perturbative contributions to $\beta_\sigma$ to
include the two-loop results. } \label{FIG.PhaseDiagram}
\end{center}
\end{figure}
%%%%%%%%%%%%%%%%%%%%%%%%%%%%%%%%%%%%%%%%%%%%%%%%%%%%%%%%%%%%%%%%%%%%%%%%%%%%%%
Figure~\ref{FIG.Renorm}b illustrates how for small values of $m$
and $n$ the instanton contributions to the $\beta$ functions
produce an infrared zero along the line $\theta=\pi$ with
$\sigma_{xx} = \mathcal{O} (1)$. This indicates that the
transition at $\theta=\pi$ now becomes a true {\em second order}
one. In Fig. ~\ref{FIG.PhaseDiagram} we have plotted the lines in
the $m$, $n$ plane that separate the regimes of {\em first order}
and {\em second order} transitions.

To proceed we first address the replica limit $m=n=0$ where the
theory describes the physics of the disordered electron gas. The
results for the $\beta$ and $\gamma$ functions can be summarized
as follows
\begin{equation}\label{RGNIEa}
\begin{array}{lcccr}
\beta_\sigma &=&  \displaystyle \frac{1}{2\pi^{2}\sigma_{xx}} & +
& \displaystyle \frac{16 \pi}{e}
\sigma_{xx}^{2}e^{-2\pi\sigma_{xx}}\cos2\pi\sigma_{xy}, \\
\beta_\theta &=&  & & \displaystyle  \frac{16 \pi}{e}
\sigma_{xx}^{2}e^{-2\pi\sigma_{xx}}\sin 2\pi\sigma_{xy}, \\
\gamma_{s} &=&- \displaystyle \frac{1}{\pi\sigma_{xx}} & - &
\displaystyle \frac{8 \pi}{e}
\sigma_{xx} e^{-2\pi\sigma_{xx}}\cos 2\pi\sigma_{xy}, \\
\gamma_{a} &=& \displaystyle \frac{1}{\pi\sigma_{xx}} & + &
\displaystyle \frac{8 \pi}{e}
\sigma_{xx} e^{-2\pi\sigma_{xx}}\cos 2\pi\sigma_{xy}, \\
\gamma_{h} &=& 0. & &
\end{array}
\end{equation}
Here we have extended the quantity $\beta_\sigma$ to include the
perturbative results obtained to two loop order.

We see that along the lines where $\sigma_{xy}$ is an integer, the
instanton contributions generally enhance the tendency of the
system toward {\em Anderson localization}. Like the theory with
large values of $m$ and $n$, this means that the $\beta_\sigma$
function renders more {\em positive} in the presence of
instantons. Unlike the previous case, however, the $\gamma_h$
function is now identically {\em zero} whereas the $\gamma_a$
function becomes manifestly {\em positive}. As already mentioned
earlier, these results are extremely important and dictated by the
physics of the problem.

\begin{figure}
\begin{center}
\includegraphics[width=110mm]{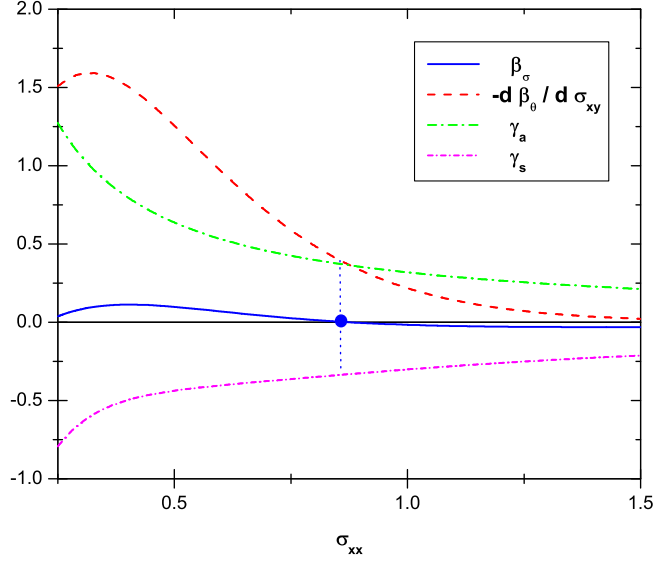}
\caption{The plot of $\beta_\sigma$,
$-d\beta_\theta/d\sigma_{xy}$, $\gamma_{a}$ and $\gamma_{s}$ as
functions of $\sigma_{xx}$ at $\sigma_{xy}=k+1/2$. Blue dot
denotes the fixed point.} \label{FIG.NIEG1}
\end{center}
\end{figure}

When $\sigma_{xy}$ is close to half-integer values the
perturbative and non-perturbative contributions generally carry
the opposite sign. The critical infrared fixed point $\beta_\sigma
=0$ at $\sigma_{xx}^* \approx 0.88$ (Fig. ~\ref{FIG.NIEG1})
indicates that the electron gas {\em de-localizes}. Notice that
the situation is in many ways identical to the {\em mobility edge}
problem in $2+\epsilon$ spatial dimensions except for the fact
that the transition is now being approached from the {\em
insulating} side only.

The consequences of this fixed point for the physics of the
electron gas can be summarized as follows.
%%%%%%%%%%%%%%%%%%%%%%%%%%%%%%%%%%%%%%%%%%%%%%%%%%%%%%%%%%%%%%%%%%%%%%%%%%%%
%%%%%%%%%%%%%%%%%%%%%%%%%%%%%%%%%%%%%%%%%%%%%%%%%%%%%%%%%%%%%%%%%%%%%%%%%%%%%%

{\em 1. Localization length.} Introducing the quantities
\begin{equation}
\Delta_\theta = \sigma_{xy}^0 - k(\nu_B^{*}) - \frac{1}{2} = \nu_B
-\nu_B^{*},\qquad \Delta_\sigma = \sigma_{xx}^0 - \sigma_{xx}^{*}
\end{equation}
which denote the linear environment of the critical fixed point
then the divergent localization length $\xi$ of the electron gas
can be expressed as
\begin{equation}
\xi =\xi_0 |\Delta_\theta |^{-\nu}.
\end{equation}
Here, $\xi_0$ is an arbitrary length scale determined by the
microscopics of the electron gas and
the localization length exponent $\nu$ is obtained as
\begin{equation}
\nu = - \left[\left (\displaystyle\frac{
d\beta_\theta}{d\sigma_{xy}} \right)^{*}\right ]^{-1} \approx 2.8
\end{equation}
%\vspace{0.5cm}

{\em 2. Transport parameters.} Next, the {\em ensemble averaged}
transport parameters of the electron gas can be written as regular
functions of two scaling variables $X$ and $Y$ (see Ref.
\cite{Pruisken4})
\begin{equation}
\sigma_{xx} = \sigma_{xx} (X,Y),\qquad  \sigma_{xy} = \sigma_{xy}
(X,Y),
\end{equation}
where
\begin{equation}
X = \left (\frac{L}{\xi_0}\right )^{y_\theta}\Delta_\theta, \qquad
Y = \left (\frac{L}{\xi_0}\right )^{y_\sigma} \Delta_\sigma
\end{equation}
with
\begin{equation}
y_\theta = \nu^{-1} \approx 0.36,\qquad y_\sigma = -\left (
\frac{d\beta_\sigma}{d\sigma_{xx}} \right )^{*} \approx -0.17.
\end{equation}
Besides the aforementioned results obtained from the large $N$
expansions there exists, until to date, no knowledge on the
explicit form of the scaling functions. Nevertheless, there
indications which tell us that the functions $\sigma_{xx} (X,Y)$
and $\sigma_{xy} (X,Y)$ are, in fact, very similar for all the
cases of interest. For example, recent experiments on quantum
criticality in the quantum Hall regime have shown that the scaling
functions for the true, interacting electron gas are given
by~\cite{PLPdV}
\begin{eqnarray}
\hspace{2.5cm}\sigma_{xx} (X,Y) &=& \frac{e^{-X}}{e^{-2X} + 1 + 2Y e^{-X}} \\
\sigma_{xy} (X,Y) &=& k(\nu) + \frac{1 +Y e^{-X}}{e^{-2X} + 1 + 2Y
e^{-X}}.
\end{eqnarray}
Notice that these results are strikingly similar to those obtained
from the large $N$ expansion, in spite of the fact that the two
systems in question are physically totally different. Both systems
are realizations of an instanton vacuum, however. The results
therefore indicate that the list of {\em super universal} features
of the theory is likely to be extended to include the actual form
of the scaling functions $\sigma_{xx} (X,Y)$ and $\sigma_{xy}
(X,Y)$.

{\em 3. Inverse participation ratio.} The electronic wave
functions $\Psi_E (\textbf{r})$ near the Fermi energy $E$ define a
quantity
\begin{equation}\label{InvPartRat}
P^{(2)} = \Bigl\langle \int d \textbf{r}
|\Psi_{E}(\textbf{r})|^{4}\Bigr\rangle,
\end{equation}
which is called the {\em inverse participation ratio}. Here, the
brackets denote the average with respect to the impurity ensemble.
$P^{(2)}$ is a measure for the inverse of the volume that is taken
by these electronic levels. It is expected to be zero for {\em
extended} states and finite for {\em localized} states. Hence,
this quantity can be used as an alternative probe for Anderson
localization. In the language of the non-linear sigma model
language we can expressed this quantity in terms of the
antisymmetric operator $O_a$ according to
\begin{equation}
P^{(2)} \propto \Bigl\langle z_h O_a (Q) \Bigr\rangle_{z_h \to 0}
= \xi^{-D_2} f_2 \left( z_h \xi^d , \Delta_\sigma \xi^{y_\sigma}
\right) |_{z_h \to 0} .
\end{equation}
Here the expectation is with respect to the non-linear sigma model
in the presence of the operator $O_h$. The exponent $D_2$ equals
\begin{equation}
D_2 = D-\gamma_a^{*} \approx 1.67
\end{equation}
with $D=2$ denoting the dimension of the system. It is interesting
to remark that the numerical value of $D_2$ is largely determined
by the perturbative contribution to $\gamma_a$, the instanton part
merely contributing an amount of roughly three percent.

{\em 4. Multifractality.} Following Wegner ~\cite{Wegner1} the
generalized {\em inverse participation ratio} is defined as
follows
\begin{equation}\label{InvPart1}
P^{(q)} = \Bigl\langle \int d \textbf{r}
|\Psi_{E}(\textbf{r})|^{2q}\Bigr\rangle .
\end{equation}
The mapping of $P^{(q)}$ onto the non-linear sigma model now
involves composite operators with $q$ matrix field variables $Q$.
On the basis of this mapping one expects a scaling behavior of the
form
\begin{equation}\label{InvPart2}
P^{(q)} \propto \xi^{-(q-1)D_q} f_q (z_h \xi^d , \Delta_\sigma
\xi^{y_\sigma}).
\end{equation}
The generalized dimension $D_q$ can be written in terms of the
anomalous dimension $\gamma_q$ of the composite operators
according to
\begin{equation}\label{Dq1}
D_q = D - \frac{\gamma_q^{*}}{q-1} .
\end{equation}
The following perturbative expression is known ~\cite{HofWegner}
\begin{equation}\label{qgamma}
\gamma_q = \frac{q(q-1)}{2\pi\sigma_{xx}}
+\mathcal{O}(\sigma_{xx}^{-2}).
\end{equation}
Based on our results, Eqs \eqref{genfinalgamma-i1} and
\eqref{tildeH}, we ready generalize the expression for $\gamma_q$
to include the effect of instantons. We obtain
\begin{equation}\label{qgammainst}
\gamma_q = \frac{q(q-1)}{2\pi\sigma_{xx}} -
\frac{8\pi}{e}\sigma_{xx} B\Bigl (\pi \sigma_{xx} q (q-1)\Bigr
)e^{-2\pi\sigma_{xx}},
\end{equation}
where function $B(x)$ is obtained from the $\beta_0\to 0$ limit of
the  ${\tilde {\mathcal H}}(\sigma_{xx})$ function and is equal to
\begin{equation}\label{Bf}
B(x) = 1 - \frac{1- e^{-x}}{x}.
\end{equation}
The validity of Eq. \eqref{qgammainst} is likely to be limited to
the case of small values of $q$ only, presumably $q \ll 10$. This
is so because the higher order terms in the perturbative series
for $\gamma_q$ are rapidly increasing with increasing values of
$q$ ~\cite{HofWegner}. The result for $D_q$ becomes
\begin{equation}\label{Dq1inst}
D_q = D - \frac{q}{2} \gamma^{*}_a \left [1 -
\frac{\gamma^{*}_a}{2q(q-1)} B\left
(\frac{q(q-1)}{\gamma^{*}_a}\right ) \right ],\notag
\end{equation}
where $\gamma^{*}_a = 1/\pi\sigma_{xx}^{*}\approx 0.36$. An
important feature of these results is that $\gamma_q^{*}
\rightarrow 0$ or $D_q \rightarrow D$ as $q$ approaches zero. This
property permits one to express the multifractal properties of
generalized inverse participation ratios in terms of the
$f(\alpha)$ singularity spectrum as follows. Introducing the
variable
\begin{equation}\label{alphaq1}
\alpha_{q} = D_{q} - (q-1)\frac{d D_{q} }{d q}
\end{equation}
then $f(\alpha_{q})$ is given by
\begin{equation}\label{fa}
f(\alpha_{q}) = q \alpha_{q} - (q-1)D_{q}.
\end{equation}
To proceed let us first consider the perturbative contributions to
the generalized dimension $D_q$, Eqs \eqref{Dq1} and
\eqref{qgamma}. The result can be cast in the familiar form
\begin{equation}\label{fa11}
f(\alpha) = D - \frac{(\alpha-\alpha_{0})^{2}}{4(\alpha_{0}-D)},
\end{equation}
where
\begin{equation}\label{quadr}
\alpha_{0} = D + \frac{1}{2}\gamma_{a}^{*}.
\end{equation}
The result of Eq. \eqref{fa11} generally describes the $f(\alpha)$
singularity spectrum near its maximum value at $\alpha
=\alpha_{0}$ only. Next, by taking into account the effects from
instantons (see Eq. \eqref{qgammainst}) we find that the
expression of Eq. \eqref{quadr} is only slightly modified in that
the quantity $\alpha_0$ is now given by
\begin{equation}\label{quadr1}
\alpha_{0} = D + \frac{3}{8}\gamma_{a}^{*}.
\end{equation}
Whereas Eq. \eqref{quadr} leads to a numerical value $\alpha_0
=2.18$, Eq. \eqref{quadr1} gives $\alpha_0 =2.14$. Again we find
that the instanton contribution is numerically a small fraction of
the final answer. We attribute important significance to these
results since they are an integral part of the final conclusion of
this paper which says that the quantum critical aspects of the
electron gas are within the range of weak coupling expansion
techniques.

As a final remark it should be mentioned that the tails of the
$f(\alpha)$ singularity spectrum, which are generally controlled
by the $D_q$ with large values of $q$, is beyond the scope of the
present analysis.

%\end{enumerate}

\subsection{Comparison with numerical work}

In Table~\ref{Tab.results} we compare our results for the critical
indices $\nu$, $y_\sigma$, $D_2$ and $\alpha_0$ with those
extracted from numerical simulations on the electron gas. The
agreement is in many respects spectacular, indicating that our
instanton analysis captures some of the most essential and
detailed physics of the problem.

%%%%%%%%%%%%%%%%%%%%%%%%%%%%%%%%%%%%%%%%%%%%%%%%%%%%%%%%%%%%%%%%%%%%%%%%%%%%%%%%%%
%%%%%%%%%%%%%%%%%%%%%%%%%%%%%%%%%%%%%%%%%%%%%%%%%%%%%%%%%%%%%%%%%%%%%%%%%%%%%%%%%%
\begin{table}
\begin{center}
\caption{\vspace{0.5cm}The critical and multifractal exponents for
the model of noninteracting electrons.}
%\begin{ruledtabular}
\begin{tabular}{||c||c|c||}\hline
Exponent & Instantons & Numerical results  \\
\hline \hline
 & & \\
 & & $2.5\pm 0.5$~\cite{NumNu1},  $2.34\pm 0.04$~\cite{NumNu2}\\
 $\nu$ & $2.8\pm 0.4$ &$2.3\pm 0.08$~\cite{NumNu3}, $2.4 \pm 0.1$~\cite{NumNu4} \\
& & $2.2\pm 0.1$~\cite{NumNu5}, $2.43\pm 0.18$~\cite{NumNu6}\\
& &\\
\hline & & \\
$y_\sigma$& $-0.17\pm 0.02$ & $-0.38 \pm 0.04$~\cite{NumY1}, $-0.4 \pm 0.1$~\cite{NumY2} \\
& & \\
\hline & & \\
$D_{2}$ & $1.67\pm 0.03$ & $1.62\pm 0.04$~\cite{NumGA1}, $1.62\pm 0.02$~\cite{NumGA2} \\
& & $1.40\pm 0.02$~\cite{NumLW}\\
\hline & & \\
 $\alpha_{0}$& $2.14\pm 0.02$ &$2.30\pm 0.07$~\cite{NumAlpha01}, $2.29\pm
 0.02$~\cite{NumAlpha02} \\
 & &  $2.260\pm 0.003$~\cite{NumY2}\\
 & &
 \\
\hline
\end{tabular}
%\end{ruledtabular}
\label{Tab.results}
\end{center}\vspace{0.5cm}
\end{table}
%%%%%%%%%%%%%%%%%%%%%%%%%%%%%%%%%%%%%%%%%%%%%%%%%%%%%%%%%%%%%%%%%%%%%%%%%%%%%%%%%%%%%
%%%%%%%%%%%%%%%%%%%%%%%%%%%%%%%%%%%%%%%%%%%%%%%%%%%%%%%%%%%%%%%%%%%%%%%%%%%%%%%%%%%%%

%%%%%%%%%%%%%%%%%%%%%%%%%%%%%%%%%%%%%%%%%%%%%%%%%%%%%%%%%%%%%%%%%%%%%%%%%%%%
\begin{figure}
\begin{center}
\includegraphics[width=100mm]{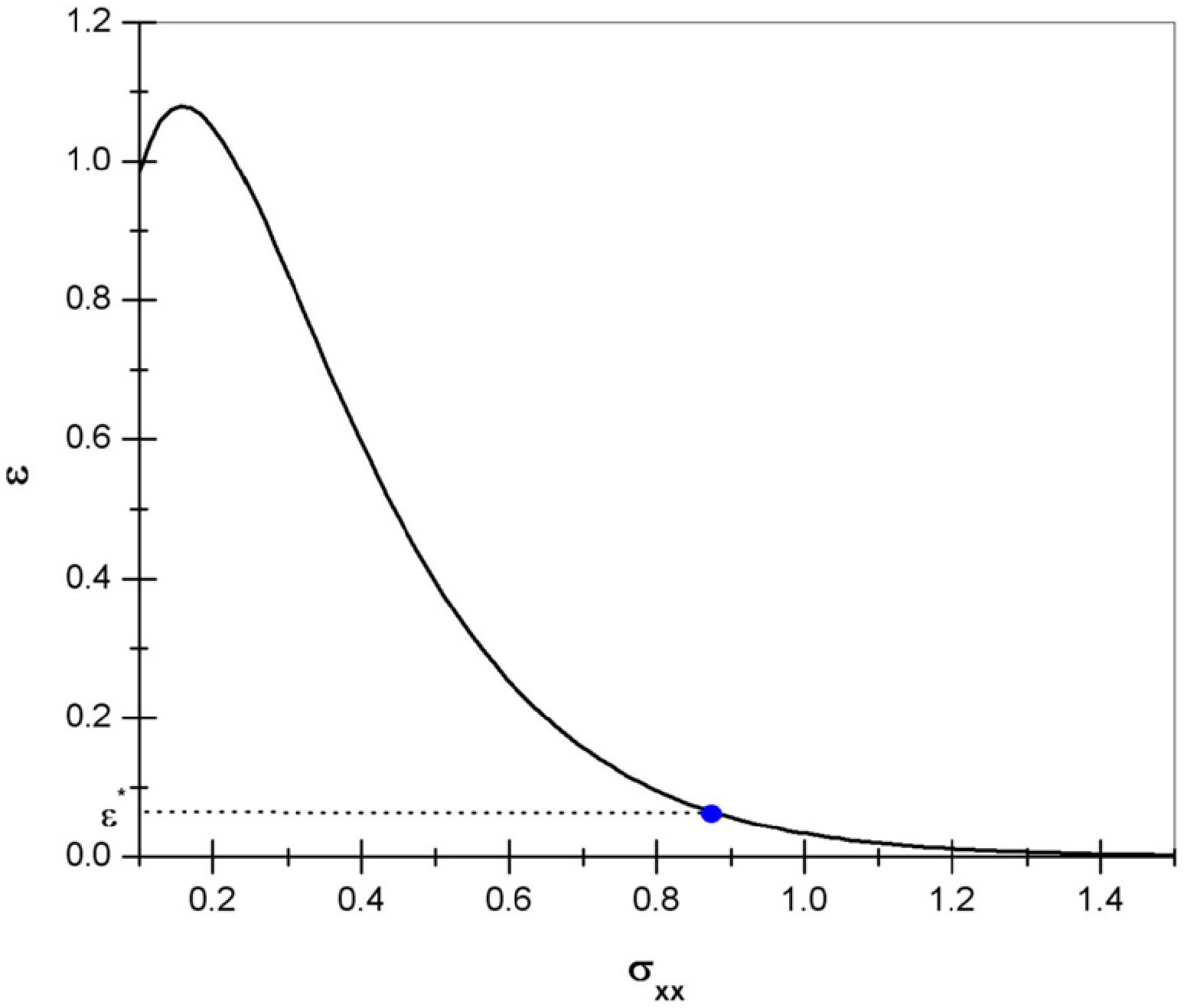}
\caption{The dependence of $\epsilon$ on the $\sigma_{xx}$ for the
model of free electrons $m=n=0$. } \label{FIG.epsA}
%\end{center}
%\end{figure}
%%%%%%%%%%%%%%%%%%%%%%%%%%%%%%%%%%%%%%%%%%%%%%%%%%%%%%%%%%%%%%%%%%%%%%%%%%%%%%
%%%%%%%%%%%%%%%%%%%%%%%%%%%%%%%%%%%%%%%%%%%%%%%%%%%%%%%%%%%%%%%%%%%%%%%%%%%%
%\begin{figure}
%\begin{center}
\includegraphics[width=80mm]{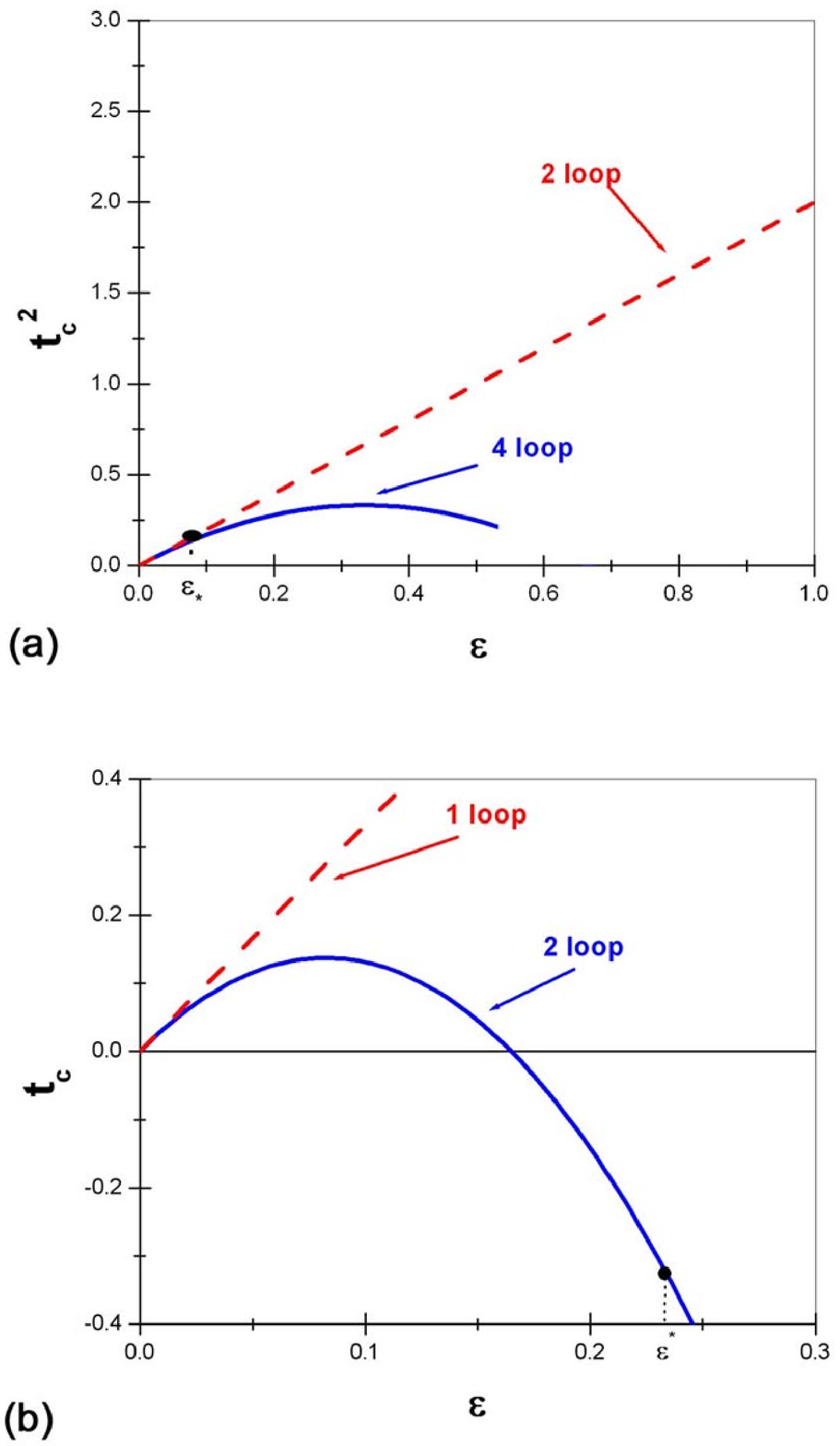}
\caption{The dependence of the critical point $t_c =
1/\pi\sigma^{*}_{xx}$ from $\epsilon$. (a) The model of free
electrons, $m=n=0$. (b) The model with  $m=n=0.3$. }
\label{FIG.eps}
\end{center}
\end{figure}
%%%%%%%%%%%%%%%%%%%%%%%%%%%%%%%%%%%%%%%%%%%%%%%%%%%%%%%%%%%%%%%%%%%%%%%%%%%%%%

The study of the multi fractal aspects of the problem - as done in
the previous Section - already indicates that higher order
instanton effects are numerically of minor importance. To
generally understand why the effects of multi-instanton
configurations do not carry much weight, however, one might argue
that the very existence of the critical fixed point at
$\sigma_{xx} = \sigma_{xx}^{*}$ and $\sigma_{xy} = k(\nu) + 1/2$
implies that the corrections due to multi-instantons typically
involve factors like $e^{-4\pi\sigma_{xx}^{*}}$ which are
negligible as compared to the leading order result
$e^{-2\pi\sigma_{xx}^{*}}$. To specify the thought we introduce
the quantity
\begin{equation}\label{epsilon}
\epsilon = \frac{16 \pi}{e} \sigma_{xx} e^{-2\pi\sigma_{xx}} .
\end{equation}
In the neighborhood of the lines $\sigma_{xy} = k(\nu) + 1/2$ the
renormalization group functions \eqref{RGNIEa} can then be written
in the form of an ordinary $\epsilon$-expansion
\begin{eqnarray}\label{RGNIE1}
%\begin{array}{lcccr}
\hspace{3cm}\beta_\sigma &=&  -\epsilon\sigma_{xx}  +
\frac{1}{2\pi^{2}\sigma_{xx}}
%+ \frac{3}{8\pi^4\sigma_{xx}^3}
+
\left[\left[ \beta_\sigma \right]\right],\\
\frac{d \beta_\theta}{d \sigma_{xy}} &=& - 2\pi
 \epsilon\sigma_{xx}
 + \left[\left[ \frac{d
\beta_\theta}{d \sigma_{xy}} \right]\right],
\notag\\
\gamma_{s} &=&  \frac{\epsilon}{2}  - \frac{1}{\pi\sigma_{xx}} +
\left[\left[\gamma_{s} \right]\right],
\notag\\
\gamma_{a} &=& - \frac{\epsilon}{2}  +
\frac{1}{\pi\sigma_{xx}}+\left[\left[\gamma_{a} \right]\right] ,
\\ \gamma_{h} &=& 0.\notag
%\end{array}
\end{eqnarray}
The brackets $[[\cdots ]]$ in the $\beta$ and $\gamma$ functions
generally stand for the higher order terms that contain the more
complex contributions from the multi-instanton configurations. It
is clear, however, that in order for the one-instanton approach to
be successful the parameter $\epsilon$ must be a small quantity.
By the same token, the expressions $[[\cdots]]$ should be well
approximated by inserting the leading order corrections as
obtained from ordinary perturbative expansions
\begin{eqnarray}\label{br}
%\begin{array}{ccr}
\hspace{3cm}\left[\left[ \beta_\sigma \right]\right] &=&
\displaystyle\frac{3}{8\pi^4\sigma_{xx}^3} + \cdots, \\
\left[\left[ \displaystyle\frac{d \beta_\theta}{d \sigma_{xy}}
\right]\right] &=& 0, \notag\\
\left[\left[\gamma_{s} \right]\right] &=&\displaystyle -\frac{3(1+c)}{8\pi^3\sigma_{xx}^3} + \cdots, \notag\\
\left[\left[\gamma_{a} \right]\right] &=&
\displaystyle\frac{3}{8\pi^3\sigma_{xx}^3}+\cdots.\notag
%\end{array}
\end{eqnarray}
Here, the constant $c$ has not yet been computed explicitly. The
critical fixed point can then be obtained formally as an expansion
in powers of $\epsilon$,
\begin{equation}\label{epsilon1}
(\pi\sigma_{xx}^{*})^{-2} = 2 \epsilon\left (1-
\frac{3}{2}\epsilon\right ).
\end{equation}
Similarly, the critical indices can be obtained as
\begin{equation}\label{epsexp}
\begin{array}{lcrcccr}
y_\theta &=&  \sqrt{2\epsilon} \Bigl (1&+& &&
\displaystyle\frac{3}{4}
\epsilon\Bigr ), \\
y_\sigma &=&  - \sqrt{2\epsilon} \Bigl (1 &-& \displaystyle\frac{3}{2} \sqrt{2\epsilon}&+&
\epsilon\Bigr ),\\
\gamma_s^{*} &=& -\sqrt{2\epsilon} \Bigl (
1&-&\displaystyle\frac{1}{4}\sqrt{2\epsilon}  & + &\displaystyle\frac{3c}{4}\epsilon\Bigr ),\\
\gamma_a^{*} &=& \sqrt{2\epsilon} \Bigl
(1&-&\displaystyle\frac{1}{4}\sqrt{2\epsilon}& &\Bigr ),\\
\gamma_h^{*}&=& 0 & & & &.
\end{array}
\end{equation}

A simple computation next tells us that at the fixed point
$\sigma_{xx}^*$ the parameter $\epsilon$ equals $\epsilon \approx
0.07$ which is indeed a ``{\em small}'' quantity in every respect.
In Fig. ~\ref{FIG.epsA} we have plotted the curve $\epsilon$ with
varying values of $\sigma_{xx}$ according to Eq.~\eqref{epsilon}.
We see that the fixed point values $\sigma_{xx}^{*}$ and
$\epsilon^{*}$ are located well inside the ``exponential tail''
region where the curve is dominated by the instanton factors
$e^{-2\pi\sigma_{xx}}$. In Fig. ~\ref{FIG.eps}(a) we have plotted
the critical value $t^2_c = (\pi\sigma_{xx}^*)^{-2}$ with varying
values of $\epsilon$ in two-loop approximation where the
correction term $\left[\left[ \beta_\sigma \right]\right]$ in
$\beta_\sigma$ (see Eqs \eqref{RGNIE1}) has been dropped, as well
as in four-loop approximation where this term is retained. The
results clearly indicate that the theory with $m=n=0$ lies well
inside the regime of validity of the weak coupling expansions as
considered in this paper.

Finally, we have used the highest order correction terms in the
series of Eq. \eqref{epsexp} as an estimate for the uncertainty in
the exponent values based on instantons (see
Table~\ref{Tab.results}).

%=====================
\subsection{Continuously varying exponents and a conjecture}
%======================

\begin{figure}
\begin{center}
\includegraphics[width=110mm]{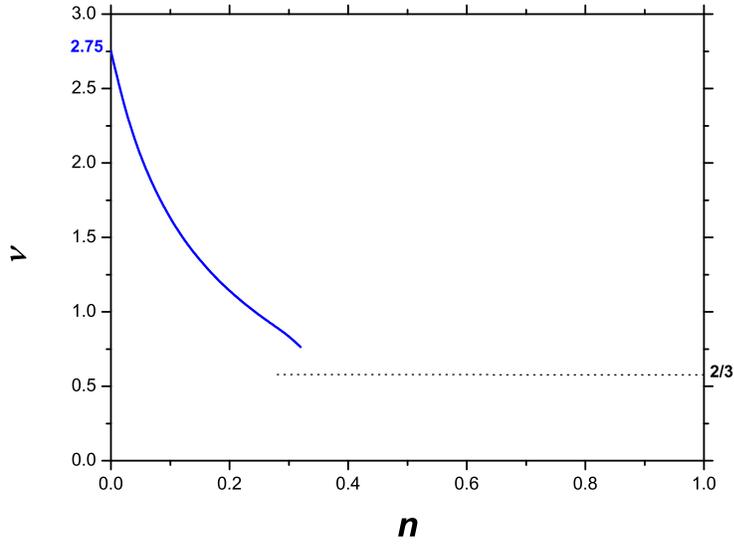}
\caption{The correlation length exponent $\nu$ with varying values
of $0 \leq m=n \lesssim 1$. For comparison we have plotted the
value $\nu =2/3$ which is known to be the exact result for
$m=n=1$, see text.} \label{FIG.IND1}
\end{center}
\end{figure}

\begin{figure}
\begin{center}
\includegraphics[width=110mm]{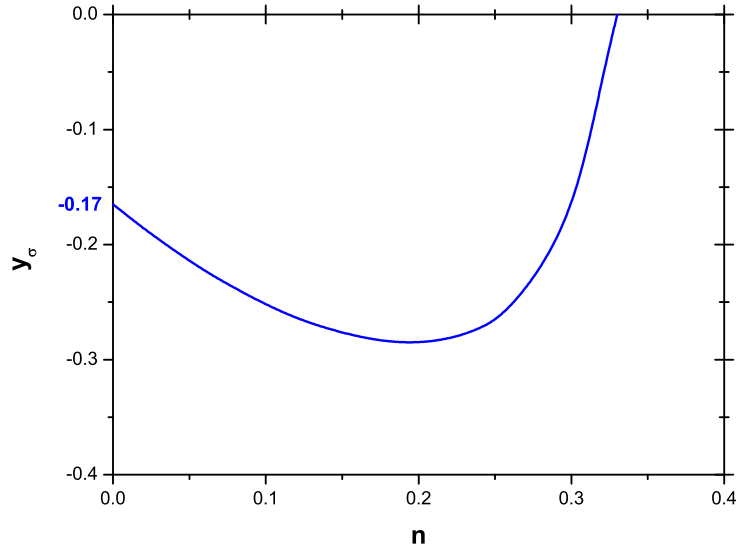}
\caption{The irrelevant exponent $y_\sigma$ with varying values of
$0 \leq m=n \lesssim 1$, see text.} \label{FIG.IND2}
\end{center}
\end{figure}

\begin{figure}
\begin{center}
\includegraphics[width=110mm]{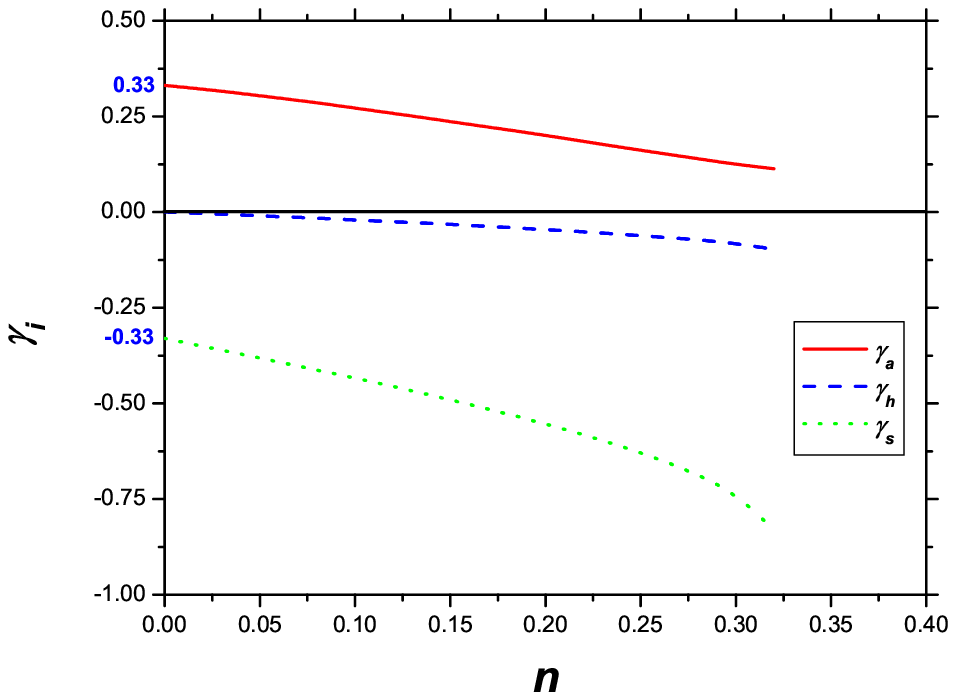}
\caption{The anomalous dimension $\gamma_s^{*}$, $\gamma_a^{*}$
and $\gamma_h^{*}$ with varying values of $0 \leq m=n \lesssim 1$,
see text.} \label{FIG.IND3}
\end{center}
\end{figure}

The critical behavior at $\theta=\pi$ changes continuously as the
value of $m$ and $n$ increases. In Figs ~\ref{FIG.IND1}
and~\ref{FIG.IND2} we plot our instanton results for the exponents
$\nu$ and $y_\sigma$ in the interval $0 \leq m=n \lesssim 0.3$. In
Fig.~\ref{FIG.IND3} the results are presented for the three
different exponents $\gamma_i^*$ with varying $m=n$. Of interest
is the critical end-point $m=n \approx 0.3$ or, more generally,
the boarder line in the $m,n$ plane that separates the regimes of
{\em second order} and {\em first order} phase transitions (Fig.
\ref{FIG.PhaseDiagram}). From the mechanism by which the critical
fixed point in $\beta_\sigma$ is generated (see Fig.
\ref{FIG.Renorm}) it is clear that the boarding line is defined by
the points $m,n$ where the exponent $y_\sigma$ renders {\em
marginal}. Along this line the expansion of the $\beta$ functions
about the fixed point values $\sigma_{xx}^{*}$ and
$\sigma_{xy}^{*} = k(\nu) + 1/2$ can be written as follows
\begin{eqnarray}
\hspace{3cm}\beta_\sigma &=& - \alpha_2 \left(\sigma_{xx} - \sigma_{xx}^{*} \right)^2 \\
\beta_\theta &=& -\nu^{-1} \left( \sigma_{xy} - k(\nu) -
\frac{1}{2} \right). \notag
\end{eqnarray}
Here, $\nu$ equals the correlation length exponent and $\alpha_2$
is a positive constant. Notice that the variable $\sigma_{xx} -
\sigma_{xx}^{*} >0 $ scales to zero in the infrared and, hence,
this quantity is {\em marginally irrelevant}. On the other hand,
the variable $\sigma_{xx} - \sigma_{xx}^{*} <0 $ increases with
increasing length scales and this quantity is therefore {\em
marginally relevant}. A characteristic feature of marginally
relevant/irrelevant scaling variables is that the critical
correlation functions of the system are no longer given by simple
power laws but, rather, they acquire logarithmic corrections.

The problem, however, is that the exact location of this line in
the $m,n$ plane is beyond the scope of the present analysis. It is
easy to see, for example, that the quantity $\epsilon$ (Eq.
\ref{epsilon}), unlike the theory with $m=n=0$, cannot be
considered as a ``small'' parameter when the values of $m$ and $n$
increase. For illustration we compare in Fig.~\ref{FIG.eps} the
critical fixed point $t_c =1/\pi\sigma_{xx}^{*}$ with varying
values of $\epsilon$ for two different values of $m=n$. Whereas
the theory with $m=n=0$ lies well inside the range of validity of
the ``$\epsilon$ expansion'', this is no longer the case when $m=n
\approx 0.3$. At the same time, the expansion is no longer
controlled by the exponential tails $e^{-2\pi\sigma_{xx}}$ when
the values of $m$ and $n$ increase. All this indicates that the
exact location of the critical boarding line in the $m,n$ plane
necessarily involves a detailed knowledge of multi-instanton
effects.

In spite of these and other complications, however, there are very
good reasons to believe that our instanton results display all the
qualitative features of quantum criticality in the theory with
small $m,n$. It is well known, for example, that the $O(3)$ model
at $\theta = \pi$ exhibits a second order phase transition. Since
the (algebraic) correlation functions turn out to have logarithmic
corrections in this case~\cite{ReadShankar} it is natural to
associate the point $m=n=1$ with the aforementioned boarder line
in the $m,n$ plane. This means that the regime of second order
phase transitions actually spans the interval $0 \leq m=n \leq 1$
in Figs \ref{FIG.IND1} - \ref{FIG.IND3}, rather than $0 \leq m=n
\leq 0.3$ as predicted by our instanton analysis. At the same time
one expects the global phase diagram (Fig. \ref{FIG.PhaseDiagram})
to be slightly modified and replaced by Fig. \ref{FIG4} in the
exact theory.


%%%%%%%%%%%%%%%%%%%%%%%
\section{Conclusion\label{Conc}}
%%%%%%%%%%%%%%%%%%%%%%%%%

The results of this paper are an integral part of the general
statement which says the physics of the quantum Hall effect is a
{\em super universal} strong coupling feature of the topological
concept of an instanton vacuum in asymptotically free field
theory. We have shown that the instanton gas unequivocally
describes the cross-over between the {\em Goldstone} phase where
perturbation theory applies and the completely non-perturbative
regime of the {\em quantum Hall effect} that generally appears in
the limit of much larger distances only.

As a major technical advance we have obtained not only the
non-perturbative $\beta$ functions of the theory but also the
anomalous dimension or $\gamma$ function associated with mass
terms. Amongst many other things, the results of this paper lay
the foundation for a non-perturbative analysis of the electron gas
that includes the effects of electron-electron interactions.

Whereas the theory with finite $m,n >1$ has been discussed in
detail in previous work, on the basis of the large $N$ expansion,
the main objective of the present analysis has been the theory
with $0 \leq m,n \leq 1$. In this case one expects a {\em second
order} phase transition at $\theta = \pi$ that is characterized by
a finite value of $\sigma_{xx}^*$ as well as continuously varying
exponents. Although much of the global phase structure of the
theory was either known or anticipated in previous work by one of
us, we have shown that the technical advances made in this paper
are nevertheless extremely important. The main difference with the
previous situation is that certain ambiguities in the theory have
been removed, notably the {\em definition} of the $\beta$
functions. Moreover, by extending the theory to include the
renormalization group $\gamma$ functions we have not only resolved
some of the outstanding problems in the instanton methodology, but
also facilitated detailed comparisons between the predictions of
the theory on the one hand, and the data known from numerical
experiments on the free electron gas on the other.

We have seen that the results of the theory with $m=n=0$ agree
remarkably well with the exponent values extracted from numerical
simulations (Table \ref{Tab.results}). Notice that this is the
first time - ever since the $\theta$ parameter was introduced in
the theory of the quantum Hall effect by one of us - that accurate
estimates for the critical indices of the plateau transitions have
been obtained {\em analytically}. The advances reported in this
paper are obviously important since they teach us something
fundamental about the strong coupling problems in QCD where the
algebra is the same, but experiments are impossible.

\section{Acknowlegements}

This research was funded in part by the Dutch National Science
Foundations \textit{NWO} and \textit{FOM}. One of us (\textit{IB})
is indebted to the Forschungszentrum J\"ulich (\textit{Landau
Scholarship}), Dynasty Foundation, Russian Foundation for Basic
Research (\textit{RFBR}), and the Russian Ministry of Science for
financial support.

\appendix
%%%%%%%%%%%%%%%%%%%%%%%%%%%%%%%%%%%%%%%%%%%%%%%%%%%%%%%%%%%%%%%%%%%%%%
%         APPENDIX A
%
%%%%%%%%%%%%%%%%%%%%%%%%%%%%%%%%%%%%%%%%%%%%%%%%%%%%%%%%%%%%%%%%%%%%
\section{\label{App0} Appendix. Observable and renormalized theories
in $2+\epsilon$ dimensions}

Br$\acute{e}$zin et.al.~\cite{BZG} originally showed that the
non-linear sigma model in $2+\epsilon$ dimensions generally
involves a renormalization of the coupling constant or
$\sigma_{xx}$ and one renormalization associated with each of the
operators $O_i$. Denoting the bare parameters of the theory by
$\sigma_{xx} = 1/g_0$ and $z_i^0$ then the relation between the
{\em bare} theory and {\em renormalized} theory $g$ and $z_i$ is
given by
\begin{equation}\label{3.12c}
\sigma_{xx} = \frac{1}{g_0} = \frac{1}{g}\mu^{\epsilon} Z (g),
\qquad z_i^0 = z_i Z_i (g)
\end{equation}
with $\mu$ an arbitrary momentum scale. The functions $Z(g)$ and
$Z_i (g)$ are usually fixed by the requirement that the theory be
finite in $\epsilon$. According to the minimum subtraction scheme,
for example, one employs the $Z$ and $Z_i$ for the purpose of
absorbing the pole terms in $\epsilon$ and nothing but the pole
terms. However, it is well known that in order to be consistent
with the infrared behavior of the theory the terms that are finite
in $\epsilon$ can play an important role. Cross-over problems, for
example, are treated incorrectly within the minimum subtraction
scheme and usually involve a very specific choice of the functions
$Z$ that includes terms that are finite in $\epsilon$. In this
Section we show that the arbitrariness in the renormalization
group is in general avoided if one employs the renormalizations
$Z$ and $Z_i$ for the purpose of identifying the {\em
renormalized} and {\em observable} theories. For simplicity we
shall present the results for the theory in the presence of the
operator $O_h$ only. It is convenient to introduce a change of
variables. Write
\begin{equation}
\sigma_{xx}^\prime= \frac{1}{g^\prime}, \qquad \sigma_{xx} =
\frac{1}{g_0},
\end{equation}
and
\begin{equation}
z_h^\prime = \frac{(h^\prime)^{2}}{g^\prime},\qquad z^0_h =
\frac{h^{2}_0}{g_0}.
\end{equation}
Since the $g^\prime, h^\prime$ fields have the same meaning for
the various Grassmannian manifolds listed in Table~\ref{TbAB} we
shall from now onward work within the general $G/H$ non-linear
sigma model. Starting from the action
\begin{equation}
S = \frac{a}{2}S_\sigma [Q] + S_h [Q],
\end{equation}
%
%%%%%%%%%%%%%%%%%%%%%%%%%%%%%%%%%%%%%%%%%%%%%%%%%%%%%%%%%%%%%%%%%
\begin{table}
\begin{center}
\caption{\vspace{0.5cm}Coefficients of two-loop computation}
\begin{tabular}{||c||c|c|c|c||}
\hline\hline
$G/H$ & $a$ & $A_1$ & $A_2$ & $B$  \\
\hline \hline & & & &\\
$ \displaystyle\frac{SO(m+n)}{S(O(m) \times O(n))}$ & $1$ & $m+n-2$ & $ 2 mn-m-n$ & $4-2m-2n$\\
& & & &\\
$ \displaystyle\frac{SU(m+n)}{S(U(m) \times U(n))}$ & $2$ & $m+n$ & $ 2 (mn+1)$ & $0$ \\
& & & &\\
 $ \displaystyle\frac{SP(m+n)}{SP(m) \times SP(n)}$ & $4$ &
$m+n+1$ & $\displaystyle 2 mn + \frac{m+n}{2}$ & $m+n+1$  \\
& & & & \\ \hline\hline
\end{tabular}\label{TbAB}
\vspace{0.5cm}
\end{center}
\end{table}
%%%%%%%%%%%%%%%%%%%%%%%%%%%%%%%%%%%%%%%%%%%%%%%%%%%%%%%%%5
then a straightforward computation to order $\epsilon^2$ of the
observable quantities $\sigma_{xx}^\prime$ (see Eq.~\eqref{sdef})
and $z_h^\prime$ (see Eq.~\eqref{rendef}) yields the following
results
\begin{eqnarray}
\frac{1}{g^\prime} &=& \frac{1}{g_0} \left( 1 + A_1 \frac{g_0
h^{\epsilon}_0}{\epsilon} + \frac{1}{2} \left( A_2 - B \right)
\frac{g^2_0 h^{2\epsilon}_0}{\epsilon} - \frac{1}{2} A_2 C g^2_0
h^{2\epsilon}_0
 \right), \label{3.10a} \\
\hspace{-2cm}\frac{(h^\prime)^2}{g^\prime} &=& \frac{h^2_0}{g_0}
\left( 1 + \left( A_1 - \frac{B}{A_1} \right) \frac{g_0
h^{\epsilon}_0}{\epsilon} - \left( A_1 - \frac{B}{A_1} \right)
\frac{B}{A_1} \; \frac{g^2_0 h^{2\epsilon}_0}{\epsilon^2} \left
(1+\frac{\epsilon}{2} \right ) \right). \label{3.10b}
\end{eqnarray}
Here, the coefficients $A_1$, $A_2$ and $B$ are listed in Table 3
and $C$ is a numerical constant. A factor $2\Gamma \left( 1 -
\epsilon/2 \right)(4\pi)^{-1-\epsilon/2}$ has been absorbed in a
redefinition of the $g_0$ and $g^\prime$.

The results of Eqs \eqref{3.10a} and \eqref{3.10b} have originally
been used in ~\cite{PruiskenWang} for the purpose of expressing
the observable parameters in $2+\epsilon$ dimensions in terms of
{\em equations of state}. Here we shall point out a slightly
different interpretation of these results which is obtained by
recognizing that the $h^\prime$ field actually plays the role of
{\em momentum scale} that is associated with the {\em observable}
quantities $g^\prime$ or $\sigma_{xx}^\prime$ and $z_h^\prime$.
This observation permits one to identify the {\em observable} and
{\em renormalized} theories in the following manner. First we
employ Eqs \eqref{3.10a} and \eqref{3.10b} and eliminate the $h_0$
field in the dimensionless combination $g_0 h_0^\epsilon$ in favor
of the induced momentum scale $h^\prime$. The results can be
written as
\begin{eqnarray}
\frac{1}{g'} &=& \frac{1}{g_0} \left( 1 + A_1 \frac{g_0
(h^\prime)^{\epsilon}}{\epsilon} + \frac{1}{2} A_2 \frac{g^2_0
(h^\prime)^{2\epsilon}}{\epsilon} - \frac{1}{2} A_2 C g^2_0
(h^\prime)^{2\epsilon}
 \right), \label{3.10c} \\
\frac{(h^\prime)^2}{g'} &=& \frac{h^2_0}{g_0} \left( 1 + \left(
A_1 - \frac{B}{A_1} \right) \frac{g_0
(h^\prime)^{\epsilon}}{\epsilon} - \left( A_1 - \frac{B}{A_1}
\right) \frac{B}{A_1} \; \frac{g^2_0
(h^\prime)^{2\epsilon}}{\epsilon^2}  \right). \label{3.10d}
\end{eqnarray}
\noindent{As} a second step we make use of Eq. \eqref{3.10c} and
eliminate the bare parameter $g_0$ in the combination $g_0
(h^\prime)^\epsilon$ in favor of the $g^\prime$. Introducing the
dimensionless quantity
\begin{equation}
\bar{g} = g^\prime (h^\prime)^{\epsilon}, \label{3.11}
\end{equation}
then we obtain from Eqs \eqref{3.10c} and \eqref{3.10d}
\begin{eqnarray}
\hspace{4cm}\frac{1}{g_0} &=&
\frac{(h^\prime)^{\epsilon}}{\bar{g}} Z
(\bar{g}), \label{3.12a} \\
\frac{h^2_0}{g_0} &=& \frac{(h^\prime)^{2+\epsilon}}{\bar{g}} \;
{Z_h (\bar{g} )}, \label{3.12b}
\end{eqnarray}
where to order ${\bar g}^2$ the $Z$ and $Z_h$ are given by
\begin{eqnarray}\label{Zbart}
\hspace{3cm}Z (\bar{g} ) &=& 1 - A_1 \frac{\bar{g}}{\epsilon} -
A_2 (1-\epsilon C) \frac{\bar{g}^2}{2 \epsilon} , \label{3.13a}
\\\label{Zhbart} {Z_h (\bar{g})} &=& 1 - \left( A_1 -
\frac{B}{A_1} \right) \frac{\bar{g}}{\epsilon}. \label{3.13b}
\end{eqnarray}
Equations \eqref{3.12a} and \eqref{3.12b} provide a natural
definition of the quantities $Z (g)$ and $Z_h (g)$ that appear in
the expressions of the {\em renormalized} theory, Eqs
\eqref{3.12c}. By fixing the renormalizations $Z$ and $Z_h$
according to Eqs \eqref{Zbart} and \eqref{Zhbart} we obtain
renormalization group $\beta$ and $\gamma$ functions in the usual
manner
\begin{eqnarray}
\beta (g) &=&  \frac{d g}{d \ln \mu} = \frac{\epsilon g}{1- g
\displaystyle\frac{d \ln Z}{d g}} = \epsilon g - A_1 g^2 - g^3
A_2( 1 - \epsilon
C ), \label{3.15} \\
\gamma_h (g) &=& -\frac{d \ln z_h}{d\ln\mu} = \beta (g) \frac{d}{d
g} \ln Z_h (g) = - \left( A_1 - \frac{B}{A_1} \right) g + O(g^3).
\label{3.16}
\end{eqnarray}
Moreover, the choice of Eqs \eqref{3.12a} and \eqref{3.12b}
implies that the observable theories at different momentum scales
$h^\prime$ and $h$ respectively can in general be expressed in
terms of the $\beta$ and $\gamma$ functions according to
\begin{eqnarray}\label{bart}
\hspace{3cm}\bar{g} &=& g(h^\prime) = g(h) + \int_{h}^{h^\prime}
\frac{d\mu}{\mu} \beta (g) \\ \label{barz} z_h^\prime &=& z_h
(h^\prime) = z_h (h) - \int_{h}^{h^\prime} \frac{d\mu}{\mu}
\gamma_h (g) z_h (\mu) .
\end{eqnarray}
The skeptical reader might want to explicitly verify the fact that
the results of Eqs \eqref{bart} and \eqref{barz} are consistent
with the original definition of the observable theory, Eqs
\eqref{3.10c} and \eqref{3.10d}. Starting from Eq. \eqref{3.10c},
for example, one proceeds by inserting ${g_0} = {h^{-\epsilon}}{g}
Z^{-1} (g)$ where $g$ is now defined for momentum scale $h$. This
leads to the following expression
\begin{equation}\label{tI}
{\bar{g}} = g({h}^\prime) = g(h) + {\mathcal{I}}\left(g(h),
\frac{h^\prime}{h}\right)
\end{equation}
where
\begin{eqnarray}
\hspace{-0.5cm}{\mathcal{I}}\left(g(h), \frac{h^\prime}{h}\right)
&=& g \left [ \left (\frac{h^\prime}{h}\right )^{\epsilon} -1
\right] - A_1 \frac{g^2}{\epsilon^2} \left\{ \epsilon + A_1 g
\left[ \left (\frac{h^\prime}{h}\right )^{\epsilon} -1
\right]\right\} \left (\frac{h^\prime}{h}\right
)^{\epsilon}\nonumber\\&\times &\left[ \left
(\frac{h^\prime}{h}\right )^{\epsilon} -1 \right] - A_2
\frac{g^3}{2 \epsilon} \left( 1 - \epsilon C \right) \left
(\frac{h^\prime}{h}\right )^{\epsilon} \left[ \left
(\frac{h^\prime}{h}\right )^{2\epsilon} -1 \right] .
\end{eqnarray}
Simple algebra next shows that to the appropriate order in $g$ the
following identity holds
\begin{equation}
\frac{d }{d\ln {h}} {\mathcal{I}}\left(g(h),
\frac{h^\prime}{h}\right) = - \beta(g(h)) .
\end{equation}
This means that Eqs \eqref{bart} and \eqref{tI} are indeed
identical expressions. Similarly, by starting from Eq.
\eqref{3.10d}, i.e.
\begin{equation}
z_h^\prime = z_h^0 \left( 1 + \left( A_1 - \frac{B}{A_1} \right)
\frac{g_0 (h^\prime)^{\epsilon}}{\epsilon} - \left( A_1 -
\frac{B}{A_1} \right) \frac{B}{A_1} \; \frac{g^2_0
(h^\prime)^{2\epsilon}}{\epsilon^2}  \right) ,
\end{equation}
one obtains the following result for the observable quantity
$z_h^\prime$
\begin{equation}\label{zJ}
z_h^\prime = z_h (h^\prime) = z_h (h) +  {\mathcal J} \left (g(h),
\frac{h^\prime}{h}\right )
\end{equation}
where
\begin{equation}
{\mathcal J} \left (g(h), \frac{h^\prime}{h}\right )= z_h
(h)\left( A_1 - \frac{B}{A_1} \right) \frac{g}{\epsilon} \left[ 1
- \frac{B}{A_1} \frac{g}{\epsilon}
\left(\frac{h^\prime}{h}\right)^{\epsilon} \right] \left[
\left(\frac{h^\prime}{h}\right)^{\epsilon} -1 \right] .
\end{equation}
Differentiating with respect to $\ln h$ leads to the following
result
\begin{equation}
\frac{d}{d\ln {h}}{\mathcal J} \left (g(h),
\frac{h^\prime}{h}\right ) = z_h (h)\gamma_h(g(h))
\end{equation}
which means that Eqs \eqref{barz} and \eqref{zJ} are identically
the same as well.
%%%%%%%%%%%%%%%%%%%%%%%%%%%%%%%%%%%%%%%%%%%%%%%%%%%%%%%%%%%%%%%%%%%%%%
%         APPENDIX A
%
%%%%%%%%%%%%%%%%%%%%%%%%%%%%%%%%%%%%%%%%%%%%%%%%%%%%%%%%%%%%%%%%%%%%
\section{\label{AppA} Appendix. Calculation of $\langle
S_h\rangle$}

In order to perform evaluation of $\langle S_h\rangle$ for general
instanton solution $T_0^{-1}R^{-1}\Lambda R T_0$ it is convenient
to use for matrix $Q$ the parameterization \eqref{top1}. Then we
have
\begin{equation}\label{s1}
S_h = z_h \int\limits_{\eta\theta} \tr  \Lambda_h T_0^{-1}R^{-1}q
R T_0,
\end{equation}
Expanding the matrix $q$ to the second order in $w$ (see
Eq.\eqref{qpar}), we obtain
\begin{equation}\label{s2}
S_h=S_h^{(0)}+S_h^{(1)}+S_h^{(2)}.
\end{equation}
Here the first term
\begin{equation}\label{s3}
S_h^{(0)}=z_h \int\limits_{\eta\theta} \tr \Lambda_h
T_0^{-1}R^{-1}\Lambda R T_0
\end{equation}
is the classical value of $S_h$. Next one
\begin{equation}\label{s4}
S_h^{(1)}=z_h \int\limits_{\eta\theta} \tr \Lambda_h T^{-1}R^{-1}w
R T,
\end{equation}
contains linear in $w$ terms. The last term in Eq.\eqref{s1} is as
follows
\begin{equation}\label{s5}
S_h^{(2)}=-\frac{z_h}{2} \int\limits_{\eta\theta} \tr \Lambda_h
T^{-1}R^{-1}\Lambda w^2 R T.
\end{equation}

\textit{Linear terms.} The matrix $(R^{-1}w R)^{\alpha\beta}$ has
the following form in the retarded-advanced space for different
values of $\alpha$ and $\beta$. If $\alpha =2,\cdots, n$ and
$\beta=2,\cdots,m$, then
\begin{equation}\label{RWR1}
(R^{-1}w R)^{\alpha\beta} = \begin{pmatrix}
 0 & \bar{\Phi}_{0,0}^{(0)} v^{\alpha\beta} \\
 \Phi_{0,0}^{(0)} v^{\dagger\alpha\beta}  &0
\end{pmatrix}.
\end{equation}
For $\alpha \neq 1$ we find
\begin{equation}\label{RWR2}
(R^{-1}w R)^{\alpha 1}=\sqrt{2\pi} \begin{pmatrix}
-\bar{\Phi}_{1,0}^{(1)} v^{\alpha 1} & \bar{\Phi}_{1,-1}^{(1)} v^{\alpha 1} \\
 \Phi_{1,-1}^{(1)} v^{\dagger \alpha 1}  &\Phi_{1,0}^{(1)} v^{\dag\alpha 1}
\end{pmatrix}
\end{equation}
and analogous for $(R^{-1}w R)^{1\alpha}$. Finally,
\begin{equation}\label{RWR3}
(R^{-1}w R)^{11} = \sqrt{\frac{4\pi}{3}} \begin{pmatrix}
-\frac{1}{\sqrt{2}}\bar{\Phi}_{1,-1}^{(2)} v^{11} -
\frac{1}{\sqrt{2}}\Phi_{1,-1}^{(2)}
v^{\dag 11}& \bar{\Phi}_{1,-2}^{(2)} v^{11} - \Phi_{1,0}^{(2)} v^{\dag 11}\\
- \bar{\Phi}_{1,0}^{(2)} v^{11}+\Phi_{1,-2}^{(2)} v^{\dag 11} &
\frac{1}{\sqrt{2}}\bar{\Phi}_{1,-1}^{(2)} v^{11} +
\frac{1}{\sqrt{2}}\Phi_{1,-1}^{(2)} v^{\dag 11}
\end{pmatrix}.
\end{equation}
From Eqs\eqref{RWR1}-\eqref{RWR3} we see that after integration
over $\eta$, $\theta$ the zero modes survive only.

\textit{Quadratic terms.} The fluctuations $w$ the quadratic term
\eqref{s5} ought to be considered as the massive modes. After
averaging over $w$ we obtain
\begin{equation}\label{s6}
\langle S_h^{(2)}\rangle = -\frac{z_h}{2} \int\limits_{\eta\theta}
\tr  \Lambda_h T^{-1}R^{-1}\Lambda G R T,
\end{equation}
where the matrix $G$ is diagonal in the replica and
retarded-advanced spaces
\begin{equation}\label{G}
G^{\alpha\beta}_{pp^{\prime}}= \delta^{\alpha\beta}_{pp^\prime}
G^{(\alpha)}_p.
\end{equation}
The diagonal elements $G^{\alpha}_p$ are defined as
\begin{equation}\label{s7}
G^{(\alpha)}_1=
  \begin{cases}
    \displaystyle\frac{m-1}{O^{(1)}}+\frac{1}{O^{(2)}}, & \alpha=1, \\
    \displaystyle\frac{m-1}{O^{(0)}}+\frac{1}{O^{(1)}}, & \alpha=2,\cdots n.
  \end{cases}
\end{equation}
\begin{equation}\label{s7_1}
G^{(\alpha)}_{-1}=
  \begin{cases}
    \displaystyle\frac{n-1}{O^{(1)}}+\frac{1}{O^{(2)}}, & \alpha=1, \\
    \displaystyle\frac{n-1}{O^{(0)}}+\frac{1}{O^{(1)}}, & \alpha=2,\cdots m.
  \end{cases}
\end{equation}
Hence the matrix $R^{-1}\Lambda GR$ can be written as
\begin{equation}\label{mat1}
R^{-1}\Lambda G R = \Lambda G + (R^{-1}\Lambda R-\Lambda)
\frac{G_1^{(1)}+G^{(1)}_{-1}}{2}.
\end{equation}
Thus the renormalization due to the quadratic fluctuations yields
\begin{gather}
\langle S_h\rangle  = z_h \int\limits_{\eta\theta}\left
(1-\frac{2}{\sigma_{xx}} \frac{G_1^{(1)}+G_{-1}^{(1)}}{2}\right
 ) \tr \Lambda_h T_0^{-1}(R^{-1}\Lambda R-\Lambda)
 T_0 \notag\\
+ z_h \int\limits_{\eta\theta} \tr \Lambda_h T_0^{-1}\Lambda \left
(1_{m+n}-\frac{2}{\sigma_{xx}} G\right ) T_0.\label{s11}
\end{gather}

\textit{Instanton vacuum.} Now we split the matrix $T_0\in U(m+n)$
as
\begin{equation}\label{split1}
T_0=W t_0,\qquad W\in U(m)\times U(n)
\end{equation}
and integrate over $W$ in order to restore the $U(m)\times U(n)$
invariance. Then we obtain
\begin{eqnarray}
\langle S_h\rangle  \to &-& z_h \frac{1}{m
n}\int\limits_{\eta\theta} e^2_0 \left (1-\frac{2}{\sigma_{xx}}
\frac{G_1^{(1)}+G_{-1}^{(1)}}{2}\right
 ) \tr\Lambda_h t_0^{-1} \Lambda X t_0 \notag\\
&+& z_h \int\limits_{\eta\theta} \tr\Lambda_h t_0^{-1}\Lambda
\Bigl [1_{m+n}-\frac{1}{m n \sigma_{xx}} \notag \\ &\times &
\left(\frac{1}{O^{(2)}}+\frac{m+n-2}{O^{(1)}}+\frac{(m-1)(n-1)}{O^{(0)}}\right
)  X \Bigr ]t_0,\label{s11.1}
\end{eqnarray}
where
\begin{equation}\label{X}
X = (m+n)1_{m+n}+(m-n) \Lambda.
\end{equation}
By using the traceless nature of the matrix $\Lambda_h$, we find
\begin{gather}
\langle S_h\rangle  = - z_h  \frac{m+n}{2mn}\left [1
-\frac{1}{4\pi\sigma_{xx}}\int\limits_{\eta\theta}\left
(\frac{2}{O^{(2)}}+\frac{m+n-2}{O^{(1)}}\right ) \right ]
\int\limits_{\eta\theta} \tr\Lambda_h t_0^{-1}\Lambda t_0 \notag\\
+ z_h\left [ 1  - \frac{1}{4\pi m n
\sigma_{xx}}\int\limits_{\eta\theta}\left
(\frac{1}{O^{(2)}}+\frac{m+n-2}{O^{(1)}}+\frac{(m-1)(n-1))}{O^{(0)}}\right
) \right ]  \notag \\ \times \int\limits_{\eta\theta} \tr\Lambda_h
t_0^{-1}\Lambda t_0.\label{s11.2}
\end{gather}

\textit{Trivial vacuum.} The same procedure of taking into account
the fluctuations about the trivial vacuum $\Lambda$ instead of
Eq.\eqref{s11} leads eventually to
\begin{equation}\label{s12}
  \langle S_h\rangle_0 = z_h \int\limits_{\eta\theta} \tr\Lambda_h T_0^{-1}\Lambda \left
(1_{m+n}-\frac{1}{\sigma_{xx}}\frac{1}{O^{(0)}} X \right ) T_0,
\end{equation}

Now it is worth mentioning that due to the traceless nature of the
matrix $\Lambda_h$ the second part of the matrix $X$ (proportional
to $\Lambda$) does not contribute to the $\langle S_h\rangle_0$.
Hence, we obtain
\begin{equation}\label{s12.1}
 \langle S_h\rangle_0 = z_h \int\limits_{\eta\theta} \left (1 - \frac{m+n}{\sigma_{xx}}\frac{1}{O^{(0)}}
\right ) \tr\Lambda_h T_0^{-1}\Lambda T_0.
\end{equation}
Finally, we find
\begin{equation}\label{s12.2}
  \langle S_h\rangle_0 = z_h \left (1 - \frac{m+n}{4\pi \sigma_{xx}}
\int\limits_{\eta\theta}\frac{1}{O^{(0)}} \right
)\int\limits_{\eta\theta}  \tr\Lambda_h T_0^{-1}\Lambda T_0.
\end{equation}

%%%%%%%%%%%%%%%%%%%%%%%%%%%%%%%%%%%%%%%%%%%%%%%%%%%%%%%%%%%%%%%%%%%%%%
%  APPENDIX B
%
%%%%%%%%%%%%%%%%%%%%%%%%%%%%%%%%%%%%%%%%%%%%%%%%%%%%%%%%%%%%%%%%%%%%%%%
\section{\label{AppB} Appendix. Matrix elements}

Let us define matrix elements of a function $f(\eta,\theta)$ as
\begin{equation}\label{APP.ME1}
{}_{(a)}\left\langle J,M|f(\eta,\theta )|M^{^{\prime
}},J^{^{\prime }}\right\rangle _{(b)} = \int\limits_{\eta\theta}
\Phi_{J,M}^{(a)}(\eta,\theta )f(\eta,\theta)\bar{\Phi}
_{J^{^{\prime}},M^{^{\prime}}}^{(b)}(\eta,\theta),\notag
\end{equation}
where $a,b=0,1,2$. By using the following relation for the
Jacobi polynomials~\cite{GR}
\begin{equation}\label{rel1}
(2n+\alpha+\beta) P_n^{(\alpha-1,\beta)}(x)=(n+\alpha+\beta)
P_n^{(\alpha,\beta)}(x) - (n+\beta) P_n^{(\alpha,\beta)}(x),
\end{equation}
and the normalization condition
\begin{gather}\label{rel2}
\int\limits_{-1}^1 d x (1-x)^{\alpha}(1+x)^{\beta}
P_n^{(\alpha,\beta)}(x) P_m^{(\alpha,\beta)}(x)\hspace{4cm} \\
\hspace{2cm} =\delta_{n,m}2^{\alpha+\beta+1}
\frac{\Gamma(\alpha+n+1)\Gamma(\beta+n+1)}{(\alpha+\beta+2n+1)
\Gamma(n+1)\Gamma(\alpha+\beta+n+1)},\notag
\end{gather}
we find the following results for diagonal matrix elements of functions $e_0^2$ and $e_0^4$
\begin{eqnarray}
\hspace{3cm}{}_{(1)}\left \langle J,M | e_0^2| M,J\right \rangle_{(1)}
&=& \frac{1}{2} \left [ 1+\frac{2M+1}{4J^2-1}\right ],\\
{}_{(2)}\left \langle J,M | e_0^2| M,J\right \rangle_{(2)}
&=& \frac{1}{2}\frac{M+1}{J(J+1)},\notag
\end{eqnarray}
\vspace{-0.6cm}
\begin{equation}
{}_{(2)}\left \langle J,M | e_0^4| M,J\right \rangle_{(2)} =
\frac{(M+1)(3M -J(J+1)(M-3))- J^2(J+1)^2}{2J(J+1)(2J-1)(2J+3)}.
\notag
\end{equation}
Performing the summation over $M$ we obtain
\begin{eqnarray}
\hspace{3cm}\sum\limits_{M=-J}^{J-1} {}_{(1)}\left \langle J,M | e_0^2|
M,J\right \rangle_{(1)} &=&\frac{2J}{2}, \notag\\
\sum\limits_{M=-J-1}^{J-1} {}_{(2)}\left \langle J,M | e_0^2|
M,J\right \rangle_{(2)} &=&\frac{2J+1}{2}.\notag \\
\sum\limits_{M=-J-1}^{J-1} {}_{(2)}\left \langle J,M | e_0^4|
M,J\right \rangle_{(2)} &=&\frac{2J+1}{3}.\notag
\end{eqnarray}

%%%%%%%%%%%%%%%%%%%%%%%%%%%%%%%%%%%%%%%%%%%%%%%%%%%%%%%%%%%%%%%%%%%%%%
%  APPENDIX C
%
%%%%%%%%%%%%%%%%%%%%%%%%%%%%%%%%%%%%%%%%%%%%%%%%%%%%%%%%%%%%%%%%%%%%%%%
\section{\label{AppC} Appendix. Renormalization around the trivial vacuum with the help of
Pauli-Villars procedure}

\textit{The $\sigma_{xx}$ renormalization.} In order to find the
renormalization of the $\sigma_{xx}$ conductivity we should
compute the average in Eq.\eqref{sdef}. Using the parameterization
$Q=T_0^{-1} q T_0$ with the global unitary matrix $T_0\in
U(m)\times U(n)$ and expanding the $q$ to the second order in $w$,
we obtain
%\begin{eqnarray}\label{sren1}
%\sigma_{xx}^{\prime} & = & \sigma_{xx}+\frac{\sigma_{xx}^2}{16 mn
%L^2}
%\int d \textbf{r}d \textbf{r}^{\prime} \\
%& \times & \tr \begin{pmatrix}
% \langle v(\textbf{r})\nabla v^\dag(\textbf{r})
%v(\textbf{r}^{\prime})\nabla v^\dag(\textbf{r}^{\prime})\rangle_0  & 0 \\
% 0 & \langle v^\dag(\textbf{r})\nabla v(\textbf{r})
%v^\dag(\textbf{r}^{\prime})\nabla v(\textbf{r}^{\prime})\rangle_0
%\end{pmatrix}.\notag
%\end{eqnarray}
% By using the Wick theorem, we obtain
\begin{gather}\label{sren2}
\sigma_{xx}^{\prime}=\sigma_{xx}+\frac{\sigma_{xx}^2}{16 mn} \int
d \textbf{r} \nabla^2
\sum\limits_{\alpha,\gamma=1}^m\sum\limits_{\beta,\delta=1}^n
\Bigl [
 \langle v^{\alpha\beta}(\textbf{r})v^{\dag\delta\alpha}(\textbf{r}^{\prime})\rangle_0
\langle v^{\gamma\delta}(\textbf{r}^{\prime})
v^{\dag\beta\gamma}(\textbf{r})\rangle_0 \\ +\langle
v^{\alpha\beta}(\textbf{r})v^{\dag\beta\gamma}(\textbf{r}^{\prime})\rangle_0
\langle v^{\gamma\delta}(\textbf{r}^{\prime})
v^{\dag\delta\alpha}(\textbf{r})\rangle_0 \Bigr ],\notag
\end{gather}
where a point $\textbf{r}^{\prime}$ can be chosen arbitrary since the averages
are depend only on the difference of the coordinates. Now by going
from $(x,y)$ to $(\eta,\theta)$ coordinates and performing the
averages, we find
\begin{equation}\label{sren3}
\sigma_{xx}^{\prime}=\sigma_{xx}-2\pi \beta_0\int\limits_{\eta\theta}
O^{(0)} \mathcal{G}_0(\eta\theta;\eta^{\prime}\theta^{\prime})
 \mathcal{G}_0(\eta^{\prime}\theta^{\prime};\eta\theta).
\end{equation}
Integrating over $\eta,\theta$ and introducing the
Pauli-Villars masses as above, we leads to the
following result
\begin{gather}\label{sren4}
\sigma_{xx}^{\prime}=\sigma_{xx}-2\pi \beta_0\left [
\sum\limits_{J=1} \frac{1}{E_J^{(0)}} + \sum\limits_{f=1}^K \hat
e_f \sum\limits_{J=0}
\frac{E_J^{(0)}}{(E_J^{(0)}+\mathcal{M}^2_f)^2}\right ]\\
\times \sum \limits_{M=-J}^{J}
\Phi_{JM}^{(0)}(\eta^{\prime},\theta^{\prime})
\bar{\Phi}_{JM}^{(0)}(\eta^{\prime},\theta^{\prime}).\notag
\end{gather}
It is worth mentioning that the Jacobi polynomial
$P_{J-M}^{M,M}(\eta)$ is proportional to the Gegenbauer polynomial
$C_{J-M}^{M+1/2}(\eta)$. By using the summation theorem~\cite{GR}
\begin{gather}
C_J^{\lambda}(\cos\phi\cos\phi^{\prime}+z
\sin\phi\sin\phi^{\prime})=
\frac{\Gamma(2\lambda-1)}{\Gamma^2(\lambda)}
\sum\limits_{M=0}^{J}
\frac{2^{2M}\Gamma(J-M+1)}{\Gamma(J+M+2\lambda)}\Gamma^2(M+\lambda)
 \notag\\ \times(2M+2\lambda-1) \sin^{M}\phi\sin^M
\phi^{\prime}
C^{M+\lambda}_{J-M}(\cos\phi)C^{M+\lambda}_{J-M}(\cos\phi^{\prime})
C_M^{\lambda-1/2}(z)\label{sumth}
\end{gather}
with $z=1$ and $\lambda=1/2$, we find that the projection operator
\begin{equation}\label{proj}
\sum \limits_{M=-J}^{J} \Phi_{JM}^{(0)}(\cos\phi,\theta)
\bar{\Phi}_{JM}^{(0)}(\cos\phi^{\prime},\theta) =
\frac{2J+1}{4\pi} C^{1/2}_{J} \left
(\cos(\phi-\phi^{\prime})\right ).
\end{equation}
Since $C^{1/2}_J(1)=1$, we obtain
\begin{equation}\label{sren5}
\sigma_{xx}^{\prime}=\sigma_{xx}-\frac{\beta_0}{2}\lim\limits_{\Lambda\to
\infty} \Biggl [ \sum\limits_{J=3/2}^\Lambda
\frac{2J(J^2-\frac{1}{4})}{(J^2-\frac{1}{4})^2}  +
\sum\limits_{f=1}^K \hat e_f \sum\limits_{J=1/2}^\Lambda \frac{
2J(J^2-\frac{1}{4})}{(J^2-\frac{1}{4}+\mathcal{M}^2_f)^2}\Biggr ].
\end{equation}
Finally, evaluation of the sums above yields
\begin{equation}\label{sreg5}
\sigma_{xx}^{\prime}=\sigma_{xx}-\frac{\beta_0}{2}\left
(Y^{(0)}_{\rm reg} + 1 \right ) = \sigma_{xx}\left (1
-\frac{\beta_0}{\sigma_{xx}} \ln \mathcal{M}e^\gamma\right )
\end{equation}

\textit{The $z_i$ renormalization.} The renormalized quantities
$z_{i}^{\prime}$ with $i=a,s,h$ are defined by Eqs~\eqref{rendef}.
Using the parameterization $Q=T_0^{-1} q T_0$ with the global
unitary matrix $T_0\in U(m)\times U(n)$ and expanding $q$ to the
second order in $w$, we find
\begin{equation}
z_{i}^{\prime} =  z_{i} \left ( 1 + \frac{\gamma^{(0)}_i\pi}{2 m n}
\sum
\limits_{\alpha=1}^{m}\sum\limits_{\beta=1}^n \langle v^{\alpha
\beta} v^{\dag\beta \alpha}\rangle_0\right ).\label{SfhP1}
\end{equation}
Here the average $\langle \cdots\rangle_0$ is defined with the
respect to the action $\delta S_\sigma^{(0)}$ (see Eqs ~\eqref{Sfh0}).
The averages yield
\begin{equation}
z_{i}^{\prime} =  z_{i} \left ( 1 + \frac{2\pi\gamma_i^{(0)}}{\sigma_{xx}}
\mathcal{G}_0(\eta\theta;\eta\theta)\right )
\end{equation}
By using Eq.\eqref{proj}, we find
\begin{equation}
z_{i}^{\prime}= z_{i} \left ( 1 +
\frac{\gamma_i^{(0)}}{2\sigma_{xx}} Y^{(0)}\right ).\label{SfhP2}
\end{equation}
With a help of the result \eqref{Y01} for the $Y_\textrm{reg}^{(0)}$ we finally obtain
\begin{equation}
z_{i}^{\prime} =  z_{i} \left ( 1 + \frac{\gamma_i^{(0)}}{\sigma_{xx}}
\ln \mathcal{M}e^{\gamma-1/2}\right ).\label{SfhP3}
\end{equation}

%%%%%%%%%%%%%%%%%%%%%%%%%%%%%%%%%%%%%%%%%%%%%%%%%%%%%%%%%%%%%%%%%%%%%%
%THE BIBLIOGRAPHY
%
%%%%%%%%%%%%%%%%%%%%%%%%%%%%%%%%%%%%%%%%%%%%%%%%%%%%%%%%%%%%%%%%%%%%%%


\begin{thebibliography}{100}

\bibitem{QHE} A.M.M. Pruisken in: R.E. Prange and S.M. Girvin (Eds), The Quantum Hall effect,
Springer-Verlag, Berlin, 1987, p. 117.
\\

\bibitem{Pruisken4} A.M.M. Pruisken, Phys. Rev. Lett. 61 (1988) 1297
\\

\bibitem{LevineLibbyPruisken} H. Levine, S. Libby, and A.M.M.
Pruisken, Phys. Rev. Lett. 51 (1983) 1915
\\

\bibitem{Pruisken1} A.M.M. Pruisken, Nucl. Phys. B 235 (1984) 277,
Phys. Rev. B 32 (1985) 2636
\\

\bibitem{PruiskenBaranov} A.M.M. Pruisken and M.A. Baranov, Europhys. Lett.
31 (1995) 543
\\

\bibitem{WeiTsuiPalaanenPruisken} H.P. Wei, D.C. Tsui, M. Palaanen,
and A.M.M. Pruisken, Phys. Rev. Lett. 61 (1988) 1294
\\

\bibitem{SchaijkdeVisserOlsthoornWeiPruisken} R.T.F. van Schaijk,
A. de Visser, S.M. Olsthoorn, H.P. Wei, and A.M.M. Pruisken, Phys.
Rev. Lett. 84 (2000) 1567; D.T.N. de Lang, L.A. Ponomarenko, A. de
Visser, C. Possanzini, S.M. Olsthoorn, and A.M.M. Pruisken,
Physica E 12 (2002) 666; L.A. Ponomarenko, D.T.N. de Lang, A. de
Visser, D. Maude, B.N. Zvonkov, R.A. Lunin and A.M.M. Pruisken,
Physica E 22 (2004) 236
\\
\bibitem{PLPdV} A.M.M. Pruisken, D.T.N. de Lang, L.A.
Ponomarenko, A. de Visser, cond-mat/0109043
\\
\bibitem{Unify1} A.M.M. Pruisken, M.A. Baranov, and B. \v{S}kori\'{c}, Phys.
Rev. B 60 (1999) 16807
\\

\bibitem{Unify2} M.A. Baranov, A.M.M. Pruisken, and B. \v{S}kori\'{c}, Phys.
Rev. B 60 (1999) 16821
\\

\bibitem{Unify3} A.M.M. Pruisken, B. \v{S}kori\'{c}, and M.A. Baranov, Phys.
Rev. B 60 (1999) 16838
\\

\bibitem{Unify4} B. \v{S}kori\'{c}, and A.M.M. Pruisken, Nucl.
Phys. B 559 (1999) 637 and references therein.
\\

\bibitem{Unify5} M.A. Baranov, I.S. Burmistrov, and A.M.M. Pruisken, Phys.
Rev. B 66 (2002) 075317 and references therein.
\\

\bibitem{PruiskenBaranovVoropaev} A.M.M. Pruisken, M.A. Baranov, and M. Voropaev,
cond-mat/0206011
\\

\bibitem{PruiskenBaranovBurmistrov} A.M.M. Pruisken, M.A. Baranov, and I.S. Burmistrov,
cond-mat/0206012
\\

\bibitem{LargeN2} E. Witten, Nucl. Phys. B 149 (1979) 285
\\

\bibitem{LargeN3} I. Affleck, Nucl. Phys. B 162 (1980) 461;
ibid 171 (1980) 420
\\

\bibitem{LargeN4} S. Coleman, Aspects of Symmetry, University Press, Cambridge, 1989
\\

\bibitem{Wilczek} See e.g.
O. Heinonen (Ed.), Composite Fermions, World Scientific, 1998
\\

\bibitem{RM} S. Edwards and P.W. Anderson, J. Phys. F 5 (1975) 965
\\

\bibitem{Pruisken2} A.M.M. Pruisken, Nucl. Phys. B 285 (1987) 719
\\

\bibitem{Pruisken3} A.M.M. Pruisken, Nucl. Phys. B 290 (1987) 61
\\

\bibitem{Thouless} D.J. Thouless in: R. Balian, R. Maynard, and G. Toulouse (Eds),
Ill-condensed Matter, North-Holland/World Scientific, 1978, p.1.
\\

\bibitem{Mishandling} H. Weidenm\"{u}ller, Nucl. Phys. B 290 (1987) 87;
 S. Xiong, N. Read, and A. Douglas Stone, Phys. Rev. B 56 (1997) 3982
\\

\bibitem{PruiskenBaranovVoropaevN} A.M.M. Pruisken, M.A. Baranov, and M. Voropaev,
cond-mat/0101003
\\

\bibitem{O3} See e.g. A. B. Zamolodchikov and Al. B. Zamolodchikov
Nucl. Phys. B 379 (1992) 602
\\

\bibitem{LargeN1} A.D'Adda, P.Di Vecchia, and M. Luescher, Nucl.
Phys. B 146 (1978) 63
\\

\bibitem{t'Hooft1} see e.g. H. Levine, S. Libby, and A.M.M.
Pruisken, Nucl. Phys. B 240 (1984) 30; ibid 240 (1984) 49; 240
(1984) 71 and references therein.
\\

\bibitem{Multifractality} C. Castellani and L. Peliti, J. Phys. A
19 (1986) L429; F. Wegner, Nucl. Phys. B 280 (1987) 210
\\

\bibitem{Wegner1} F. Wegner, Z. Phys. B 36 (1979) 209
\\

\bibitem{Pruisken5} A.M.M. Pruisken, Phys. Rev. B 31 (1985) 416
\\

\bibitem{NumNu1} J.T. Chalker and P.D. Coddington, J. Phys. C 21 (1988) 2665
\\

\bibitem{NumNu2} B. Huckestein and B. Kramer, Phys. Rev. Lett. 64 (1990) 1437
\\

\bibitem{NumNu3} B. Mieck, Europhys. Lett. 13 (1990) 453
\\

\bibitem{NumNu4} Y. Huo and R.N. Bhatt, Phys. Rev. Lett. 68 (1992) 1375
\\

\bibitem{NumNu5} T. Ando, J. Phys. Soc. Japan 61 (1992) 415
\\

\bibitem{NumNu6} D.-H. Lee, Z. Wang and S. Kivelson, Phys. Rev.
Lett. 70 (1993) 4130
\\
\bibitem{NumGA1} J.T. Chalker and C.J. Daniell, Phys. Rev. Lett.
61 (1988) 593
\\

\bibitem{NumGA2} B. Huckestein and L. Schweitzer, Phys.
Rev. Lett. 72 (1994) 713
\\

\bibitem{NumPi2} S. Hikami, Prog. Theor. Phys. 76 (1986) 1210
\\

\bibitem{NumLW} D.-H. Lee and Z. Wang, Phil.
Mag. Lett. 73 (1996) 145
\\

\bibitem{NumAlpha01} W. Pook and M Jan$\ss$en, Z. Phys. B
82 (1991) 295
\\

\bibitem{NumAlpha02} B. Huckestein, B. Kramer, L. Schweitzer, Surf. Sci. 263 (1992) 125
\\

\bibitem{NumY1} J.T. Chalker and J.P.G. Eastmond, unpublished;
B. Huckestein, Phys. Rev. Lett. 72 (1994) 1080
\\

\bibitem{NumY2} F. Evers, A. Mildenberger,
and A.D. Mirlin, Phys. Rev. B 64 (2001) 241303
\\

\bibitem{Interactions} A.M.M. Pruisken and I.S. Burmistrov, in
preparation.
\\

\bibitem{Affleck1} I. Affleck, Nucl. Phys. B 191 (1981) 429;
M. Nielsen and N.K. Nielsen, Phys. Rev. D 61 (2000) 105020  and
references therein.
\\

\bibitem{BZG} E. Br\'{e}zin, S. Hikami and J. Zinn-Justin, Nucl.
Phys. B 165 (1980) 528
\\

\bibitem{HofWegner} D. H$\ddot{o}$f and F. Wegner, Nucl. Phys. B
275 (1986) 561; F. Wegner, ibid 280 (1987) 193; ibid 280 (1987)
210
\\
\bibitem{ReadShankar} R. Shankar and N. Read, Nucl. Phys. B 336
(1990) 457
\\
\bibitem{PruiskenWang} A.M.M. Pruisken and Z. Wang, Nucl. Phys. B
322 (1989) 721
\\
\bibitem{GR} I.S. Gradshteyn and I.M. Ryzhik, Table of integrals,
series, and products, 4th ed., Academic Press, 1980

%....................................................................

\end{thebibliography}
\end{document}